\documentclass[11pt]{article}

\usepackage{amsmath,amssymb,epsfig}

\usepackage{sint}

\font\sixrm=cmr6

\newcommand{\rmO}{{\rm O}}
\newcommand{\rmd}{{\rm d}}
\newcommand{\rme}{{\rm e}}
\newcommand{\msbar}{{\rm \overline{MS\kern-0.14em}\kern0.14em}}
\newcommand{\ms}{{\rm MS}}
\newcommand{\Zr}{Z_{{\hbox{\sixrm R}}}}
\newcommand{\Zor}{Z_{1{\hbox{\sixrm R}}}}

\begin{document}

\begin{titlepage}

  \begin{flushright}
    MPP-2009-40\\
    May 2009
  \end{flushright}

  \vskip 0.20 true cm

  \begin{center}
    {\Large\bf 
      The puzzle of apparent linear lattice artifacts in the 
      2d non-linear $\sigma$--model and Symanzik's solution} 
  \end{center}
  \vskip 1 true cm
  \centerline{\large Janos Balog}
  \vskip1ex
  \centerline{Research Institute for Particle and Nuclear Physics}
  \centerline{1525 Budapest 114, Pf. 49, Hungary}
  \vskip 1 true cm
  \centerline{\large Ferenc Niedermayer}
  \vskip1ex
  \centerline{Institute for Theoretical Physics, University of Bern}
  \centerline{CH-3012 Bern, Switzerland}
  \vskip 1 true cm
  \centerline{\large Peter Weisz}
  \vskip1ex
  \centerline{Max-Planck-Institut f\"ur Physik}
  \centerline{F\"ohringer Ring 6, D-80805 M\"unchen, Germany}
  \vskip 1 true cm
  \centerline{\bf Abstract}
  \vskip 1.0ex
  Lattice artifacts in the 2d O($n$) non-linear $\sigma$--model are 
  expected to be of the form $\rmO(a^2)$, and hence it was 
  (when first observed) disturbing that some quantities in the O(3) 
  model with various actions show parametrically stronger cutoff dependence, 
  apparently $\rmO(a)$, up to very large correlation lengths. 
  In a previous letter \cite{BNWletter} we described the solution 
  to this puzzle.
  Based on the conventional framework of Symanzik's effective action, 
  we showed that there are logarithmic corrections to the $\rmO(a^2)$ 
  artifacts which are especially large ($\ln^3 a$) for $n=3$ 
  and that such artifacts are consistent with the data. 
  In this paper we supply the technical details of this computation.
  Results of Monte Carlo simulations using various lattice actions for
  O$(3)$ and O$(4)$ are also presented.

  \vfill
  \eject

\end{titlepage}

\section{Introduction} \label{intro.tex}

In a previous letter \cite{BNWletter} we presented results on 
logarithmic corrections to $\rmO(a^2)$ lattice artifacts for a class
of lattice actions for the non-linear $\rmO(n)$ sigma-model in two
dimensions. It is the purpose of this paper to supply the technical
details of the computation. The main results, which are summarized in
subsection~3.5, are that the generic leading artifacts are of the form
$a^2\left[\ln(a^2)\right]^{n/(n-2)}$, and this result together with 
the next-to-leading expressions describe well the lattice artifacts in
the step scaling function \cite{BNWletter}, 
which are for $n=3$ in a large range of the
cutoff apparently of the form $\rmO(a)$ \cite{HHNSW}. 
The goal of our work was indeed to explain this 
long-standing scientific puzzle which was 
mentioned by Hasenfratz in his lattice plenary talk in 2001 \cite{PeterH}.

Most of our knowledge concerning renormalization of quantum field theories
stems from perturbation theory. Although there are no rigorous proofs 
in general, many of the results are structural and 
hence considered to carry over to non-perturbative formulations. 
Indeed there is supporting evidence from various studies, e.g. of soluble 
models in 2 dimensions and of $1/n$ expansions of some theories.
The same situation holds concerning cutoff artifacts in lattice 
regularized theories. 

Thus we start our paper with perturbative considerations.
In section~2 we briefly summarize perturbative renormalization 
of the sigma model in the framework of dimensional regularization,
including the renormalization of isoscalar composite operators of 
dimension 4. Next we discuss a large class of lattice regularizations.
We summarize the results for connected 2-- and 4--point functions
to 1-loop order and also the relations of the corresponding renormalized 
functions to those of the dimensionally regularized theory.

In the early 80's Symanzik was working on the nature of lattice artifacts,
in particular with respect to his improvement program 
\cite{Sym1,Sym2,BMMS,SymCarg,SymBerl}.
In this paper the general theory is not discussed; we only consider 
in section~3 Symanzik's theory applied to the 2-dimensional O$(n)$ 
$\sigma$-model\footnote{Concerning Symanzik's program for QCD, 
see e.g. \cite{fen} and references therein.}. 
Nevertheless the spirit of the general theory
can already be understood by studying this example.
Symanzik's main conclusion is that leading artifacts
are summarized in an effective action.

In this framework generic lattice artifacts are, in particular for
asymptotically free (or trivial) theories, expected to be 
integer powers in the lattice spacing $\rmO(a^p)\,,p=1,2,\dots$ up to 
possible multiplicative logarithmic corrections. 
In particular this framework explains
why in the 2-dimensional $\sigma$-model quadratic artifacts are expected.
To obtain the effective action in lowest orders of perturbation
theory we compute the leading lattice artifacts in the 
2- and 4-point functions to 1-loop order and express them
in terms of insertions of composite operators of dimension 4 in
the continuum. This is a rather lengthy computation and
thus many technical details are relegated to the Appendices.

In section~4 we reanalyze the presently available Monte Carlo
data on the step scaling function for cases $n=3$ and $n=4$.
In particular for the case $n=3$ we study three different lattice actions, 
and see how the particular logarithmic corrections predicted by 
Symanzik's theory solves the puzzle of apparent linear artifacts.
Similar conclusions that logarithmic corrections may 
explain the puzzle have been reached previously in studies of the 
$\rmO(n)$ model in the first orders of the $1/n$ expansion 
\cite{KLW,BKLW,CarPel}. 
In Appendix E we describe a modified version of Hasenbusch's improved
estimator which was essential to obtain a sufficiently small error for
small $a/L$ values.

There is a vast literature on the subject of this paper, however we feel 
that this technical report is not the appropriate place 
to properly review all important contributions. Hence we restrict 
ourselves here to citing a selection of papers where the reader can find 
further references.
2-dimensional O$(n)$ spin systems, including the Ising model and the large
$n$ limit, are reviewed in \cite{PeVi}. Various aspects (on the lattice and
in the continuum) of the exactly solvable $n\to\infty$ limit, using a
number of different lattice actions, are reviewed in \cite{CaRo}. Symanzik's
improvement program for the 2-dimensional O$(n)$ models are studied
in \cite{AlPe,CaPe} and some exact results about the lattice artifacts in the
large $n$ limit are given in \cite{CarPel}.

\section{The non-linear O($n$) sigma model in two dimensions}
\label{Sigma_model.tex}

The non-linear O($n$) sigma model describes the interaction of
spin fields $S^a(x)\,,a=1,\dots,n$ with the constraint $S(x)^2=1$
in two space-time dimensions. Here we consider the space-time to be
Euclidean. Formally the Lagrangian is that of a free massless field
and the interaction derives entirely from the constraint. 
The field theory requires an ultra-violet regularization; 
in the next subsection we will consider dimensional regularization  
and then in subsect.~2.2 a class of lattice regularizations. 

\subsection{Dimensional regularization} \label{dimreg.tex}

The action of the O$(n)$ model in $D=2-\varepsilon$ dimensions (including
source terms) is
\begin{equation}
  {\cal A}=\frac{1}{g_0^2}\int\rmd^Dx\,\left\{\frac{1}{2}\partial_\mu S\cdot
    \partial_\mu S-I\cdot S\right\}\,,
  \label{A}
\end{equation}
where $g_0$ is the bare coupling. To satisfy the constraint it is usual
to  parameterize the bare spin field by
\begin{equation}
  S^i=g_0\pi^i,\qquad i=1,2,\dots,n-1;\qquad\qquad
  S^n=\sigma=\sqrt{1-g_0^2\pi^2}\,. 
\end{equation}
The source dependent action (\ref{A}) appears in the generating functional
\begin{equation}
  {\cal Z}[I]=\int\left({\cal D}\pi\right)\,\rme^{-{\cal A}}\,,
  \label{genfunc}
\end{equation}
which can be used to obtain bare correlation functions of the field $S^a(x)$
using the formula
\begin{equation}
  {\cal G}^{a_1\dots a_r}(x_1,\dots,x_r)=g_0^{2r}{\cal Z}^{-1}[I_0]
  \frac{\delta}{\delta I^{a_1}(x_1)}\dots \frac{\delta}{\delta I^{a_r}(x_r)}
  \Big\vert_{I_0}\,{\cal Z}[I]\,,
  \label{corrfun}
\end{equation}
where the functional derivative is taken at
\begin{equation}
  I^a_0(x)=m^2\delta^{an}\,,
\end{equation}
i.e. a mass term (external magnetic field) is introduced to avoid infrared
singularities. 
As proven by David \cite{David}, O($n$) invariant correlators
are infrared finite order by order in perturbation theory and hence for
these the limit $m\to0$ can be taken at the end of the calculation.

In their seminal paper \cite{BZJG} Br\'ezin, Zinn-Justin and Le Guillou
prove the renormalizability of the O$(n)$ model using functional methods.
In this paper they show that the generating functional ${\cal Z}^{-1}[I_0] 
{\cal Z}[I]$ is finite as function of the renormalized quantities
$j^a(x),g,\mu,m_R$ if we write
\begin{equation}
  \begin{split}
    I^a(x)   & = Z_1(g,\varepsilon)Z^{-1/2}(g,\varepsilon)g^2j^a(x)+I^a_0,\\
    g_0^2 & = \mu^\varepsilon Z_1(g,\varepsilon)g^2\,,\\
    m^2 & = Z_1(g,\varepsilon)Z^{-1/2}(g,\varepsilon)m^2_R\,,
  \end{split}
\end{equation}
where the renormalization constants contain only pole terms,
\begin{equation}
  \begin{split}
    Z_1(g,\varepsilon) & = 1-
    \frac{2\beta_0g^2}{\varepsilon}-
    \frac{\beta_1g^4}{\varepsilon}
    +\frac{4\beta_0^2g^4}{\varepsilon^2}+\dots,\\
    Z(g,\varepsilon) & = 1-
    \frac{\gamma_0g^2}{\varepsilon}-
    \frac{\gamma_1g^4}{2\varepsilon}+
    \frac{g^4(\gamma_0^2+2\beta_0\gamma_0)}{2\varepsilon^2}+\dots
  \end{split}
\end{equation}
with
\begin{equation}
  \begin{split}
    \beta_0 & = \frac{n-2}{4\pi}\,,\\
    \gamma_0 & = \frac{n-1}{2\pi}\,,
  \end{split}
  \qquad\qquad\quad
  \begin{split}
    \beta_1 & = \frac{n-2}{8\pi^2}\,,\\
    \gamma_1 & = 0\,,
  \end{split}
  \qquad\qquad
  \begin{split}
    \beta_2 & = \frac{(n+2)(n-2)}{64\pi^3}\,,\\
    & \phantom{\gamma_1}
  \end{split}
\end{equation}
where the 3-loop coefficient $\beta_2$ will appear in (\ref{213}) below.
Functional derivation with respect to the source $j^a(x)$ gives renormalized
correlation functions, i.e. correlation functions of the renormalized fields
$S^a_R=Z^{-1/2}S^a$. The relation between bare and renormalized correlation
functions is given by
\begin{equation}
  \widehat{\cal G}^X_{(R)}(g,\mu,\varepsilon)=Z^{-r/2}(g,\varepsilon)
  {\cal G}^X(g_0,\varepsilon)\,,
\end{equation}
where the upper index $X$ symbolizes any $r$-point
correlation function (in $x$-space or in Fourier space). We assume that
$X$ is O$(n)$ invariant and that the $m\to0$ limit has been taken.
Finiteness means that the limit
\begin{equation}
  {\cal G}^X_{(R)}(g,\mu)=
  \widehat{\cal G}^X_{(R)}(g,\mu,0)
\end{equation}
exists and defines the renormalized correlation function in two dimensions.

The renormalization group (RG) equations express the fact that the bare
correlation functions are independent of the renormalization scale $\mu$.
In terms of the renormalized correlation functions this is expressed as
\begin{equation}
  \left\{{\cal D}+\frac{r}{2}\gamma(g)\right\}{\cal G}^X_{(R)}(g,\mu)=0\,,
  \label{RG1}
\end{equation}
where the RG differential operator is
\begin{equation}
  {\cal D}=\mu\frac{\partial}{\partial\mu}+\beta(g)\frac{\partial}{\partial g}\,,
\end{equation}
and the RG beta and gamma functions are defined as
\begin{equation}
  \beta(g)=\frac{\varepsilon g}{2}-\frac{\varepsilon g}
  {2+g\frac{\partial\ln Z_1(g,\varepsilon)}{\partial g}}=
  -\beta_0g^3-\beta_1g^5-\beta_2g^7+\dots
\label{213}
\end{equation}
and
\begin{equation}
  \gamma(g)=\left\{\beta(g)-\frac{\varepsilon g}{2}\right\}
  \frac{\partial\ln Z(g,\varepsilon)}{\partial g}=
  \gamma_0g^2+\gamma_1g^4+\dots
\end{equation}
We also introduce the RG invariant $\Lambda$--parameter 
in the $\msbar$ scheme by the formula
\begin{equation}
  \Lambda_{\msbar}=\rme^{\frac{\gamma}{2}}\Lambda_{\ms}\,,\,\,\,\,\,\,
  \Lambda_{\ms}=\mu\left(
    2\beta_0g^2\right)^{-\chi}\rme^{-\frac{1}{2\beta_0g^2}}\rme^{k(g)}\,,
\end{equation}
where
\begin{equation}
  \chi=\frac{\beta_1}{2\beta_0^2}=\frac{1}{n-2} \,, \qquad\quad
  \gamma=\ln4\pi+\Gamma^\prime(1)\,,
\end{equation}
and
\begin{equation}
  k(g)=\int_0^g{\rm d}y\left\{\frac{2\chi}{y}-\frac{1}{\beta_0y^3}
    -\frac{1}{\beta(y)}\right\}=\rmO\left(g^2\right)\,.
\end{equation}
In the O$(n)$ model, instead of $\Lambda_{\overline{\rm MS}}$
we could also use $M$, the physical mass of the particles, since its
relation to the Lambda parameter is known \cite{HMN}:
\begin{equation}
  \frac{M}{\Lambda_{\overline{\rm MS}}}=\left(\frac{8}{{\rm e}}\right)^\chi
  \frac{1}{\Gamma(1+\chi)}\,.
  \label{MoverLambdaMS}
\end{equation}

\subsubsection{Local operators}

We now turn to the renormalization of local operators. In our Euclidean
framework operators ${\cal O}_i$ correspond to insertions into the 
generating functional:
\begin{equation}
  {\cal O}_i\quad\longrightarrow\quad{\cal Z}_i[I]=
  \int\left({\cal D}\pi\right)\,{\cal O}_i\,\rme^{-{\cal A}}.
  \label{opgenfunc}
\end{equation}
Beyond operators corresponding to local expressions of the O$(n)$ field
$S^a(x)$ and its derivatives we will also consider here operators 
${\cal O}_i$ depending on the source $I^a(x)$. 

Bare correlation functions with ${\cal O}_i$ operator insertion can be obtained
by using ${\cal Z}_i[I]$ in (\ref{corrfun}) in place of ${\cal Z}[I]$.
We will denote these symbolically as ${\cal G}^X_i$.

Operators are mixed with other operators under renormalization. 
Renormalized operators ${\cal O}_{i(R)}$ are given by 
\begin{equation}
  {\cal O}_{i(R)}=Z_{ij}(g,\varepsilon){\cal O}_j,
\end{equation}
which is a symbolical expression of the fact that
\begin{equation}
  {\cal Z}^{-1}[I_0]Z_{ij}(g,\varepsilon){\cal Z}_j[I]
\end{equation}
is finite in terms of $j^a(x),g,\mu$ and $m_R$. In terms of correlation
functions with operator insertion we have
\begin{equation}
  \widehat{\cal G}^X_{i(R)}(g,\mu,\varepsilon)=
  Z^{-r/2}(g,\varepsilon)Z_{ij}(g,\varepsilon){\cal G}^X_j(g_0,\varepsilon)
\end{equation}
and finiteness means that the limiting correlation functions 
\begin{equation}
  {\cal G}^X_{i(R)}(g,\mu)=\widehat{\cal G}^X_{i(R)}(g,\mu,0)
\end{equation}
exist. They satisfy the RG equation
\begin{equation}
  \left\{{\cal D}+\frac{r}{2}\gamma(g)\right\}{\cal G}^X_{i(R)}(g,\mu)
  +\nu_{ij}(g){\cal G}^X_{j(R)}(g,\mu)=0\,,
\end{equation}
where the anomalous dimension matrix is defined by
\begin{equation}
  \nu_{ij}(g)=Z_{is}(g,\varepsilon)\left(\beta(g)-\frac{\varepsilon g}{2}
  \right)\frac{\partial W_{sj}(g,\varepsilon)}{\partial g}
\end{equation}
with $W_{ij}$ the matrix inverse of $Z_{ij}$.

In perturbation theory (PT) we have
\begin{equation}
  Z_{ij}(g,\varepsilon)=\delta_{ij}-\frac{g^2}{\varepsilon}w_{ij}+
  \frac{g^4}{2\varepsilon}p_{ij}+\frac{g^4}{2\varepsilon^2}
  \left(w_{is}w_{sj}+2\beta_0w_{ij}\right)+\dots 
  \label{Zijgeneral}
\end{equation}
and
\begin{equation}
  \nu_{ij}(g)=-w_{ij}g^2+p_{ij}g^4+\dots
\end{equation}

\subsubsection{Dimension 4 operators}

Br\'ezin et al. prove in \cite{BZJG} that the following set of mass dimension 
four, O$(n)$ invariant, Lorentz scalar operators is closed under 
renormalization.
\begin{equation}
  \begin{split}
    {\cal O}_1 & = \frac{1}{8}\left(\partial_\mu S\cdot\partial_\mu S\right)^2\,,\\
    {\cal O}_2 & = \frac{1}{8}\left(\partial_\mu S\cdot\partial_\nu S\right)
    \left(\partial_\mu S\cdot\partial_\nu S\right)\,,\\
    {\cal O}_3 & = \frac{1}{2}\square S\cdot\square S\,,\\
    {\cal O}_4 & = \frac{1}{2}\alpha
    \partial_\mu S\cdot\partial_\mu S\,,\\
    {\cal O}_5 & = \frac{1}{8}\,\alpha^2\,,
  \end{split}
  \label{calOibasis}
\end{equation}
where
\begin{equation}
  \alpha=\frac{\square \sigma+I^n(x)}{\sigma}\,.
\end{equation}
Although the source dependent terms 
${\cal O}_4$ and  ${\cal O}_5$ look O$(n)$ non-invariant it is 
demonstrated in Appendix~A that in fact all (otherwise O($n$) 
invariant) correlators containing an insertion of these operators are 
O($n$) invariant. Br\'ezin et al. \cite{BZJG} showed that 
they must be included in the operator renormalization scheme
for consistency for $m\ne0$ and off shell. 
The one-loop mixing matrix for dimension four invariant
Lorentz scalars is calculated in the paper using this basis.
This result and further considerations on the 
$5\times 5$ operator renormalization problem 
are discussed in Appendix~A.  

In our work we also need to consider 
four-index symmetric tensor operators:
\begin{align}
  t_{\mu\nu\rho\sigma} & = 
  S\cdot\partial_\mu\partial_\nu\partial_\rho\partial_\sigma S\,,
  \label{ttensor}
  \\
  k_{\mu\nu\rho\sigma} & = \frac{1}{3}\Big\{
  \left(\partial_\mu S\cdot\partial_\nu S\right)
  \left(\partial_\rho S\cdot\partial_\sigma S\right)
  \nonumber\\
  & + \left(\partial_\mu S\cdot\partial_\rho S\right)
  \left(\partial_\nu S\cdot\partial_\sigma S\right)+
  \left(\partial_\mu S\cdot\partial_\sigma S\right)
  \left(\partial_\nu S\cdot\partial_\rho S\right)\Big\}\,.
  \label{ktensor}
\end{align}
In particular in the Symanzik effective Lagrangian 
we will encounter the following operators $A,B$ which are defined 
from the totally traceless parts $\widehat{t},\widehat{k}$ of $t,k$:
\begin{align}
  A & = \sum_{\mu=1}^D\widehat t_{\mu\mu\mu\mu}\,,
  \label{Aop}
  \\
  B & = \sum_{\mu=1}^D\widehat k_{\mu\mu\mu\mu}\,.
  \label{Bop}
\end{align}
Note that these operators are invariant
only under discrete ($D$-dimensional lattice) rotations. 
Their precise definitions and a discussion of their
renormalization are given in subsection~(A.2).

\subsection{Lattice regularization} \label{lattice.tex}

In the lattice regularization the fields are restricted to the sites
of a regular square lattice. With the usual assumption of universality
an infinite class of local lattice actions could be invoked. 
In this paper we will only consider O($n$) lattice actions quadratic in
the spins: 
\begin{equation}
  {\cal A}=\frac{\beta}{2}a^4\sum_{x,y}\sum_b S^b(x)K(x-y)S^b(y)\,,
  \label{action}
\end{equation}
with $K$ short range and demanding
\begin{equation}
  \sum_x K(x)=0\,,
\end{equation}
and
\begin{equation}
  K(z)=K({\cal R}z)\,,
\end{equation}
where ${\cal R}$ is a lattice rotation or reflection. 

Let $K_p$ be the Fourier transform of $K(x)$:   
\begin{equation}
  K_p=a^2\sum_x\rme^{-ipx}K(x).
  \label{Kp}
\end{equation}
The behavior of $K_p$ for small $a$ is assumed to take the form
\begin{equation}
  K_p=p^2\left[1+a^2r(p)+\rmO(a^4)\right]\,,
\end{equation}
with
\begin{align}
  r(p) & = p^2\left[\kappa_1 R(p)+\kappa_2\right]\,,
  \label{rp}
  \\
  R(p) & \equiv \frac{p^4}{(p^2)^2}\,,
\end{align}
where we have introduced the notation $p^r=\sum_\mu p_\mu^r$. 

For the familiar case of the standard action (ST) 
\begin{equation}
  K(z)=K_{{\rm ST}}(z)\equiv\sum_\mu 
  a^{-4}\left[2\delta_{z,0}-\delta_{z,a\hat{\mu}}
    -\delta_{z,-a\hat{\mu}}\right]\,,
\end{equation}
where $\hat{\mu}$ is the unit vector in the $\mu$ direction, 
with Fourier transform 
\begin{equation}
  K_{p;{\rm ST}}=\hat{p}^2\,,\,\,\,\,\,\,\,\,\,\,
  \hat{p}_\mu=\frac{2}{a}\sin\left(\frac{ap_\mu}{2}\right)\,,
\end{equation}
and so in this case $\kappa_1=-1/12\,,\kappa_2=0$.

Expectation values of an arbitrary Euclidean observable ${\cal O}[S]$
are given by
\begin{equation}
  \langle 
  {\cal O}\rangle=\frac{1}{Z}\int [\rmd S]\,\rme^{-{\cal A}}\,{\cal O}[S]\,,
\end{equation}
where
\begin{equation}
  [\rmd S]=\prod_x\,\rmd\mu(S(x))\,.
\end{equation}
Here $\rmd\mu(S)$ denotes the O($n$) invariant single spin
distribution normalized to 1, 
\begin{equation}
  \rmd\mu(S)=\delta\left(S^2-1\right)\prod_a\rmd S^a \,
\end{equation}
and the partition function $Z$ is such that
$\langle 1\rangle=1$.

Perturbation theory is an expansion for $\beta\to\infty$
and so for this purpose we set
\begin{equation}
  \beta\equiv\frac{1}{\lambda_0^2}\,.
\end{equation}
As in the case of dimensional regularization, to avoid 
intermediate divergent expressions (from some Feynman diagram contributions) 
it is useful to introduce an infrared regulator e.g. work in finite volume
or add a coupling of the spins to an external magnetic field. 
These aspects are discussed further in Appendix~B.
In this section we give results for infinite volume and zero magnetic field.

In our work we consider correlation functions of scalar products of the
spin field, and for this purpose it is convenient to define
\begin{equation}
  \theta(x,y)\equiv S(x)\cdot S(y)-1\,.
\end{equation}

\subsubsection{One $\theta$ correlation functions}

One $\theta$ correlation functions have a perturbative expansion of the 
form:
\begin{align}
  C(x,y) & \equiv \langle\theta(x,y)\rangle\,
  \\
  & = (n-1)\lambda_0^2\sum_{r=0}\lambda_0^{2r}C_r(x,y)\,.
\end{align}
In the lowest order (for $p\ne0$) the Fourier transform is given by
\begin{equation}
  \widetilde{C}_0(p)=\frac{1}{K_p}=\frac{1}{p^2}\left[1-a^2 
    r(p)+\dots\right]\,.
\end{equation}
In the next order we have
\begin{equation}
  \widetilde{C}_1(p)=K_p^{-1}\left[\frac12 F(p)+\frac{k_p}{K_p}\right]\,,
\end{equation}
where $F(p)$ corresponds to an eye diagram
\begin{equation}
  F(p)=\int_s \frac{K_p-K_s-K_{p+s}}{K_{p+s}K_s}\,,
  \label{Fp}
\end{equation}
and $k_p$ to a tadpole
\begin{equation}
  k_p=\int_s\frac{1}{K_s}\left(K_p+K_s-K_{p+s}\right)\,,
  \label{littlek}
\end{equation}
where we have introduced the shorthand notation
\begin{equation}
  \int_s=\int_{-\pi/a}^{\pi/a}\frac{\rmd^2s}{(2\pi)^2}\,.
\end{equation}

\subsubsection{Two $\theta$ correlation functions}

Consider the $\theta$-connected correlation function
\begin{equation}
  C(x_1,y_1;x_2,y_2)\equiv\langle\theta(x_1,y_1)\theta(x_2,y_2)\rangle  
  -\langle\theta(x_1,y_1)\rangle\langle\theta(x_2,y_2)\rangle\,.
\end{equation}
This has a perturbative expansion of the form:
\begin{equation}
  C(x_1,y_1;x_2,y_2)=(n-1)\lambda_0^4
  \sum_{r=0}\lambda_0^{2r}C_r(x_1,y_1;x_2,y_2)\,.
\end{equation}

Define its Fourier transform by 
\begin{equation} \widetilde{C}_r(p_1,q_1;p_2,q_2) 
  =a^6\sum_{x_1,y_1,x_2}\rme^{-i(p_1x_1+q_1y_1+p_2x_2+q_2y_2)} 
  C_r(x_1,y_1;x_2,y_2)\,. 
\end{equation} 
To avoid disconnected contributions we restrict attention to momentum 
configurations such that 
\begin{align} 
  & p_1+q_1+p_2+q_2=0\,, 
  \nonumber\\ 
  & p_1\ne0\,,\,\,\, p_2\ne0\,, \,\,\, q_1\ne0\,, \,\,\, q_2\ne0\,, 
  \nonumber\\ 
  & p_1+p_2\ne0\,,\,\,\,\,\,\,\,p_1+q_2\ne0\,, 
  \label{momconfig} 
\end{align} 
i.e. all momenta unequal zero and the sum of pairs of momenta 
associated with different $\theta$'s unequal to zero. Then 
\begin{equation} 
  \widetilde{C}_0(p_1,q_1;p_2,q_2)=0\,. 
\end{equation}

In the next order we have simply
\begin{multline}
  \widetilde{C}_1(p_1,q_1;p_2,q_2) = 
  \left[\frac{1}{K_{p_2}K_{q_2}}+\frac{1}{K_{p_1}K_{q_1}}\right]
  \left[\frac{1}{K_{p_1+p_2}}+\frac{1}{K_{p_1+q_2}}\right] \\
  -\frac{1}{K_{p_1}K_{q_1}K_{p_2}K_{q_2}}
  \left[(n-1)K_{p_1+q_1}+K_{p_1+p_2}+K_{p_1+q_2}\right]\,.
\end{multline}
The expression for the next order which is rather lengthy is given
in Appendix~B. The continuum limits of these results and the leading
lattice artifacts will also be considered there.

\subsection{Relation between the renormalization schemes}

Renormalized lattice correlation functions which have a finite 
continuum limit ($a\to0$) can be obtained by applying wave function
and coupling renormalization. In particular the correlation functions
in the $\ms$ scheme are obtained by
\begin{equation}
  \widetilde{C}^{\ms}(p_1,q_1 ; \dots ; p_k,q_k)
  =\Zr^{-k}\widetilde{C}(p_1,q_1 ; \dots ; p_k,q_k)+\rmO(a^2)\,,
\end{equation}
provided one expresses the lattice functions in terms of the renormalized
$\ms$ coupling related to the bare lattice coupling through
\begin{equation}
  g^2=\Zor^{-1}\lambda_0^2\,.
\end{equation}
We of course checked explicitly that our computations confirmed this claim.

The renormalization constants have a perturbative expansion of the form 
\begin{align}
  \Zr & = 1+\Zr^{(1)}g^2+\Zr^{(2)}g^4+\dots\,,
  \\
  \Zor & = 1+\Zor^{(1)}g^2+\Zor^{(2)}g^4+\dots\,.
\end{align}
At 1-loop one obtains: 
\begin{align}
  \Zr^{(1)} & = (n-1)\frac{1}{2\pi}\left[\ln(a\mu)
    +\frac12(c_1+\gamma)\right]\,,
  \\
  \Zor^{(1)} & = (n-2)\frac{1}{2\pi}\left[\ln(a\mu)+\frac12(c_1+\gamma)\right]
  -A_0\,,
  \label{Zor1}
\end{align}
where the constants $A_0,c_1$ are given in (\ref{defA0}),(\ref{defc1}) 
respectively.
Defining the lattice $\Lambda$--parameter by
\begin{equation}
  \Lambda_{\rm latt}
  =a^{-1}(2\beta_0\lambda_0^2)^{-\chi}\rme^{-\frac{1}{2\beta_0\lambda_0^2}}
  \left[1+\rmO(\lambda_0^2)\right]\,,
  \label{Lambdalatt}
\end{equation}
one gets the relation
\begin{equation}
  \frac{\Lambda_{\rm latt}}{\Lambda_{\ms}}
  =\exp\left[\frac12(c_1+\gamma)-\frac{2\pi A_0}{n-2}\right]\,.
  \label{LambdalattoverMS}
\end{equation}
For the standard action this result was first obtained by 
Parisi \cite{Parisi}.

\section{Symanzik's theory of lattice artifacts}   
\label{symanzik_theory.tex}

\subsection{The effective Lagrangian}  

We write the lattice Lagrangian including the source terms symbolically as
\begin{equation}
  {\cal L}_{\rm latt}=\frac{1}{2\lambda_0^2}
  \left(\partial_\mu S\cdot\partial_\mu S\right)^{\rm latt}-J\cdot S\,,
  \qquad\qquad S^2=1\,,
  \label{lattA2}
\end{equation}
with some lattice regularization of the kinetic term. We use the 
corresponding action in the generating functional for bare lattice
$S$-field correlation functions, which, after Fourier transformation,
become functions of the momenta, the bare lattice coupling $\lambda_0$
and the lattice spacing $a$. Performing, for fixed momenta and coupling,
a small $a$ expansion we can write
\begin{equation}
  {\cal G}^X_{\rm latt}(\lambda_0,a)={\cal G}^{X(0)}(\lambda_0,a)+
  a^2\,{\cal G}^{X(1)}(\lambda_0,a)+\rmO\left(a^4\right)\,,
  \label{cutoff}
\end{equation}
where both the scaling functions ${\cal G}^{X(0)}$ and the
leading cutoff corrections ${\cal G}^{X(1)}$ are still weakly
(logarithmically) depending on $a$. 

The separation of the full lattice correlation function into a scaling
piece and cutoff corrections is unambiguous and straightforward in PT.
In fact, in PT at $\ell$ loop order both terms are finite (order $\ell$) 
polynomials in $\ln a$. For example (using the results in Appendix~B)
for the 2-spin correlation function at 1--loop one has
\begin{multline}  \label{C1pa}
  \widetilde{C}_1(p) = \frac{1}{4\pi p^2}\Bigl[ L(ap)+c_1+4\pi A_0 \\
  + a^2p^2\left\{\kappa_1\left(\frac34-R(p)\right)L(ap)
    +c_3 R(p)+c_4\right\}+\rmO(a^4p^4)\Bigr]\,,
\end{multline}

where 
\begin{equation}
  L(ap)\equiv\ln(a^2p^2)\,,
\end{equation}
and all undefined constants are given in Appendix~B.

One of our main assumptions here is that the 
expansion (\ref{cutoff}) makes sense also beyond PT. Usual renormalization 
theory deals with the scaling part ${\cal G}^{X(0)}$ and in later subsections 
we will analyze the next term ${\cal G}^{X(1)}$ using Symanzik's method.

Symanzik's local effective Lagrangian ${\cal L}_{\rm eff}$ which is 
defined in the continuum in $D$~dimensions, and is written in terms 
of the bare $S$-field normalized to unity, depends nonlinearly on the 
source $I$ and has the property that the generating functional obtained 
from it completely reproduces the correlation functions corresponding to 
(\ref{lattA2}) up to terms of order $a^4$. It is of the form 
\begin{equation}
  \begin{split}
    -{\cal L}_{\rm eff}&=-\frac{1}{2g_0^2}\left(\partial_\mu S\cdot
      \partial_\mu S\right)+\frac{1}{g_0^2} I\cdot S\\
    & + \frac{a^2}{g_0^2}\left\{\widetilde Y_1{\cal O}_1+
      \widetilde Y_2{\cal O}_2+\widetilde Y_3{\cal O}_3+
      \widetilde Y_AA+\widetilde Y_BB\right\}\\
    & + \frac{a^2}{g_0^2}\left\{\widetilde W_1 I\cdot\square S+
      \widetilde W_2 I\cdot S
      \left(\partial_\mu S\cdot\partial_\mu S\right)\right\}\\
    & + \frac{a^2}{g_0^2}\left\{\widetilde X_1 I^2+\widetilde
      X_2\left(I\cdot S\right)^2
    \right\}.
  \end{split}
  \label{LEL2}
\end{equation}
Here the operators ${\cal O}_1$, ${\cal O}_2$ and ${\cal O}_3$ are defined
in (\ref{calOibasis}), the operators $A$ and $B$ in (\ref{Aop}),(\ref{Bop})
and the constants $\widetilde Y_1,\dots,\widetilde X_2$
can (in principle) be determined by comparing correlation functions
calculated directly on the lattice and from (\ref{LEL2}).

Actually, this form of the local effective action for the O$(n)$ nonlinear
sigma model cannot be found in Symanzik's published papers \cite{Sym1,Sym2}.
But in the case of the $\phi^4$ model an analogous local effective Lagrangian
is introduced in \cite{Sym1} and in \cite{Sym2} an improved lattice action
is presented for the O$(n)$ sigma model. (\ref{LEL2}) is constructed
using the $\phi^4$ model analogy and has exactly the same terms as the
improved action in \cite{Sym2}. It is not unique: as discussed in \cite{Sym2}
there is an ambiguity in choosing the coefficients which can be used to
put (for example) $\widetilde W_2$ and $\widetilde X_2$ equal to zero.
To be able to identify the lattice correlation functions with the ones 
obtained using the effective action we also have to relate the couplings 
$\lambda_0$ and $g$ and the sources $J^a$ and $j^a$, as will be discussed 
below.
We can eliminate the terms quadratic in the source fields by using the
identities derived in Appendix~A. We get
\begin{equation}
  -{\cal L}_{\rm eff}=-\frac{1}{2g_0^2}\left(\partial_\mu S\cdot
    \partial_\mu S\right)+\frac{1}{g_0^2} I\cdot S
  +\frac{a^2}{g_0^2}\left\{\overline Y_AA+\overline Y_BB+
    \sum_{i=1}^5 \overline Y_i{\cal O}_i\right\}\,,
  \label{LEL3}
\end{equation}
where
\begin{equation}
  \begin{split}
    \overline Y_A&=\widetilde Y_A\,,\\
    \overline Y_B&=\widetilde Y_B\,,
  \end{split}\qquad\qquad
  \begin{split}
    \overline Y_1&=\widetilde Y_1+8\widetilde W_2+8\widetilde X_2\,,\\
    \overline Y_2&=\widetilde Y_2\,,\\
    \overline Y_3&=\widetilde Y_3+2\widetilde X_1-2\widetilde W_1\,,\\
    \overline Y_4&=2\widetilde W_2+4\widetilde X_2+4\widetilde X_1
    -2\widetilde W_1\,,\\
    \overline Y_5&=8\widetilde X_2+8\widetilde X_1\,.
  \end{split}
  \label{barY}
\end{equation}
One can see that $\overline Y_i$ are invariant under the transformations 
\begin{equation}
  \delta\widetilde X_1=-\omega\,,\qquad
  \delta\widetilde X_2=\omega\,,\qquad
  \delta\widetilde Y_1=-8\omega\,,\qquad
  \delta\widetilde Y_3=2\omega
\end{equation}
and
\begin{equation}
  \delta\widetilde W_1=\rho\,,\qquad
  \delta\widetilde W_2=\rho\,,\qquad
  \delta\widetilde Y_1=-8\rho\,,\qquad
  \delta\widetilde Y_3=2\rho\,.
\end{equation}
These correspond to the ambiguities of the original effective
Lagrangian (\ref{LEL2}) discussed above.

Finally we introduce the operator basis consisting of 
\begin{equation}
  U_i=\frac{1}{g_0^2}\,\overline{U}_i\,,\,\,\,\,i=1,\dots,5\,,
  \label{redef}
\end{equation}
together with $U_6$ and $U_7$, where these and 
the operators $\overline{U}_i\,,i=1,\dots,5$, are defined in Appendix~A.
In this basis we have
\begin{equation}
  -{\cal L}_{\rm eff}=-\frac{1}{2g_0^2}\left(\partial_\mu S\cdot
    \partial_\mu S\right)+\frac{1}{g_0^2} I\cdot S
  +a^2\sum_{i=1}^7 Y_i(g,\varepsilon)U_i\,,
  \label{LEL4}
\end{equation}
where the coefficients $Y_i$ are appropriate linear combinations
of $\overline Y_i$ and $\overline Y_{A,B}$. 
This particular basis of operators is chosen such that they are
renormalized according to
\begin{equation} \label{UiR}
  U_{i(R)}=K_{ij}(g,\varepsilon)\,U_j\,,
\end{equation}
where the matrix of renormalization constants is block diagonal consisting
of the $5\times5$ block for $i,j=1,2,3,4,5$ discussed in subsection~(A.1)
and the $2\times2$ block for $i,j=6,7$ discussed in subsection~(A.2).
Recalling the redefinition (\ref{redef}) we have for the first block
\begin{equation}
  K_{ij}=Z_1\,\overline{K}_{ij}\,,
\end{equation}
where $\overline{K}$ is the renormalization matrix for the 
$\overline{U}_i$.

We now define
\begin{equation}
  \widehat{c}_j(g,\varepsilon)=\sum_{i=1}^7Y_i(g,\varepsilon)
  K^{-1}_{ij}(g,\varepsilon)
\end{equation}
and using this definition we can write
\begin{equation}
  \sum_{i=1}^7Y_iU_i=\sum_{i=1}^7\widehat{c}_iU_{i(R)}\,,
\end{equation}
which shows that the limit 
\begin{equation}
  c_i(g)=\widehat c_i(g,0)
\end{equation}
must exist.

\subsection{Relations between correlation functions}  

We now write down the precise relation between the lattice correlation
functions and the ones obtained by using the effective action. We
identify (for simplicity) the scale parameter $\mu$ of dimensional
regularization with the inverse of the lattice spacing $a$. In this case
we can write
\begin{equation}
  \begin{split}
    {\cal G}^X_{\rm latt}(\lambda_0,a)&=
    y^r(g){\cal G}^X_{(R)}(g,a^{-1})\\
    &+a^2 y^r(g)\sum_{i=1}^7 c_i(g)
    {\cal G}^X_{i(R)}(g,a^{-1})+\dots\,,
  \end{split}
  \label{Sym0}
\end{equation}
where the finite wave function renormalization constant $y(g)$ comes from
the relation between the lattice source $J$ and the (renormalized) dimensional
regularization source $j$:
\begin{equation}
  j^a(x)=y(g)\,J^a(x)
\end{equation}
and from universality it follows that a relation of the form
$g=G(\lambda_0)$ must exist between the couplings of the theory.

Comparing (\ref{Sym0}) with the expansion (\ref{cutoff}) we see that
\begin{equation}
  \begin{split}
    {\cal G}^{X(0)}(\lambda_0,a)&=y^r(g){\cal G}^X_{(R)}(g,a^{-1})\,,\\
    {\cal G}^{X(1)}(\lambda_0,a)&=y^r(g)\sum_{i=1}^7c_i(g)
    {\cal G}^X_{i(R)}(g,a^{-1})\,.
  \end{split}
  \label{Sym01}
\end{equation}
This result can also be written as
\begin{equation}
  {\cal G}^X_{\rm latt}(\lambda_0,a)=
  {\cal G}^{X(0)}(\lambda_0,a)\left\{
    1+a^2\delta^X(\lambda_0,a)\right\}+\rmO\left(a^4\right)\,,
\end{equation}
where
\begin{equation}
  \delta^X(\lambda_0,a)=\sum_{i=1}^7c_i(g)
  \delta^X_i(g,a)
\end{equation}
and
\begin{equation}
  \delta^X_i(g,a)=\frac{{\cal G}^X_{i(R)}(g,a^{-1})}
  {{\cal G}^X_{(R)}(g,a^{-1})}\,.
  \label{cuti}
\end{equation}

\subsection{Tree level effective action}  

Let us write down the tree level effective action. If we start from the
standard lattice regularization of the sigma model with action
\begin{equation}
  -{\cal A}_{\rm ST}=\frac{1}{\lambda_0^2}\sum_{x,\mu}
  \left(S_x\cdot S_{x+\hat\mu}-1\right)=
  -\frac{a^2}{2\lambda_0^2}\sum_{x,\mu}\left(\frac
    {S_{x+\hat\mu}-S_x}{a}\right)^2\,,
\end{equation}
we get classically in the continuum
\begin{equation}
  -{\cal A}_{\rm ST}=-\frac{1}{2\lambda_0^2}\sum_\mu\int{\rm d}^2x
  \left(\partial_\mu S\right)^2 +\frac{a^2}{24\lambda_0^2}
  \sum_\mu\int\rmd^2x\,\left(S\cdot\partial^4_\mu S\right)
  +\rmO\left(a^4\right)\,.
\end{equation}
Written in the basis of operators introduced in subsection~(2.1) 
this corresponds to
\begin{equation}
  -{\cal L}_{\rm eff}^{(0)}=-\frac{1}{2\lambda_0^2}\left(\partial_\mu S\cdot
    \partial_\mu S\right)+\frac{a^2}{\lambda_0^2}\left\{\frac{1}{24}A+
    \frac{1}{16}{\cal O}_3\right\}+\rmO\left(a^4\right)\,.
\end{equation}

If we start from a different lattice action in our class 
(which is still quadratic in $S$),
we will find in the tree level effective Lagrangian a different linear 
combination of the two operators $A$ and ${\cal O}_3$, including the
possibility of the vanishing of both coefficients (which occurs for 
example for the improved action). In general we have
\begin{equation}
  -{\cal L}_{\rm eff}^{(0)}=-\frac{1}{2\lambda_0^2}\left(\partial_\mu S\cdot
    \partial_\mu S\right)+\frac{a^2}{\lambda_0^2}\left\{
    \frac{e_4}{24}A+
    \frac{e_0}{16}
    {\cal O}_3\right\}+\rmO\left(a^4\right)\,,
  \label{L0}
\end{equation}
where
\begin{equation}
  e_0=e_4=1
\end{equation}
for the standard action ST.
For the general case
\begin{equation} \label{e0_e4}
  e_0 = -12\kappa_1  \,, \qquad 
  e_4 = -12\kappa_1 - 16\kappa_2 \,.
\end{equation}

We can read off the leading coefficients $c_i^{(0)}$ of the
operators in our basis. This is the leading term in the expansion
\begin{equation}
  c_i(g)=\sum_{\ell=0}^\infty c_i^{(\ell)}g^{2\ell}\,.
\end{equation}
Using (the third row of) (\ref{UtoO}) and (\ref{L0}) we find
\begin{equation}
  \begin{split}
    c^{(0)}_1&=\frac{e_0}{4(n-1)}\,,\\
    c^{(0)}_2&=-\frac{e_0}{4(n-1)}\,,\\
    c^{(0)}_3&=\frac{e_0(n^2-2n-4)}{4n(n-2)}\,,\\
    c^{(0)}_4&=\frac{9e_0}{4(2n-1)}\,,\\
    c^{(0)}_5&=\frac{(3-n)(3+n)e_0}{16n(n-1)}\,,
  \end{split}\qquad\qquad\qquad
  \begin{split}
    c^{(0)}_6&=\frac{e_4}{24}\,,\\
    c^{(0)}_7&=0\,.
  \end{split}
\end{equation}
Note that $c^{(0)}_5=0$ for $n=3$. 

\subsection{Renormalization group considerations} 

In this section we will analyze the structure of the 
lattice artifacts using RG methods. We recall that operators 
are renormalized in our basis according to \eqref{UiR}
where the operator renormalization matrix is of the form
\begin{equation}
  K_{ij}(g,\varepsilon)=\delta_{ij}-\frac{g^2}{\varepsilon}\,k_{ij}+
  \frac{g^4}{2\varepsilon}\,\nu_{ij}^{(2)}+\frac{g^4}{2\varepsilon^2}
  \left(k_{is}k_{sj}+2\beta_0k_{ij}\right)+\dots
\end{equation}
and the anomalous dimension matrix is defined as
\begin{equation}
  \nu_{ij}(g)=K_{is}(g,\varepsilon)\left(\beta(g)-
    \frac{\varepsilon g}{2}\right)\frac{\partial M_{sj}(g,\varepsilon)}
  {\partial g}=-k_{ij}g^2+\nu_{ij}^{(2)}g^4+\nu_{ij}^{(3)}g^6+\dots,
  \label{nuijg}
\end{equation}
where $M_{ij}$ is the matrix inverse of $K_{ij}$. It is clear that
if the matrix $K$ is (block) diagonal, then so is the matrix $\nu$ and
if the matrix $K$ is (block) triangular, then so is the matrix $\nu$. 

$k_{ij}$ (in (\ref{nuijg})) is a diagonal matrix of the form
\begin{equation}
  k_{ij}=2\beta_0\Delta_i\delta_{ij}
\end{equation}
with eigenvalues corresponding to
\begin{equation}
  \Delta_i=\left\{
    \frac{n}{n-2}\,;-1\,;0\,;\frac{1-n}{n-2}\,;
    \frac{1}{n-2}\,;0\,;-1\right\}\,.
  \label{spec}
\end{equation}

It is easy to see that the functions $\delta^X_i$ 
(defined by (\ref{cuti})) satisfy the RG equation
\begin{equation}
  {\cal D}\,\delta^X_i(g,a)=-\nu_{ij}(g)\,\delta^X_j(g,a)\,,
  \label{rg}
\end{equation}
where
\begin{equation}
  {\cal D}=-a\frac{\partial}{\partial a}+\beta(g)\frac{\partial}{\partial g}\,.
\end{equation}

To solve this partial differential equation we introduce the matrix 
$U_{ij}(g)$, which solves the ordinary differential equation
\begin{equation}
  U_{ij}^\prime(g)=-\rho_{is}(g)\,U_{sj}(g)\,,
  \label{Uijeqtn}
\end{equation}
where
\begin{equation}
  \rho_{ij}(g)=\frac{\nu_{ij}(g)}{\beta(g)}\,.
\end{equation}
This has the expansion
\begin{equation}
  \rho_{ij}(g)=\frac{2\Delta_i}{g}\,\delta_{ij}+R_{ij}(g)\,,
\end{equation}
where
\begin{equation}
  R_{ij}(g)=\sum_{\ell=2}^\infty\rho^{(\ell)}_{ij}\,g^{2\ell-3}\,.
\end{equation}
The two-loop and three-loop coefficients are:
\begin{equation}
  \begin{split}
    \rho^{(2)}_{ij} & =-\frac{1}{\beta_0}\,\nu_{ij}^{(2)}-
    \frac{\beta_1}{\beta_0^2}\,k_{ij} \,, \\
    \rho^{(3)}_{ij} & =-\frac{1}{\beta_0}\,\nu_{ij}^{(3)}+
    \frac{\beta_1}{\beta_0^2}\,\nu^{(2)}_{ij}+
    \left(\frac{\beta_1^2}{\beta_0^3}-\frac{\beta_2}{\beta_0^2}\right)
    \,k_{ij}\,.
  \end{split}
\end{equation}
If we find the solution of (\ref{Uijeqtn}) we can write the general 
solution
of (\ref{rg}) as
\begin{equation}
  \delta^X_i(g,a)=U_{ij}(g)\,D^X_j(\Lambda)\,,
\end{equation}
and the lattice artifacts are of the form
\begin{equation}
  \delta^X(\lambda_0,a)=\sum_{i=1}^7\,\hat v_i(g)\,D^X_i(\Lambda)\,,
  \label{arti1}
\end{equation}
where
\begin{equation}
  \hat v_i(g)=\sum_{s=1}^7\,c_s(g)\,U_{si}(g)\,.
\end{equation}
The functions $D^X_j$ depend only on $\Lambda$, the RG invariant 
combination of $g$ and $a$. These functions are non-perturbative and 
depend on the quantity ($X$) we are considering. On the other hand, the
coefficients $\hat v_i(g)$ are perturbative and universal in the sense that 
they remain the same for all physical quantities (but depend on the 
lattice action we started with).

We take the following ansatz: 
\begin{equation}
  U_{ij}(g)=\left\{\delta_{ij}+Q_{ij}(g)\right\}\,g^{-2\Delta_j}\,,
  \label{Uij}
\end{equation}
where $Q_{ij}(g)$ has the loop expansion
\begin{equation}
  Q_{ij}(g)=\sum_{\ell=2}^\infty\,k^{(\ell)}_{ij}\,g^{2\ell-2}\,.
  \label{Qij}
\end{equation}
Here the coefficients $k^{(\ell)}_{ij}$ may still weakly (logarithmically) 
depend on the coupling. This can arise if the difference between two
eigenvalues $\Delta_i-\Delta_j$ is a non-zero integer, which is
possible in our case, i.e. for $n=3$. We can however ignore this subtlety,
because we verify in Appendix~D that for the quantities we need here 
it plays no role.

We can now write the lattice artifacts as
\begin{equation}
  \delta^X(\lambda_0,a)=\sum_{i=1}^7v_i(g)\,g^{-2\Delta_i}\,D^X_i(\Lambda)\,,
  \label{arti2}
\end{equation}
where
\begin{equation}
  v_i(g)=c_i(g)+\sum_sc_s(g)Q_{si}(g)=\sum_{\ell=0}^\infty 
  v_i^{(\ell)} \,g^{2\ell}\,.
  \label{vig}
\end{equation}
The spectrum of one-loop eigenvalues given by (\ref{spec}) plays a crucial
role in our considerations. The leading term corresponds to
\begin{equation}
  \Delta_1=\frac{n}{n-2}=n\chi=1+2\chi\,,
\end{equation}
and the sub-leading one to
\begin{equation}
  \Delta_5=\frac{1}{n-2}=\chi\,.
\end{equation}
We thus have the leading expansion
\begin{equation}
  \begin{split}
    \delta^X(\lambda_0,a)&=v_1\,D^X_1\,\left(g^{-2}\right)^{1+2\chi}+
    v_5\,D^X_5\,\left(g^{-2}\right)^\chi+\dots\\
    &=D^X_1\,\left\{\left(g^{-2}\right)^{1+2\chi}v_1^{(0)}+
      v_1^{(1)}\left(g^{-2}\right)^{2\chi}\right\}+
    \rmO\left(\left(g^{-2}\right)^\chi\right)\,.
  \end{split}
  \label{arti3}
\end{equation}
We have already noted that for the $n$=3 case 
$v^{(0)}_5=c^{(0)}_5=0$. This means that the corrections start one power 
later and in this case we have the leading expansion
\begin{equation}
  \delta^X(\lambda_0,a)=D^X_1\left\{
    v^{(0)}_1\,g^{-6}+v^{(1)}_1\,g^{-4}+v_1^{(2)}\,g^{-2}\right\}
  +\rmO(1)\,.
\label{352}
\end{equation}
The expansion coefficients are
\begin{equation}
  \begin{split}
    v^{(0)}_1&=c^{(0)}_1=\frac{e_0}{4(n-1)}\,,\\
    v^{(1)}_1&=c^{(1)}_1+\sum_sc^{(0)}_sk^{(2)}_{s1}\,,\\
    v^{(2)}_1&=c^{(2)}_1+\sum_sc^{(1)}_sk^{(2)}_{s1}+
    \sum_sc^{(0)}_sk^{(3)}_{s1}\,.
  \end{split}
  \label{expcoffs}
\end{equation}
Concretely we have
\begin{equation}
  k^{(2)}_{s1}=\frac{1}{2(\Delta_1-1-\Delta_s)}
  \rho^{(2)}_{s1}\,,
\end{equation}
which is different from zero for $s=1,2$ only.

We first note that the functional form (\ref{arti3}), (\ref{352}) is only valid
for actions where the leading coefficient $v^{(0)}_1=c^{(0)}_1$ does
not vanish. An important special case is Symanzik's (tree) improved
action, where the above condition is not satisfied.
We also note that, as can be seen from (\ref{A12}), the connected correlation
functions of the operator $U_5$ are pure contact terms and therefore do not
contribute to on-shell physical quantities. For such observables the
sub-subleading corrections in (\ref{arti3}) come from the operators $U_3$ and
$U_6$ and are O$(1)$ (for $n\geq4$).

We also need the connection between the lattice coupling $\lambda_0$ and $g$:
\begin{equation}
  g^2=\lambda_0^2+\frac{\psi}{2\pi}\lambda_0^4+\dots,
  \label{psi}
\end{equation}
which can be calculated from (\ref{Zor1}) by setting $a\mu=1$. We find
\begin{equation} \label{psi_eq}
  \psi = -\frac12 (c_1+\gamma)(n-2) + 2\pi A_0 \,.
\end{equation}
We will also use the inverse coupling $\widetilde\beta=2\pi/\lambda_0^2$.

\subsection{The final result} 

The information above is all we need to write down the final result:
\begin{equation}
  \delta^X(\lambda_0,a)=\frac{e_0}{4(n-1)\left(2\pi\right)^{n\chi}}\,
  D^X_1(\Lambda)\left\{
    \left(\widetilde\beta\right)^{1+2\chi}+r^{(2)}
    \left(\widetilde\beta\right)^{2\chi}\right\}+
  \rmO\left(\widetilde\beta^\chi\right)
  \label{final}
\end{equation}
for $n\geq4$\footnote{
For on-shell physical quantities the corrections are O$(1)$ only.}, and
\begin{equation}
  \delta^X(\lambda_0,a)=\frac{e_0}{(4\pi)^3}\,
  D^X_1(\Lambda)\left\{
    \widetilde\beta^3+r^{(2)}\widetilde\beta^2+r^{(3)}\widetilde\beta\right\}+
  \rmO(1)
  \label{final3}
\end{equation}
for $n=3$. We write the two-loop coefficient as a sum of three terms:
\begin{equation}
  r^{(2)}=r^{(2)}_{\rm I}+r^{(2)}_{\rm II}+r^{(2)}_{\rm III}\,,
  \label{r2formula}
\end{equation} 
with
\begin{align}
  r^{(2)}_{\rm I} & = \frac{8\pi(n-1)}{e_0}c^{(1)}_1\,,
  \\
  r^{(2)}_{\rm II} & = \frac{n}{n-2}(1-\psi)\,,
  \\
  r^{(2)}_{\rm III} & = (2\pi)^2\left[
    \frac{1}{n-2}\nu^{(2)}_{11}+\frac{1}{n}\nu^{(2)}_{21}\right]\,.
\end{align}
The above form (\ref{final}),(\ref{final3}) of artifacts
is general in the sense that it holds for all observables: the leading
perturbative coefficients are the same for all observables (but depend
on the lattice action chosen). Only the overall coefficient of this
universal series depends on the quantity in question.
This coefficient is nonperturbative and for any physical quantity $X$
it can be parameterized by
\begin{equation}
  D_1^X(\Lambda)=\Lambda^2 f^X(z)\,,
\end{equation}
where $z=L\Lambda$ or a vector with components $z_j=L_j\Lambda$,
where $L$ (or the $L_j$-s) are the relevant length scale(s) in the problem.

If we use the RG relation between the lattice spacing $a$ and the
inverse coupling, we can write the functional form of the leading
artifacts as 
\begin{equation}
  a^2\left[\ln(a^2)\right]^{n/(n-2)}\,.
\end{equation}
Note that our final result is completely consistent with 
the results of refs.~\cite{KLW,BKLW,CarPel}, who found leading artifacts
$\propto a^2\ln(a^2)$ in the leading and next-to-leading orders
of the $1/n$ expansion.

We have calculated the three terms contributing to the sub-leading coefficient
(\ref{r2formula}). For the calculation of the first term
we need the 1-loop coefficients of the effective action.
We obtained these using the computations described in section~2
of the 2- and 4-point spin correlation functions both on the lattice and 
in the continuum. More precisely we need the scaling part and the 
O$(a^2)$ piece of the lattice correlation functions
and in the continuum we need 
the original correlation functions as well as the ones where those dimension
four operators that appear in the tree level effective action are inserted.
These are both long computations and are discussed in more detail in
Appendices~B and ~C respectively. 

For the spin-four tensor operators we find
\begin{equation}
  c^{(1)}_6=\frac{1}{2}A_1-\frac{\kappa_1}{2}A_0,\qquad\quad
  c^{(1)}_7=\frac{1}{16}F^{(A)}_{17}\,.
\end{equation}
The definition and numerical values of these and all other lattice
integrals can be found in Appendix~B.

The coefficients of the scalar operators are simplest in the original
basis. We will list the coefficients $d^{(1)}_A$, where
\begin{equation}
  \sum_{i=1}^5c^{(1)}_i\,U_i=\frac{1}{g_0^2}
  \sum_{A=1}^5d^{(1)}_A\, {\cal O}_A\,.
\end{equation}
The result is
\begin{equation}
  \begin{split}
    d^{(1)}_1&=\frac{1}{8}F_{17}^{(A)}+\frac{1}{2}F_{17}^{(C)}
    -F_{28}^{(B)}+F_{30}^{(C)}
    -\frac{3\kappa_1}{16\pi}+\frac{3\kappa_3}{2\pi}+
    \frac{\gamma\kappa_3}{\pi}\,,\\
    d^{(1)}_2&=F_{10}+\frac{1}{4}F_{17}^{(A)}+\frac{1}{2}F_{17}^{(B)}
    +F_{17}^{(C)}+F_{28}^{(B)}
    -F_{30}^{(C)}+2F_6\\
    &\qquad+\frac{3\kappa_1}{16\pi}-\frac{7\kappa_3}{2\pi}
    -\frac{4\gamma\kappa_3}{\pi},\\
    d^{(1)}_3&=\frac{3}{4}A_1+A_2-\kappa_3A_0,\\
    d^{(1)}_4&=\frac{1}{4}F_{10}+F_6-\frac{(1+\gamma)\kappa_3}{2\pi},\\
    d^{(1)}_5&=(n-1)\left[2F_6+\frac{\gamma\kappa_3}{\pi}\right]\,,
  \end{split}
\end{equation}
where
\begin{equation}
  \kappa_3=\kappa_2+\frac34\kappa_1\,.
\end{equation}

For the calculation of the second term in (\ref{r2formula}) we just need
$\psi$ given by (\ref{psi_eq}).  One then gets: 
\begin{equation}
  r^{(2)}_{\rm II}=\frac{n(1-2\pi A_0)}{n-2}+\frac{n}{2}(\gamma+c_1)\,.
\end{equation}

Finally the third term can be obtained from the $5\times5$ two-loop anomalous
dimension matrix of the dimensionally regularized scalar operators.
The computation is described in the last subsection of Appendix~A,
and leads to the simple result
\begin{equation}
  r^{(2)}_{\rm III}=-2-\frac{9}{2(n-2)}\,.
\end{equation}
For $n=3$ it is $-13/2$.

Putting everything together, we find for the sub-leading coefficient
\begin{equation}
  r^{(2)}=\alpha_1+n\alpha_2+\frac{\alpha_3}{n-2}\,,
  \label{r2result}
\end{equation}
where
\begin{equation}
  \begin{split}
    \alpha_1=-\frac{19}{8}&+\frac{c_1}{2}+\frac{27\kappa_1}{64\kappa_3}-
    \frac{3c_1\kappa_1}{8\kappa_3}+\frac{\pi}{\kappa_3}\Big\{
    -\frac{3}{2}A_1+\frac{1}{4}F_{10}+\frac{1}{4}F^{(B)}_{15}\\
    &
    -\frac{1}{8}F^{(A)}_{17}
    -\frac{1}{8}F^{(B)}_{17}
    -\frac{1}{2}F^{(C)}_{17}
    +\frac{1}{4}F^{(B)}_{28}
    -\frac{1}{4}F^{(C)}_{30}
    +\frac{5}{2}F_6\Big\}\,,
  \end{split}
\end{equation}
\begin{equation}
  \alpha_2=\frac{c_1}{2}-\frac{\pi F_6}{\kappa_3}
\end{equation}
and
\begin{equation}
  \alpha_3=-\frac{5}{2}-4\pi A_0\,.
\end{equation}

Finally we remark that
in the $n=3$ case it would be nice to know also the three-loop coefficient 
$r^{(3)}$, however it would require a lot more effort to compute. 
This is built from (among other things) the two-loop coefficient $c^{(2)}_1$ 
appearing in the effective action and the three-loop anomalous dimension 
matrix elements $\nu^{(3)}_{ij}$.

\section{Results from Monte-Carlo simulations} \label{numerical.tex}

We study here the lattice artifacts of the step scaling function
introduced in \cite{LWW}.
It is defined as follows.
One considers the model confined to a finite
(1-dimensional) box of extension $L$ with periodic boundary
conditions. 
The dimensionless LWW coupling is defined as 
\begin{equation} \label{u0}
  u_0 = L m(L) \,,
\end{equation}
where $m(L)$ is the mass gap of the theory in finite volume.
Next one measures $u_1$ defined similarly with doubled box size. 
In the continuum $u_1$ is a function of $u_0$, called the step scaling
function $u_1=\sigma(2,u_0)$.

In the lattice formulation one considers
an infinitely long strip with $L/a$ sites in the spatial direction
and tunes the parameter $\beta=1/\lambda_0^2$ so that the measured
value of $u_0$ equals to a prescribed value. (In our case, following
\cite{LWW}, we used $u_0 = 1.0595$.)
Then with the same $\beta$ one measures
\begin{equation} \label{u1}
  u_1 = 2L m(2L) = \Sigma(2,u_0,a/L) \,.
\end{equation}
Finally, repeating these steps with finer resolution, one extrapolates 
to the continuum limit
\begin{equation} \label{sigma_cont}
  \sigma(2,u_0) = \lim_{a/L\to 0} \Sigma(2,u_0,a/L) \,.
\end{equation}

The deviation from the continuum limit is denoted below by
\begin{equation} \label{dSigma}
  \delta\Sigma(2,u_0,a/L) = \Sigma(2,u_0,a/L) - \sigma(2,u_0) \,.
\end{equation}

The advantage of this measurement for the purpose of studying
lattice artifacts is that there is no need to 
know the box size $L$ or the mass gap $m(L)$ in physical units,
i.e. one does not need to measure the correlation length
in an infinite volume.
Moreover the continuum 2-dimensional O($n$) model is exactly integrable and the
finite volume mass gap (and hence also the step scaling function) is exactly
calculable using Bethe Ansatz techniques \cite{BH}.

\begin{figure}
  \begin{center}
    \psfig{figure=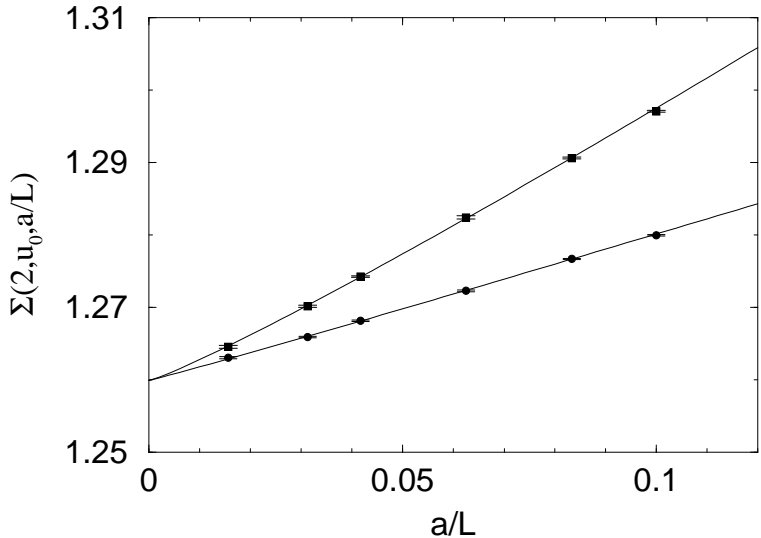,width=10cm}
  \end{center}
  \vspace{-0.5cm}
  \caption{Monte Carlo measurements of the step scaling function at
    $u_0=1.0595$ for two lattice actions in the O(3) model. 
    The data for the larger artifacts correspond to a modified
    action. The fit contains $a$ and $a\ln a$ terms.}
  \label{lww}
\end{figure}

In the MC simulations we used a modified version of the improved
estimator of Hasenbusch \cite{Hasenbusch}. This is described in Appendix E.
The results of the MC measurements for $\Sigma(2,u_0,a/L)$ for the O(3) case
are shown in Fig.~\ref{lww} for two different lattice actions,
for the standard one (ST) and an action D(1/3) defined below.
One can see that the lattice artifacts (cutoff effects) are very nearly 
linear as function of the lattice spacing $a$ both for the case of the 
ST and D(1/3) action. 
Although the effects are in this case relatively very small, 
they seem not of the theoretically expected form. 
Note however the encouraging feature that computations with different 
lattice actions are consistent with the same continuum limit, supporting 
the crucial concept of universality, even if both extrapolations would miss
the exact continuum limit by about 0.002.

\subsection{The lattice actions}

In the simulations we used a one-parameter lattice action D($\alpha$) 
which includes diagonal interaction: 
\begin{equation} \label{action_diag}
  \mathcal{A}(S) = -(1-2\alpha) \sum_{x, \mu} S(x)\cdot S(x+a\hat{\mu})
  - \alpha \sum_{x,d} S(x)\cdot S(x+ad) \,,
\end{equation}
where in the second term the summation is over the two diagonal directions 
$d=(1,1)$ and $(-1,1)$.

In the analytic calculations we considered a more general class of
lattice actions including those corresponding to the kernel with two parameters
\begin{equation} \label{Kp_g}
  K(p) = \hat{p}^2 + g_1 a^2\hat{p}^4 + g_2  a^2\left( \hat{p}^2 \right)^2 \,.
\end{equation}
This includes besides the diagonal interaction also the 
on-axis second-neighbor interaction as well.
The action D($\alpha$) in \eqref{action_diag} is a special case with
\begin{equation} \label{g_alpha}
  g_1 = -g_2 = \frac12 \alpha\,.
\end{equation}
The tree-level Symanzik improved action (SYM) corresponds to
$g_1=1/12$, $g_2=0$. For the $\kappa_i$ in (\ref{rp}) we have
\begin{equation} \label{kappa_g}
  \kappa_1 = g_1 - \frac{1}{12}\,, \qquad \kappa_2 = g_2\,.
\end{equation}

The coefficients of the tree-level effective action are given by
\begin{equation} \label{e0_e4_alpha}
  e_0 = -12\kappa_1 - 16\kappa_2 = 1 + 2\alpha \,,
  \qquad 
  e_4 = -12\kappa_1 = 1 -6\alpha \,.
\end{equation}

Results of the numerical simulations for the O(3) and O(4) cases 
using different actions are collected 
in Tables~\ref{res_o3_ST}-\ref{res_o4_D3}.

\renewcommand\arraystretch{1.5}

\begin{table}[ht] 
  \centering 
  \begin{tabular}[t]{c|c|l|l|l|l} 
    \hline 
    $\beta$  &$L/a$& \phantom{xxxx}$u_0$ & \phantom{xxxx}$u_1$ & $\Sigma(2,\bar{u}_0, a/L)$ & \phantom{xxx}$\delta\Sigma$\\
    \hline \hline 
    $1.56923$& $ 5$& $1.05947(4)$  & $1.29375(5)$  & $1.293798(81)$& $0.032588(81)$ \\
    $1.60476$& $ 6$& $1.05952(5)$  & $1.29018(4)$  & $1.290148(89)$& $0.028938(89)$ \\
    $1.69850$& $10$& $1.059484(21)$& $1.279805(33)$& $1.279830(47)$& $0.018620(47)$ \\
    $1.73020$& $12$& $1.059495(12)$& $1.276719(29)$& $1.276727(35)$& $0.015517(35)$ \\
    $1.77900$& $16$& $1.059492(10)$& $1.272430(15)$& $1.272442(22)$& $0.011233(22)$ \\
    $1.84603$& $24$& $1.059492(8)$ & $1.268074(27)$& $1.268087(30)$& $0.006877(30)$ \\
    $1.89295$& $32$& $1.059498(14)$& $1.265932(29)$& $1.265935(37)$& $0.004725(37)$ \\
    $2.00553$& $64$& $1.059516(6)$ & $1.262938(14)$& $1.262912(17)$& $0.001703(17)$ \\
    \hline
  \end{tabular} 
  \caption{Results of MC simulations for the O(3) standard action. 
    The last two columns are the
    values extrapolated to $\bar{u}_0=1.0595$ and the deviation from
    the continuum value $\sigma(2,\bar{u}_0)=1.261210$ \protect \cite{BH}.
  }
  \label{res_o3_ST} 
\end{table}

\begin{table}[ht] 
  \centering 
  \begin{tabular}[t]{c|c|l|l|l|l} 
    \hline
    $\beta$ & $L/a$ & \phantom{xxxx}$u_0$ & \phantom{xxxx}$u_1$ & $\Sigma(2,\bar{u}_0, a/L)$ & \phantom{xxx}$\delta\Sigma$\\
    \hline \hline 
    $2.00460$& $10$& $1.059521(26)$& $1.297390(45)$& $1.297357(61)$& $0.036147(61)$ \\
    $2.03654$& $12$& $1.059578(22)$& $1.290733(48)$& $1.290609(59)$& $0.029399(59)$ \\
    $2.08453$& $16$& $1.059485(25)$& $1.282422(47)$& $1.282446(61)$& $0.021236(61)$ \\
    $2.14870$& $24$& $1.059544(24)$& $1.274298(46)$& $1.274228(60)$& $0.013018(60)$ \\
    $2.19293$& $32$& $1.059526(23)$& $1.270176(47)$& $1.270135(59)$& $0.008925(59)$ \\
    $2.29813$& $64$& $1.059515(30)$& $1.264571(59)$& $1.264547(76)$& $0.003337(76)$ \\
    \hline
  \end{tabular} 
  \caption{Results of MC simulations for the O(3) D(1/3) action,
  }
  \label{res_o3_D3} 
\end{table}

\begin{table}[ht]
  \centering 
  \begin{tabular}[t]{c|c|l|l|l|l} 
    \hline 
    $\beta$   &  $L/a$  &  \phantom{xxxx}$u_0$  &  \phantom{xxxx}$u_1$ %
    &  $\Sigma(2,\bar{u}_0, a/L)$&\phantom{xxx}$\delta\Sigma$\\
    \hline \hline 
    $1.60705$& $12$ & $1.059571(17)$& $1.270557(37) $& $1.270444(46 ) $& $0.009234(46) $ \\
    $1.65572$& $16$ & $1.059461(26)$& $1.268140(33) $& $1.268202(53 ) $& $0.006992(53) $ \\
    $1.72330$& $24$ & $1.059469(24)$& $1.265479(35) $& $1.265528(52 ) $& $0.004318(52) $ \\
    $1.83791$& $48$ & $1.059450(20)$& $1.262886(101)$& $1.262965(106) $& $0.001755(106)$ \\
    \hline
  \end{tabular} 
  \caption{Results of MC simulations for the O(3) D(-1/4) action.
  }
  \label{res_o3_D4} 
\end{table}

\begin{table}[ht]
  \centering 
  \begin{tabular}[t]{c|c|l|l|l|l} 
    \hline 
    $\beta$ & $L/a$ & \phantom{xxxx}$u_0$ & \phantom{xxxx}$u_1$ %
    & $\Sigma(2,\bar{u}_0, a/L)$ & \phantom{xxx}$\delta\Sigma$\\
    \hline \hline 
    $2.4772$ & $ 8$ & $1.00003(5)$ & $1.21496(5)$ & $1.21491(9) $ & $0.00625(9) $\\
    $2.5532$ & $10$ & $1.00021(6)$ & $1.21357(5)$ & $1.21325(10)$ & $0.00459(10)$\\
    $2.6152$ & $12$ & $0.99994(3)$ & $1.21209(4)$ & $1.21218(6) $ & $0.00352(6) $\\
    $2.7117$ & $16$ & $0.99997(4)$ & $1.21100(5)$ & $1.21105(9) $ & $0.00239(9) $\\
    $2.8470$ & $24$ & $1.00006(4)$ & $1.20995(5)$ & $1.20986(8) $ & $0.00129(8) $\\
    $2.9428$ & $32$ & $0.99992(3)$ & $1.20925(7)$ & $1.20937(9) $ & $0.00071(9) $\\
    $3.1726$ & $64$ & $1.00002(4)$ & $1.20887(6)$ & $1.20884(8) $ & $0.00018(8) $\\
    \hline
  \end{tabular} 
  \caption{Results of MC simulations for the O(4) standard action. 
    $\sigma(2,\bar{u}_0)=1.208658$ at $\bar{u}_0=1$ \protect \cite{BH}.
  }
  \label{res_o4_ST} 
\end{table}

\begin{table}[ht]
  \centering 
  \begin{tabular}[t]{c|c|l|l|l|l} 
    \hline 
    $\beta$    &$L/a$ &  \phantom{xxxx}$u_0$       &  \phantom{xxxx}$u_1$& %
    $\Sigma(2,\bar{u}_0, a/L)$ & \phantom{xxx}$\delta\Sigma$\\
    \hline \hline 
    $2.8512$ & $ 8$ & $0.99992(5)$ & $1.22615(7)$ & $1.22628(11)$ & $0.01762(11)$\\
    $3.0845$ & $16$ & $0.99996(4)$ & $1.21452(6)$ & $1.21458(9) $ & $0.00592(9) $\\
    $3.3095$ & $32$ & $1.00001(5)$ & $1.21034(6)$ & $1.21033(9) $ & $0.00167(9) $\\
    $3.5332$ & $64$ & $0.99988(5)$ & $1.20898(7)$ & $1.20916(11)$ & $0.00050(11)$\\
    \hline
  \end{tabular} 
  \caption{Results of MC simulations for the O(4) D(1/3) action
    for $\bar{u}_0=1$.
  }
  \label{res_o4_D3} 
\end{table}

The analytic expression for the lattice artifacts are expressed in
terms of the inverse lattice coupling, $\beta=1/\lambda_0^2$.
To relate the results for different actions it is convenient
to introduce $\beta_\mathrm{eff}$ from the two-loop formula
\eqref{Lambdalatt}.
For the ST action one can get this
using the relations (\ref{LambdalattoverMS}), (\ref{MoverLambdaMS}) 
and the results
\begin{equation}  \label{ML0}
  \begin{split}
    ML & =0.2671536 \qquad \text{for O(3), at } u_0=1.0595\,,\\
    ML & =0.2390557 \qquad \text{for O(4), at } u_0=1\,, \\
    ML & =0.3408255 \qquad \text{for O(8), at } u_0=1.0595\,, \\
  \end{split}
\end{equation}
which can be calculated using TBA techniques \cite{BH}.
In the range of our simulations $\beta_\mathrm{eff}$ differs 
only slightly, by $\sim 0.04$ from the actual $\beta$.

According to \eqref{psi} for other actions  one has 
\begin{equation} \label{beta_beta_ST}
  \beta_{\mathrm{eff}} =\beta_{\mathrm{eff}}^{\mathrm{ST}}
  + \frac{1}{2\pi}(\psi-\psi_{\mathrm{ST}}) \,.
\end{equation}
Values of $\psi/(2\pi)$ for various actions are given in Table~\ref{psi_r_rho}.

\begin{figure}
  \begin{center}
    \psfig{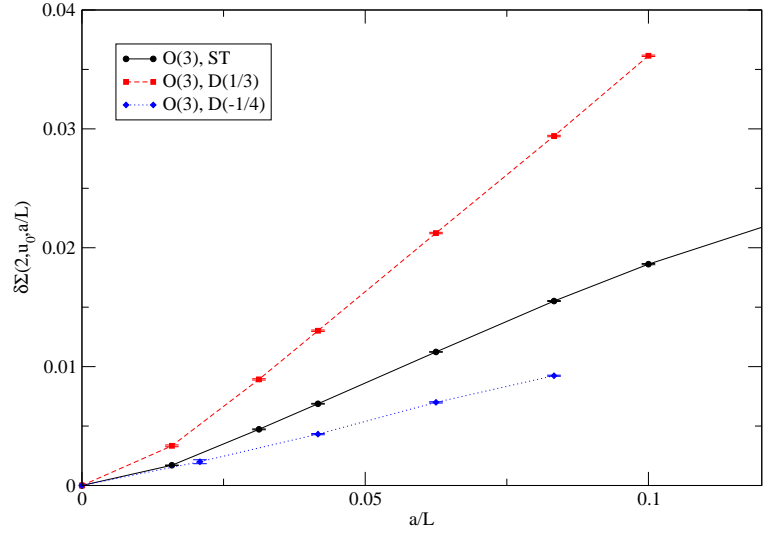}
  \end{center}
  \vspace{-0.5cm}
  \caption{$\delta\Sigma(2,u_0,a/L)$ is plotted for the O(3) 
    ST, D(1/3) and D(-1/4) lattice actions.}
  \label{all_o3}
\end{figure}

\begin{figure}
  \begin{center}
    \psfig{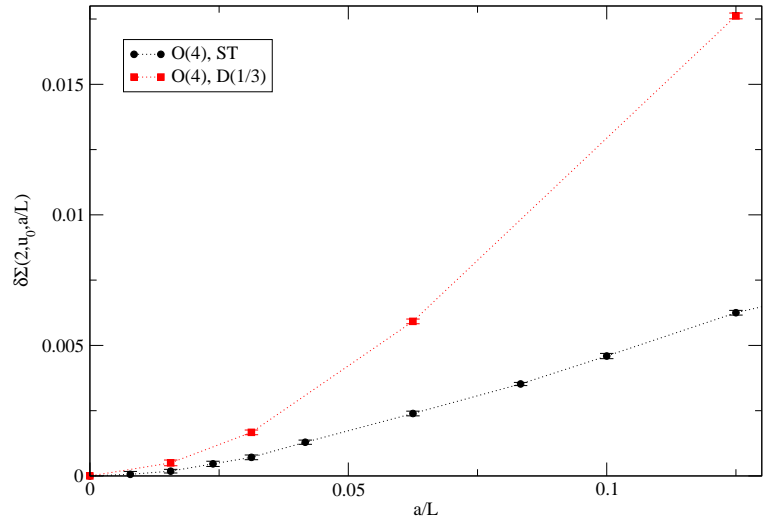}
  \end{center}
  \vspace{-0.5cm}
  \caption{$\delta\Sigma(2,u_0,a/L)$ for the O(4) ST and D(1/3) actions. 
  }
  \label{all_o4}
\end{figure}

Figs.~\ref{all_o3} and ~\ref{all_o4}
show the deviations from the continuum limit,
$\delta\Sigma(2,u_0,a/L)$, as a function of $a/L$ for 
simulations using different actions in the O(3) and O(4)
case, respectively.

In the ratio of artifacts for different actions
the unknown non-perturbative coefficient
$D_1^X(\Lambda)$ in \eqref{final},\eqref{final3}  cancels.
Expressing the ratio in terms of $\beta_{\mathrm{eff}}^{\mathrm{ST}}$ 
one has
\begin{equation} \label{delta_X_ratio}
  \frac{\delta^X(\lambda_0,a)}{\delta^X_\mathrm{ST}(\lambda_{0\mathrm{ST}},a)}
  =e_0 \left( 1 + (\rho-\rho_\mathrm{ST}) 
    \frac{1}{\beta_{\mathrm{eff}}^{\mathrm{ST}}}
    + \ldots  \right) \,,
\end{equation}
where 
\begin{equation} \label{rho_def}
  \rho=\frac{1}{2\pi}\left( r^{(2)} + \frac{n}{n-2}\psi \right) \,.
\end{equation}

The values for these constants for various actions 
are summarized in Table \ref{psi_r_rho}.

\begin{table}[ht] 
  \centering 
  \begin{tabular}[t]{c|c|c|c|c|c} 
    \hline
    $n$ & Action & $\psi/(2 \pi)$ & $r^{(2)}/(2 \pi)$ & $\rho$ & 
    $\rho-\rho_\mathrm{ST}$ \\
    \hline \hline
    3  & ST      & $0.37031535$ & $-1.13861509$ & $-0.02766903$&  \\
    & D(1/3)  & $0.58672731$ & $-1.56413634$ & $\phantom{-}0.19604560$& $0.22371463$ \\
    &D($-1/4$)& $0.26551678$ & $-0.13328534$ & $\phantom{-}0.66326502$& $0.69093405$ \\
    & SYM     & $0.24344345$ &               &              &      \\
    \hline
    4  & ST      & $0.49063070$ & $-0.72946015$ & $\phantom{-}0.25180125$& $$ \\
    & D(1/3)  & $0.79446745$ & $-0.96233225$ & $\phantom{-}0.62660266$& $0.37480141$ \\
    &D($-1/4$)& $0.35356625$ & $\phantom{-}0.22986274$ & $\phantom{-}0.93699525$& $0.68519400$ \\
    & SYM     & $0.31658887$ &               &              &      \\
    \hline
    8  & ST      & $0.97189212$ & $-0.58931931$ & $\phantom{-}0.70653685$ &  \\
    \hline
  \end{tabular} 
  \caption{Values of constants appearing in the description of the
    cutoff effects for different actions.
  }
  \label{psi_r_rho} 
\end{table}

\begin{figure}
  \begin{center}
    \psfig{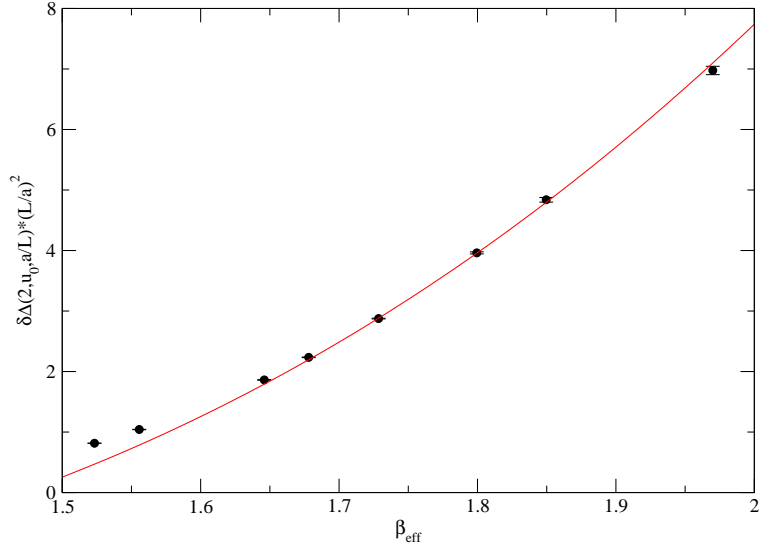}
  \end{center}
  \vspace{-0.5cm}
  \caption{$\delta\Sigma(2,u_0,a/L) (L/a)^2$ for O(3), ST action.
    The solid line is the fit
    $3.1320(\beta^3 + c \beta^2 -0.4883\beta)$,
    with $c=r^{(2)}/(2\pi)=-1.1386$ fixed.
  }
  \label{o3_ST_per_asq}
\end{figure}

\begin{figure}
  \begin{center}
    \psfig{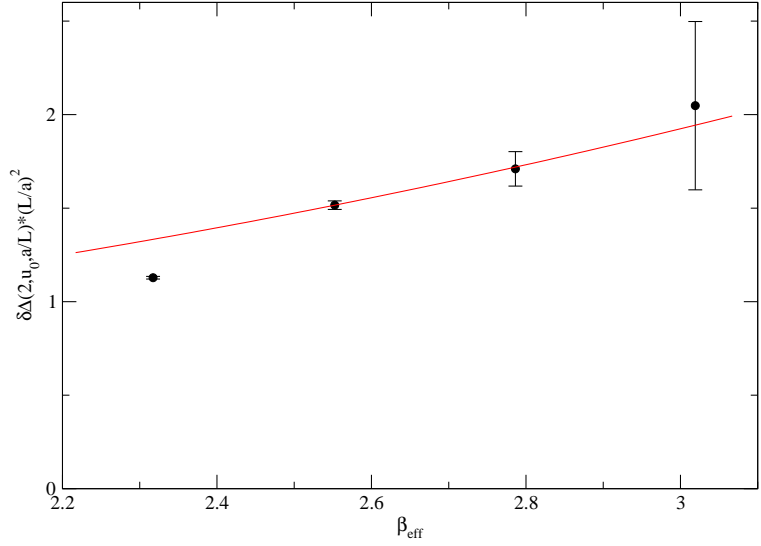}
  \end{center}
  \vspace{-0.5cm}
  \caption{$\delta\Sigma(2,u_0,a/L) (L/a)^2$  for O(4), D(1/3) action.
    The solid line is the fit
    $0.199(\beta^2 + c \beta + 3.566)$,
    with $c=r^{(2)}/(2\pi)=-0.9623$ fixed.
  }
  \label{o4_D3_per_asq}
\end{figure}

Figures \ref{o3_ST_per_asq}.-\ref{o4_D3_per_asq}. 
show the values of $\delta\Sigma(2,u_0,a/L) (L/a)^2$
vs. the corresponding $\beta$ values.
The 2-parameter fits are the predictions 
from \eqref{final},\eqref{final3},
where the overall constant and the sub-subleading correction coefficient 
are fitted, while the value of $r^{(2)}$ is fixed to the known value.

\begin{figure}
  \begin{center}
    \psfig{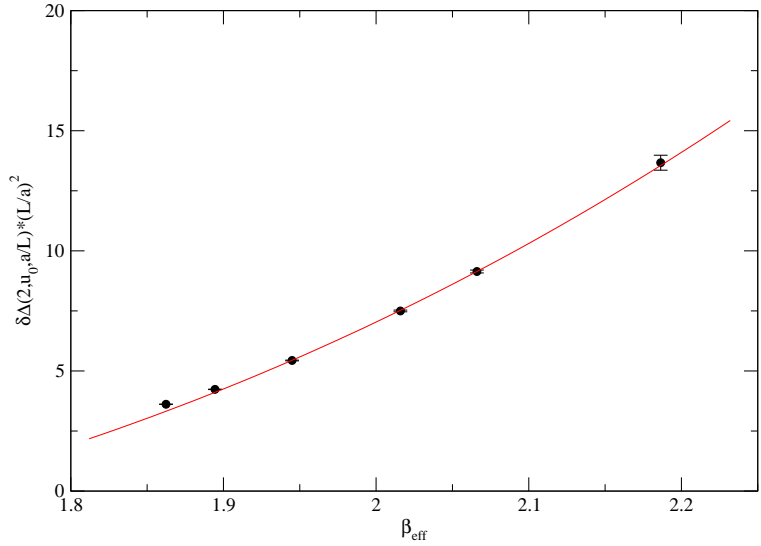}
  \end{center}
  \vspace{-0.5cm}
  \caption{$\delta\Sigma(2,u_0,a/L) (L/a)^2$  for O(3), D(1/3) action.
    The solid line is the fit
    $5.4803(\beta^3 + c \beta^2 -0.2309\beta)$,
    with $c=r^{(2)}/(2\pi)=-1.5641$ fixed.
  }
  \label{o3_D3_per_asq}
\end{figure}

\begin{figure}
  \begin{center}
    \psfig{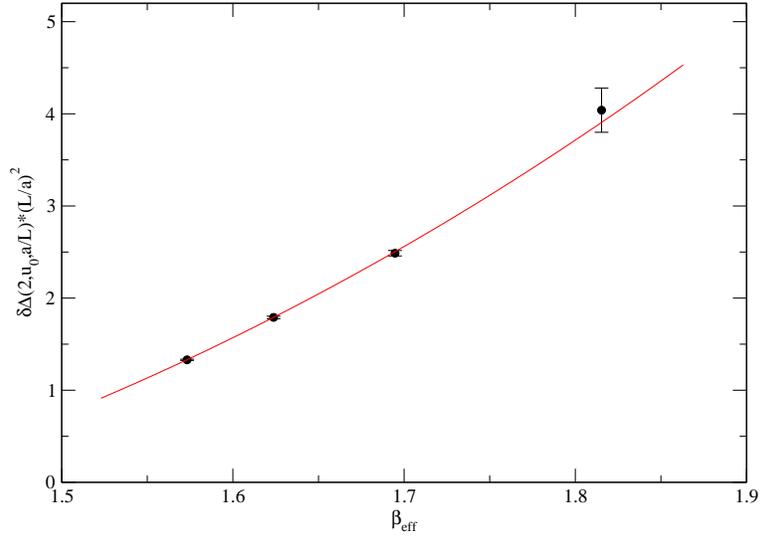}
  \end{center}
  \vspace{-0.5cm}
  \caption{$\delta\Sigma(2,u_0,a/L) (L/a)^2$  for O(3), D($-1/4$) action.
    The solid line is the fit
    $1.6566(\beta^3 + c \beta^2 -1.7554\beta)$,
    with $c=r^{(2)}/(2\pi)=-0.1333$ fixed.
  }
  \label{o3_D4_per_asq}
\end{figure}

\begin{figure}
  \begin{center}
    \psfig{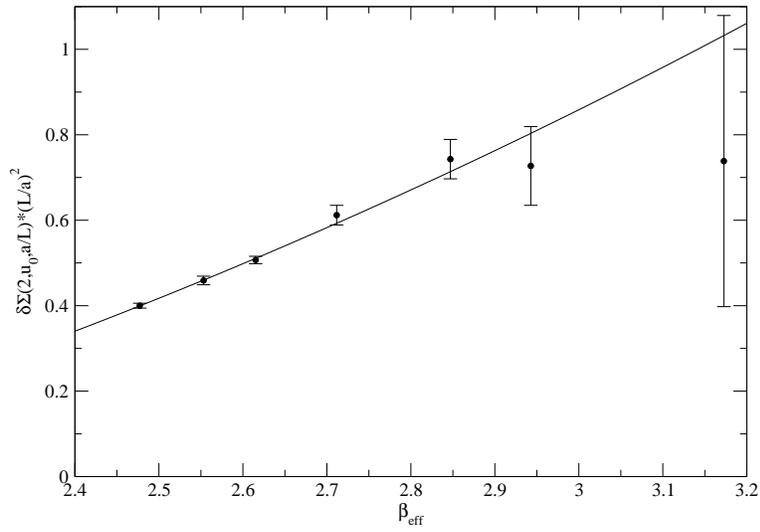}
  \end{center}
  \vspace{-0.5cm}
  \caption{$\delta\Sigma(2,u_0,a/L) (L/a)^2$ for O(4), ST action.
    The solid line is the fit
    $0.185(\beta^2 + c \beta -2.170)$,
    with $c=r^{(2)}/(2\pi)=-0.7295$ fixed.
  }
  \label{o4_ST_per_asq}
\end{figure}

\begin{figure}
  \begin{center}
    \psfig{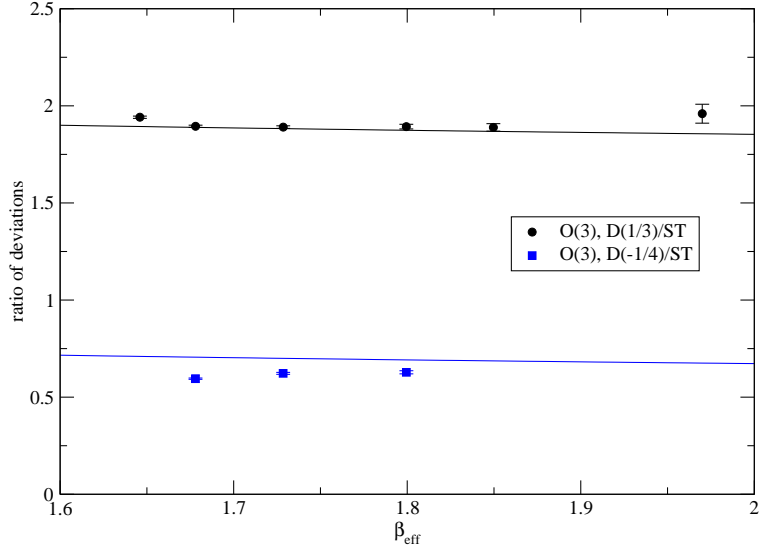}
  \end{center}
  \vspace{-0.5cm}
  \caption{Ratio of deviations $\delta\Sigma(2,u_0,a/L)$
    for the D(1/3)/ST and D($-1/4$)/ST actions vs. 
    $\beta_{\mathrm{eff}}^{\mathrm{ST}}$ for the O(3) model.
    The solid lines are the predictions 
    $5/3(1+0.224/\beta_{\mathrm{eff}}^{\mathrm{ST}})$,
    and 
    $1/2 (1 + 0.691/\beta_{\mathrm{eff}}^{\mathrm{ST}})$.
  }
  \label{rdev_o3_D3_D4_ST}
\end{figure}

A further check is provided by the ratio of artifacts. 
For the O(3) case the ratios for D(1/3)/ST and D($-1/4$)/ST actions 
are shown in Fig.~\ref{rdev_o3_D3_D4_ST} where
the data show a remarkable agreement with the parameter-free prediction
in \eqref{delta_X_ratio}. (Note however that it is possible that this few
percent agreement is due to an accidentally very small coefficient of the
next, $1/\beta^2$, term.)

\begin{figure}
  \begin{center}
    \psfig{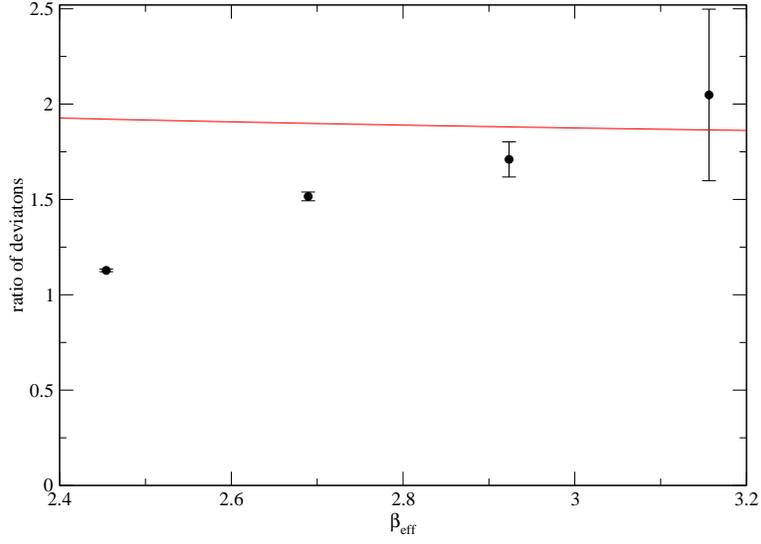}
  \end{center}
  \vspace{-0.5cm}
  \caption{Ratio of deviations $\delta\Sigma(2,u_0,a/L)$
    for the D(1/3) and ST actions vs. $\beta_{\mathrm{eff}}$
    for O(4).
    The solid line is the prediction 
    $5/3(1+0.375/\beta_{\mathrm{eff}}^{\mathrm{ST}})$.
  }
  \label{rdev_o4_D3_ST}
\end{figure}

The corresponding ratio for the O(4) case is shown in
Fig.~\ref{rdev_o4_D3_ST}. Although the agreement is poorer in this
case, at larger $\beta$ (for $L/a=32,\, 64$) the data seem
to approach the prediction.

\begin{figure}
  \begin{center}
    \psfig{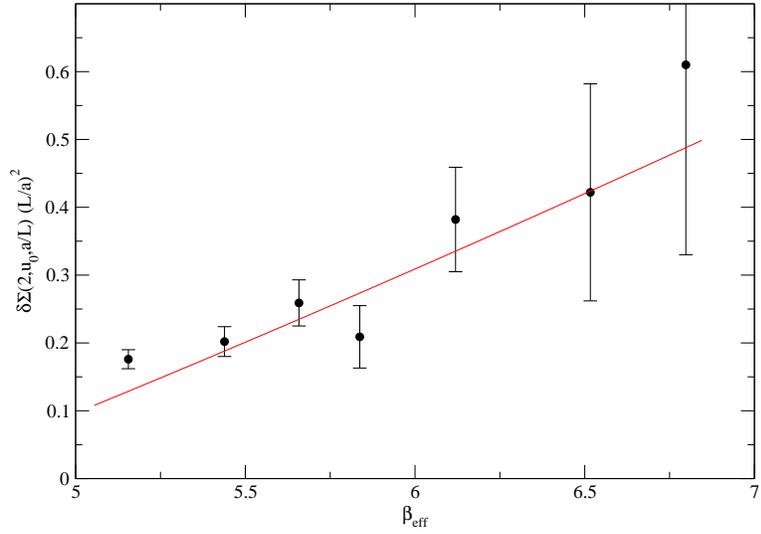}
  \end{center}
  \vspace{-0.5cm}
  \caption{$\delta\Sigma(2,u_0,a/L) (L/a)^2$ for O(8), ST action.
    The solid line is the fit
    $0.093(\beta^{4/3} + c \beta^{1/3} - 6.50)$,
    with $c=r^{(2)}/(2\pi)=-0.5893$ fixed.
  }
  \label{o8_ST_per_asq}
\end{figure}

In \cite{Leder},\cite{KLW} 
the cut-off effects for the step scaling
function were measured for O(4) and O(8) at $u_0=1.0595$ with the ST action.
Since our errors for the O(4) case are considerably smaller, we analyse here
only the O(8) data from these papers. 
Fig.\ref{o8_ST_per_asq} shows that the MC data of \cite{Leder},\cite{KLW} 
are also consistent with the analytic predictions.
(For O(8) one has $\sigma(2,u_0)=1.345757$ at $u_0=1.0595$.)

\section{Conclusions}

The main motivation of this work was to explain how the apparently
linear artifacts found in ref. \cite{HHNSW} can be reconciled with standard 
expectations. We have shown that the data, although astonishingly 
linear as function of the lattice spacing, can equally well be described
by a more complicated formula we calculated using Symanzik's theory of
lattice artifacts.

Although both type of fits describe the data well, we think that by now
there is no doubt that the conservative description of lattice artifacts
based on Symanzik's effective action is correct and there is no unexpected
new physics behind the phenomenon (as originally suspected). Our arguments
can be summarized as follows.

1) Since ref. \cite{HHNSW} appeared, the exact continuum limit has been 
   calculated \cite{BH} by bootstrap techniques and using this extra piece
   of information we see that the \lq\lq linear" fits of Fig.~\ref{lww}
   must be bent
   a little, as seen in Fig.~\ref{all_o3}. In other words, 
   continuum extrapolations
   based on \lq\lq linear" fits would miss the exact result by about 0.002.

2) Lattice artifacts normally decrease very rapidly as $a^2$, but here 
   this is partially compensated by the cubic logarithmic increase of
   the $\beta^3$ factor for O$(3)$. This increase is further enhanced by the
   relatively large negative coefficient of the subleading term. As can
   be seen in Fig.~\ref{o3_ST_per_asq}, 
   the logarithmic correction factor increases by about
   an order of magnitude between $\beta=1.6$ and $\beta=2.0$. This can 
   mimic a $1/a$-like increase in our limited range of $\beta$. 

3) The correctness of the effective action description is further
   corroborated by its parameter-free prediction for the ratios of
   artifacts, which agrees very well with the measured values 
   (see Fig.~\ref{rdev_o3_D3_D4_ST}
   and Fig.~\ref{rdev_o4_D3_ST} 
   for the less spectacular O$(4)$ case).

4) In the large $n$ limit our formula reproduces the results of
   refs. \cite{KLW,BKLW,CarPel}. 

After completing this long computation we arrived at the sobering
conclusion: there are no linear artifacts and Symanzik's theory describes
the data well. But it was nevertheless useful to go through the steps of
the calculation because it provided us with the opportunity to learn about
Symanzik's theory of artifacts and improvement.
Similar computations should, in our opinion, accompany 
precision lattice studies of QCD in order to control and better
estimate systematic errors arising from lattice artifacts for 
extrapolations to the continuum limit.

\clearpage

\section*{Acknowledgments}

J.~B. is grateful to the Max--Planck--Institut f\"ur Physik, where most
of this investigation was carried out, for its
hospitality. The authors are grateful to P. Hasenfratz, who initiated this
work and participated in the early stages of this project.
This research was supported in part by the Hungarian
National Science Fund OTKA (under T049495).


\appendix
\renewcommand{\thesection}{Appendix~A: 
  Anomalous dimensions of the dimension 4 operators}
\section{}
\renewcommand{\thesection}{A}

\subsection{Scalar operators}

\begin{figure}[htb]
  \begin{center}
    \leavevmode
    \epsfxsize=1.4cm
    \epsfbox{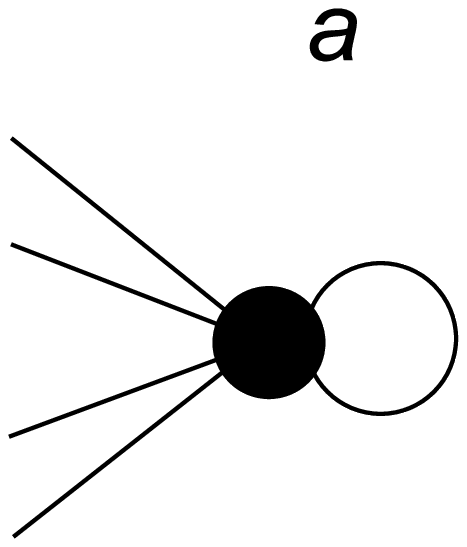}
    \hspace{4mm} 
    \epsfxsize=3.3cm
    \epsfbox{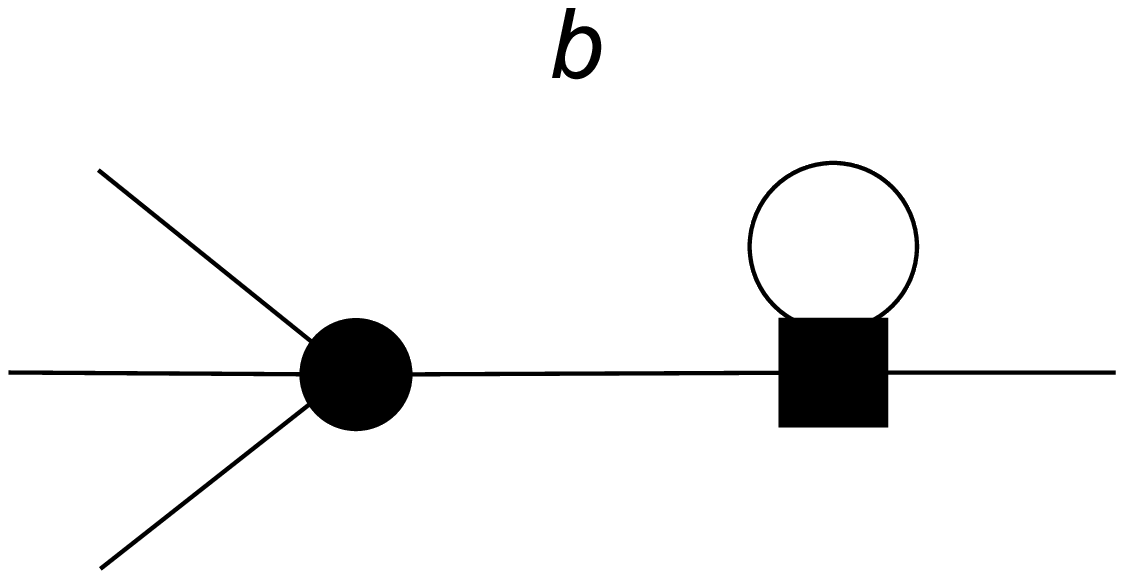}
    \hspace{2mm} 
    \epsfxsize=3.3cm
    \epsfbox{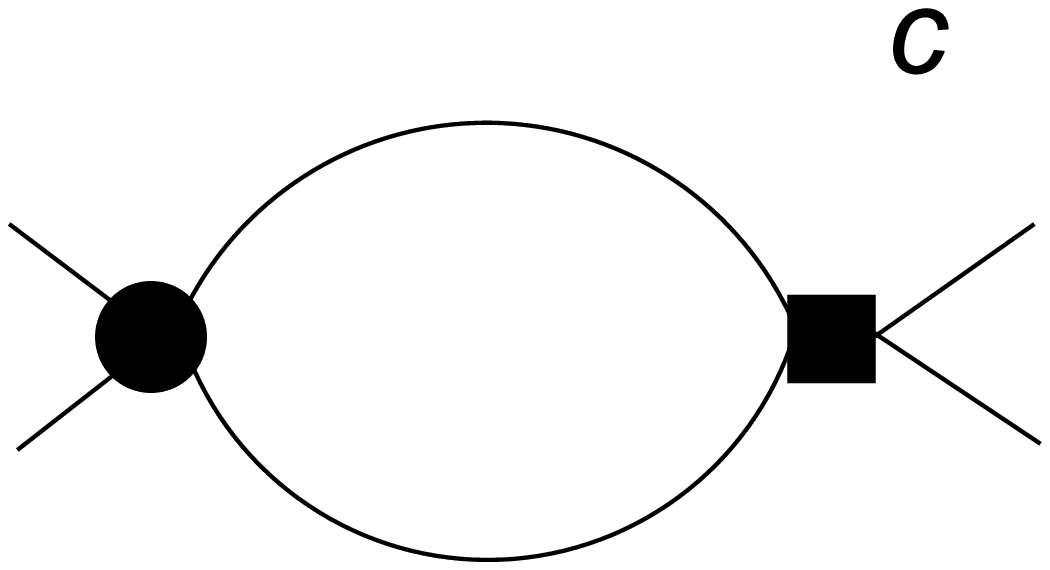}
    \hspace{4mm}
    \epsfxsize=3.3cm
    \epsfbox{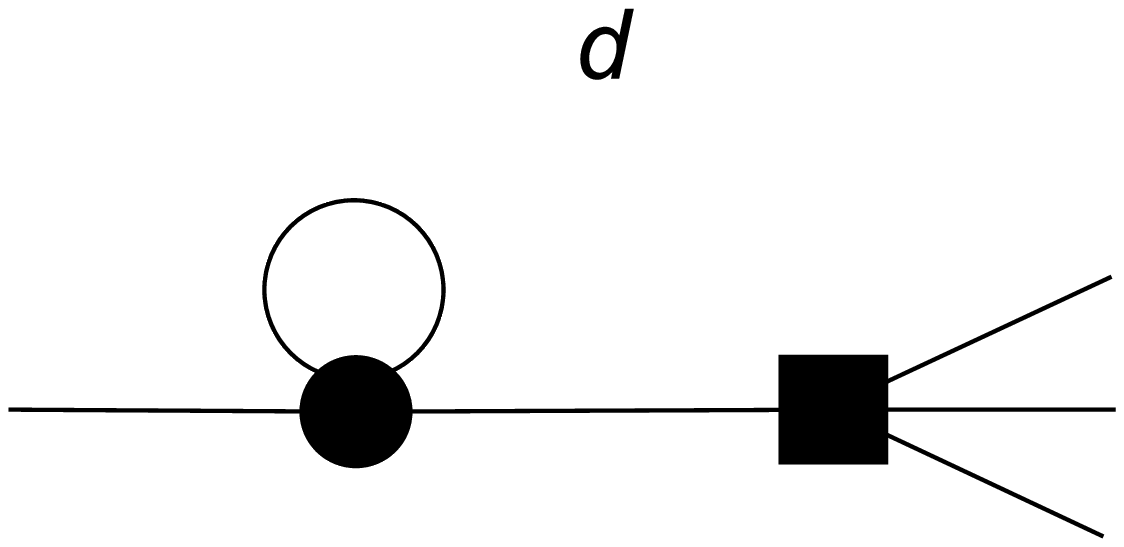}
  \end{center}
  \caption{
    1-loop contributions to the renormalization of operators ${\cal O}_1$
    and ${\cal O}_2$. Full circles
    represent the (4 or 6-leg part) of the operators, 
    full squares stand for the 4-point interaction vertex of the model. 
  }
  \label{1LoopGraphs}
\end{figure}

The one-loop mixing matrix for dimension four invariant
Lorentz scalars has been first computed by Br\'ezin et al. \cite{BZJG} 
using the basis (\ref{calOibasis})
\footnote{where, for simplicity, 
  all operators can be taken at zero momentum (space integrals)}.
It is given by (\ref{Zijgeneral}) with
\footnote{(and we have checked their result)} 
\begin{equation}
  2\pi w_{ij}=
  \begin{pmatrix}4-2n&-4&0&-\frac{n}{2}&0\\
    -2&2-2n&0&-\frac{n+2}{4}&0\\
    4&-16&2-n&-2&4(n-1)\\
    -4&16&0&4-n&4(1-n)\\
    1&-4&\frac{1}{4}&0&2
  \end{pmatrix}\,.
  \label{mix}
\end{equation}
In Fig.~\ref{1LoopGraphs} we show the diagrams which are needed 
for the renormalization of operators ${\cal O}_1$
and ${\cal O}_2$ to 1-loop order.

We will now simplify the mixing problem using some operator identities. 
The identities can be derived by making in (\ref{genfunc}) the 
infinitesimal change of variables
\begin{equation}
  \delta \pi^i=\delta\omega Q^i\,,
\end{equation}
where $\delta\omega$ is an infinitesimal parameter and $Q^i$ is an arbitrary
local expression of the fields and sources. 
The corresponding change in the action is
\begin{equation}
  \delta {\cal A}=-\delta\omega\int{\rm d}^Dx\left(\square\pi^i-\alpha \pi^i+
    \frac{1}{g_0} I^i\right)Q^i
\end{equation}
and since the measure $\left({\cal D}\pi\right)$ is invariant in dimensional
regularization under any local change of variables we have 
\begin{equation}
  {\cal Z}[I]=\int\left({\cal D}\pi\right)\,
  {\rm e}^{-{\cal A}-\delta {\cal A}}=
  {\cal Z}[I]-\int\left({\cal D}\pi\right)\,
  \delta {\cal A}{\rm e}^{-{\cal A}}\,,
\end{equation}
leading to the operator identity $\delta {\cal A}=0$.
Thus we have
\begin{equation}
  \left(\square\pi^i-\alpha \pi^i+\frac{1}{g_0} I^i\right)Q^i\approx0\,,
\end{equation}
where $\approx$ means operator identity for the corresponding space integrals.

We now consider the infinitesimal transformations corresponding to
respectively
\begin{equation}
  Q^i=\left\{g_0^2\left(\partial_\mu S\cdot\partial_\mu S\right)\pi^i\,,
    g_0^2\left(I\cdot S+\bar\alpha\right)\pi^i,g_0 I^i+
    g_0^2\bar\alpha\pi^i,g_0^2\square\pi^i\right\}
\end{equation}
and find the identities
\begin{equation}
  \begin{split}
    \left(I\cdot S\right)\left(\partial_\mu S\cdot\partial_\mu S\right)&\approx
    \bar\alpha\,\partial_\mu S\cdot\partial_\mu S\,,\\
    \left(I\cdot S\right)^2&\approx\bar\alpha^2\,,\\
    I^2+I\cdot\square S+\left(I\cdot S\right)\,(\partial_\mu S\cdot\partial_\mu S)
    &\approx\bar\alpha^2\,,\\
    I\cdot\square S-(\partial_\mu S\cdot\partial_\mu S)^2+
    \square S\cdot\square S&\approx
    -\bar\alpha (\partial_\mu S\cdot\partial_\mu S)\,,
  \end{split}
  \label{A7}
\end{equation}
respectively.
Here we have introduced the operator
\begin{equation}
  \bar\alpha=\alpha +\partial_\mu S\cdot\partial_\mu S\,. 
\end{equation}
We see that insertions of the apparently O$(n)$ non-invariant operators 
${\cal O}_4$ and ${\cal O}_5$ are equivalent to inserting the manifestly 
invariant ones in (\ref{A7}).
There are no further identities independent of the ones in (\ref{A7}). This
can be shown by using the continuum analog of the lattice considerations
of ref. \cite{Sym2}.

It will be convenient to use a new basis of operators. We keep
the \lq\lq hard" operators ${\cal O}_1$ and ${\cal O}_2$ but instead of the
rest we first introduce $W,U$ and $V$, where
\begin{equation}
  \begin{split}
    4W&=\bar\alpha\,\partial_\mu S\cdot\partial_\mu S\,,\\
    8U&=\bar\alpha^2\,,\\
    2V&=\square S\cdot\square S-(\partial_\mu S\cdot\partial_\mu S)^2
    +\bar\alpha^2\,.
  \end{split}
\end{equation}
This basis change corresponds to
\begin{equation}
  \begin{split}
    {\cal O}_3&=V-4U+4{\cal O}_1\,,\\
    {\cal O}_4&=2W-4{\cal O}_1\,,\\
    {\cal O}_5&=U-W+{\cal O}_1\,.
  \end{split}
\end{equation}
Further we introduce the combinations
\begin{equation}
  \overline{U}_3=W-2U+\frac{1}{2}V\,,\qquad\qquad 
  \overline{U}_4=U-\frac{1}{4n}V\,,\qquad\qquad 
  \overline{U}_5=V\,.
\end{equation}
Now the operator identities can be rewritten as
\begin{equation}
  \begin{split}
    \overline{U}_3&\approx -\frac{1}{4}I\cdot\square S\,,\\
    \overline{U}_4&\approx\frac{1}{8}I^aI^b\tau^{ab}\,,\\
    \overline{U}_5&\approx \frac{1}{2}I^2\,,
  \end{split}
\label{A12}
\end{equation}
where
\begin{equation}
  \tau^{ab}=S^aS^b-\frac{1}{n}\delta^{ab}\,.
\end{equation}

Inserting the above identities into the generating functional 
(\ref{opgenfunc}) we can derive useful identities for the correlation
functions of the operators $\overline{U}_j\,,j=3,4,5$. These
are best formulated in Fourier space. We define, as usual,
\begin{equation}
  \begin{split}
    \int{\rm d}^Dx\,\rme^{i(p_1x_1+\dots+p_rx_r)}\,
    &{\cal G}^{a_1\dots a_r}_{[\bar U]}(x_1,\dots,x_r)\\
    &=(2\pi)^D\delta(p_1+\dots +p_r)\,\widetilde{\cal G}^{a_1\dots a_r}_{[\bar U]}
    (p_1,\dots,p_r)
  \end{split}
\end{equation}
for correlation functions with and without operator insertions. With
this notation we have
\begin{equation}
  \widetilde{\cal G}_{\overline{U}_3}^{a_1\dots a_r}(p_1,\dots,p_r)=
  \frac{g_0^2}{4}\,\left(\sum_{k=1}^rp_k^2\right)\,
  \widetilde{\cal G}^{a_1\dots a_r}(p_1,\dots,p_r)\,.
\end{equation}
From this formula it is clear that the operator $\overline{U}_3$ is
renormalized multiplicatively with renormalization constant
\begin{equation}
  Z_{\overline{U}_3}=\frac{1}{Z_1}=1-\frac{(2-n)g^2}{2\pi\varepsilon}+\dots
  \label{renW}
\end{equation}

For $\overline{U}_4$ we have
\begin{equation}
  \begin{split}
    \widetilde{\cal G}_{\overline{U}_4}^{a_1\dots a_r}(p_1,\dots,p_r)&=
    \frac{g_0^4}{4}\,
    \widetilde{\cal G}_\tau^{a_1a_2;a_3\dots a_r}(p_1+p_2;p_3,\dots,p_r)
    +\left[\begin{pmatrix}r\\2\end{pmatrix}-1\right] {\rm perms}\\
    &+\frac{m^2g_0^2}{4}\,
    \widetilde{\cal G}_\tau^{a_1n;a_2\dots a_r}(p_1;p_2,\dots,p_r)
    +\left[r-1\right] {\rm perms}\\
    &+\frac{m^4}{8}\,
    \widetilde{\cal G}_\tau^{nn;a_1\dots a_r}(0;p_1,\dots,p_r)\,,
  \end{split}
  \label{barU}
\end{equation}
where
\begin{equation}
  \widetilde{\cal G}_\tau^{ab;a_1\dots a_r}(p;p_1,\dots,p_r)
\end{equation}
is the Fourier space correlation function of the local operator 
$\tau^{ab}$ at momentum $p$. Actually, (\ref{barU}) is valid for
$r>2$ only. For $r=2$ we have
\begin{equation}
  \begin{split}
    \widetilde{\cal G}_{\overline{U}_4}^{a_1a_2}(p,-p)&=
    \frac{g_0^4}{4}\langle\tau^{a_1a_2}\rangle
    +\frac{m^2g_0^2}{4}\widetilde{\cal G}_\tau^{a_1n;a_2}(p;-p)\\
    &+\frac{m^2g_0^2}{4}\widetilde{\cal G}_\tau^{a_2n;a_1}(-p;p)
    +\frac{m^4}{8}\widetilde{\cal G}_\tau^{nn;a_1a_2}(0;p,-p)\,.
  \end{split}
  \label{barU2}
\end{equation}
Since in (\ref{barU}) (and in (\ref{barU2})) all terms require the
same overall renormalization constant, the operator $\overline{U}_4$
renormalizes multiplicatively with renormalization constant
\begin{equation}
  Z_{\overline{U}_4}=\frac{ZZ_\tau}{Z_1^2}=
  1-\frac{(3-2n)g^2}{2\pi\varepsilon}+\dots,
  \label{renU}
\end{equation}
where we have used the result
\begin{equation}
  Z_{\tau}=1+\frac{ng^2}{2\pi\varepsilon}+\dots
\end{equation}

Finally for $\overline{U}_5$ we find
\begin{equation}
  \begin{split}
    \widetilde{\cal G}_{\overline{U}_5}^{a_1\dots a_r}(p_1,\dots,p_r)&=
    (2\pi)^D\delta(p_1+p_2)g_0^4\,
    \widetilde{\cal G}^{a_3\dots a_r}(p_3,\dots,p_r)\delta^{a_1a_2}
    +\left[\begin{pmatrix}r\\2\end{pmatrix}-1\right] {\rm perms}\\
    &+(2\pi)^D\delta(p_1)\delta^{a_1n}m^2g_0^2\,
    \widetilde{\cal G}^{a_2\dots a_r}(p_2,\dots,p_r)
    +\left[r-1\right] {\rm perms}
  \end{split}
\end{equation}
for $r>2$ and
\begin{equation}
  \widetilde{\cal G}_{\overline{U}_5}^{a_1a_2}(p,-p)=g_0^4\delta^{a_1a_2}+2m^2g_0^2
  \delta^{a_1n}\delta^{a_2n}\langle\sigma\rangle\delta(p)(2\pi)^D
\end{equation}
for $r=2$. Again, $\overline{U}_5$ renormalizes multiplicatively with
\begin{equation}
  Z_{\overline{U}_5}=\frac{Z}{Z_1^2}=
  1-\frac{(3-n)g^2}{2\pi\varepsilon}+\dots
  \label{renV}
\end{equation}
We can check the results (\ref{renW}), (\ref{renU}) and (\ref{renV}) by using
the mixing matrix (\ref{mix}). We define the linear operator
\begin{equation}
  C({\cal O}_i)=2\pi w_{ij}{\cal O}_j
\end{equation}
and then find 
\begin{equation}
  C(\overline{U}_3)=(2-n)\overline{U}_3\,,\qquad
  C(\overline{U}_4)=(3-2n)\overline{U}_4\quad{\rm and}\quad
  C(\overline{U}_5)=(3-n)\overline{U}_5\,.
\end{equation}
There are two other linear combinations $\overline{U}_i\,,i=1,2$ 
diagonalizing the one-loop mixing matrix (\ref{mix}) with
\begin{equation}
  C(\overline{U}_1)=2\overline{U}_1\qquad\qquad{\rm and}\qquad\qquad 
  C(\overline{U}_2)=(4-2n)\overline{U}_2\,.
\end{equation}
The relations between the bases are given by
\begin{equation}
  \overline{U}_i={\cal S}_{ij}{\cal O}_j\,,
\end{equation}
with the matrix ${\cal S}$ given by
\begin{equation}
  {\cal S}=\begin{pmatrix}
    \frac{n+4}{2n-1}&-2&\frac{(n-2)(n+1)}{2n(2n-1)}
    &\frac{(n-2)(n+1)}{2n(2n-1)}&\frac{2(n-2)(n+1)}{2n-1}\\
    5&-2&-\frac{1}{n-2}&\frac{2n-5}{n-2}&4\\
    0&0&\frac12&\frac12&0\\
    1&0&-\frac{1}{4n}&\frac{n-1}{2n}&\frac{n-1}{n}\\
    0&0&1&2&4
  \end{pmatrix}\,.
  \label{OtoU}
\end{equation}
For later use we also write down the inverse of this matrix
\begin{equation}
  {\cal S}^{-1}=\begin{pmatrix}
    \frac{1}{n-1}&-\frac{1}{n-1}&\frac{n^2-2n-4}{n(n-2)}&
    \frac{2(n+4)}{2n-1}&-\frac{(n^2-n-4)}{2n(n-1)}\\
    \frac{1}{2(n-1)}&-\frac{n}{2(n-1)}&
    \frac{n^2-4n-4}{2n(n-2)}&
    \frac{5n+2}{2n-1}&-\frac{(n^2-3n-2)}{4n(n-1)}\\
    \frac{4}{n-1}&-\frac{4}{n-1}&
    \frac{4(n^2-2n-4)}{n(n-2)}&
    \frac{36}{2n-1}&-\frac{(n^2-9)}{n(n-1)}\\
    -\frac{4}{n-1}&\frac{4}{n-1}&
    -\frac{2(n^2-2n-8)}{n(n-2)}&
    -\frac{36}{2n-1}&\frac{n^2-9}{n(n-1)}\\
    \frac{1}{n-1}&-\frac{1}{n-1}&-\frac{4}{n(n-2)}&
    \frac{9}{2n-1}&-\frac{(n-9)}{4n(n-1)}
  \end{pmatrix}\,.
  \label{UtoO}
\end{equation}

\subsection{Tensor operators}

Here we discuss some properties of the four-index symmetric tensor operators
defined in (\ref{ttensor}) and (\ref{ktensor}). 

Let $X_{\mu\nu\rho\sigma}$ be a tensor operator, totally symmetric in its 
indices $\mu$, $\nu$, $\rho$, $\sigma$. We can construct the corresponding 
totally symmetric, traceless tensor operator $\widehat X_{\mu\nu\rho\sigma}$
by subtracting trace terms as follows.
\begin{equation}
  \begin{split}
    \widehat X_{\mu\nu\rho\sigma}&=X_{\mu\nu\rho\sigma}-\frac{1}{D+4}
    \Big\{g_{\mu\nu}\xi_{\rho\sigma}+g_{\rho\sigma}\xi_{\mu\nu}+
    g_{\mu\rho}\xi_{\nu\sigma}+g_{\nu\sigma}\xi_{\mu\rho}\\
    &+g_{\mu\sigma}\xi_{\nu\rho}+g_{\nu\rho}\xi_{\mu\sigma}\Big\}+
    \frac{\bar\xi}{(D+2)(D+4)}\Big\{
    g_{\mu\nu}g_{\rho\sigma}+
    g_{\mu\rho}g_{\nu\sigma}+
    g_{\mu\sigma}g_{\nu\rho}\Big\}\,,
  \end{split}
  \label{trless}
\end{equation}
where
\begin{equation}
  \xi_{\rho\sigma}=X_{\alpha\alpha\rho\sigma},\qquad\qquad
  {\rm and}\qquad\qquad \bar\xi=X_{\alpha\alpha\beta\beta}\,.
\end{equation}
In the effective Lagrangian there are operators of the form 
$\sum_{\mu=1}^DX_{\mu\mu\mu\mu}$, which can be rewritten, 
using (\ref{trless}) as
\begin{equation}
  \sum_{\mu=1}^DX_{\mu\mu\mu\mu}=\sum_{\mu=1}^D\widehat X_{\mu\mu\mu\mu}+
  \frac{3}{4-\varepsilon}X_{\alpha\alpha\beta\beta}\,.
  \label{tensor40}
\end{equation}

If we apply this procedure to the operator $t_{\mu\nu\rho\sigma}$
in (\ref{ttensor})we get
\begin{equation}
  \sum_{\mu=1}^DS\cdot \partial^4_\mu\,S=A+
  \frac{6}{4-\varepsilon}{\cal O}_3\,,
\end{equation}
where $A$ is defined in (\ref{Aop}).
The traceless symmetric tensor operator 
$\widehat t_{\mu\nu\rho\sigma}$ can mix under renormalization with 
other traceless symmetric tensor operators only (which must have the
same values for other quantum numbers). All tensor components renormalize
the same way and using this property we can simplify the renormalization
problem by considering the $\widehat t_{++++}$ tensor component, where 
\begin{equation}
  V_+=V_1+iV_2
\end{equation}
for any vector. Great simplification occurs in PT calculations for
this tensor component since $g_{++}=0$ and many diagrams vanish,
and it is thus easier to renormalize
\begin{equation}
  \bar A=\widehat t_{++++}=\partial^2_+S\cdot \partial^2_+ S
\end{equation}
instead of $A$ itself. (Note that we are considering operators at
zero momentum, i.e. their space integrals.)

$\widehat t_{\mu\nu\rho\sigma}$ can mix under renormalization with
$\widehat k_{\mu\nu\rho\sigma}$ given in (\ref{ktensor}).
For this operator we find
\begin{equation}
  \sum_{\mu=1}^D\left(\partial_\mu S\cdot \partial_\mu S\right)^2=B+
  \frac{8}{4-\varepsilon}\left({\cal O}_1+2{\cal O}_2\right)\,,
\end{equation}
where $B$ is defined in (\ref{Bop}).
For convenience we define
\begin{equation}
  \bar B=\widehat k_{++++}=
  \left(\partial_+S\cdot \partial_+ S\right)^2\,.
\end{equation}

There are no other spin--four, dimension--four O$(n)$ invariant operators,
therefore we have a $2\times2$ renormalization problem here. 
By calculating the matrix elements of $\bar A$, $\bar B$
we find that the mixing matrix is diagonal,
\begin{equation}
  K_{ij}=\delta_{ij}+\frac{g^2(n-2)}{2\pi\varepsilon}
  \begin{pmatrix}0&0\\0&1\end{pmatrix}+\dots,
\end{equation}
for the operators
\begin{equation}
  U_6=\frac{1}{g_0^2}A\,,\qquad\qquad
  U_7=\frac{1}{g_0^2}B\,.
\end{equation}

\subsection{2-loop Mixing matrix elements}

For the computation of the mixing coefficients $\nu^{(2)}_{ij}$,
we worked in the Br\'ezin et al basis (\ref{calOibasis})
and computed the corresponding $p_{ij}$ coefficients of the 
$g^4/\varepsilon$ terms in (\ref{Zijgeneral}). 
From the relation between the bases
(\ref{OtoU}) the $\nu_{ij}^{(2)}$ are then given by
\begin{equation} 
  \nu_{ij}^{(2)}=-\frac{(n-2)}{(2\pi)^2}\delta_{ij}
  +{\cal S}_{ik}p_{kl}({\cal S}^{-1})_{lj}\,. 
\end{equation}
For the computation of $r^{(2)}_{\rm III}$
we only require $\nu_{ij}^{(2)}$ for $i,j=1,2$. For this purpose
we only need the $p_{ij}$ for $i=1,2$ because using the fact
that $K_{ij}=0$ for $i=3,4,5$ and $j=1,2$ we have
\begin{equation}
  \sum_{k=1}^5{\cal S}_{ik}(p{\cal S}^{-1})_{kj}
  =\sum_{k=1}^2 {\cal T}_{ik}(p{\cal S}^{-1})_{kj}\,,\,\,\,\,i,j\in\{1,2\}\,,
\end{equation}
with
\begin{equation}
  {\cal T}_{ik}={\cal S}_{ik}-\sum_{r,s=3}^5{\cal S}_{ir}\sigma_{rs}{\cal 
    S}_{sk}
  \,,\,\,\,\,i,k\in\{1,2\}\,,
\end{equation}
where $\sigma$ is the inverse of the lower $3\times3$ diagonal block 
of ${\cal S}$ (i.e. $\sum_{s=3}^5 \sigma_{rs}{\cal S}_{st}=\delta_{rt}$). 
For the computation of $p_{ij}$ for $j=3,4,5$ we need to compute 
the divergent parts of insertions of ${\cal O}_i$ in the 2-point pion 
correlation functions (with just up to one interaction vertex), and 
for the computation of $p_{ij}$ for $1,2$ we need the analogous
computation for 
the insertions of ${\cal O}_i$ in the 4-point pion correlation functions
(with up to two interaction vertices). 

\begin{figure}[ht]
  \begin{center}
    \leavevmode
    \epsfxsize=3.3cm
    \epsfbox{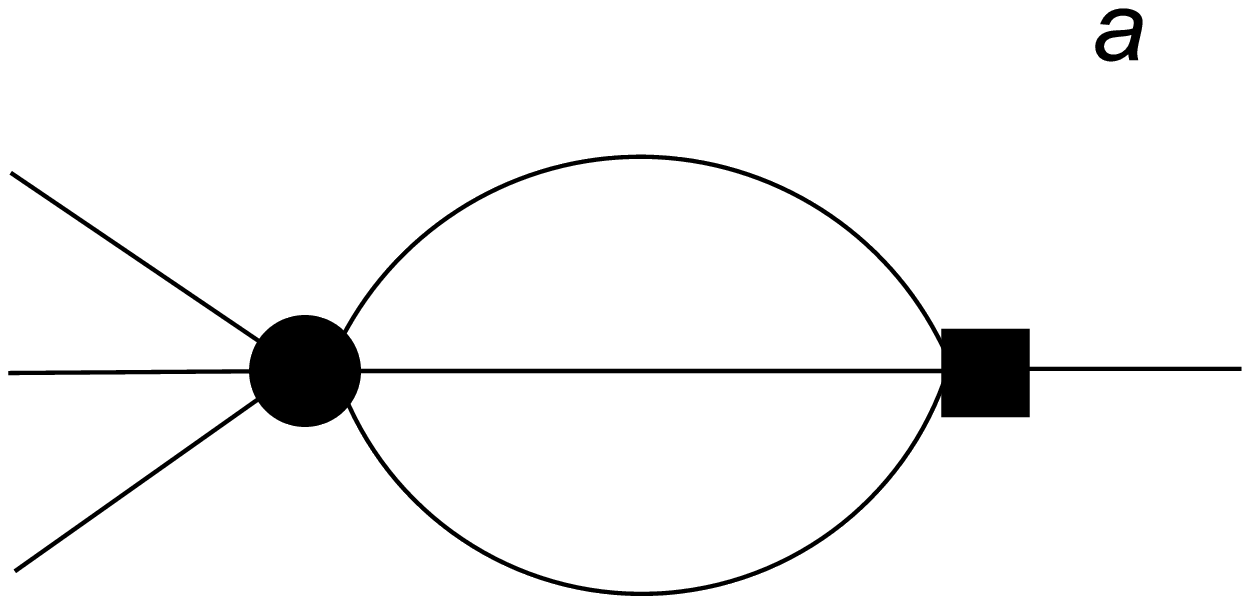}
    \hspace{3mm} 
    \epsfxsize=3.3cm
    \epsfbox{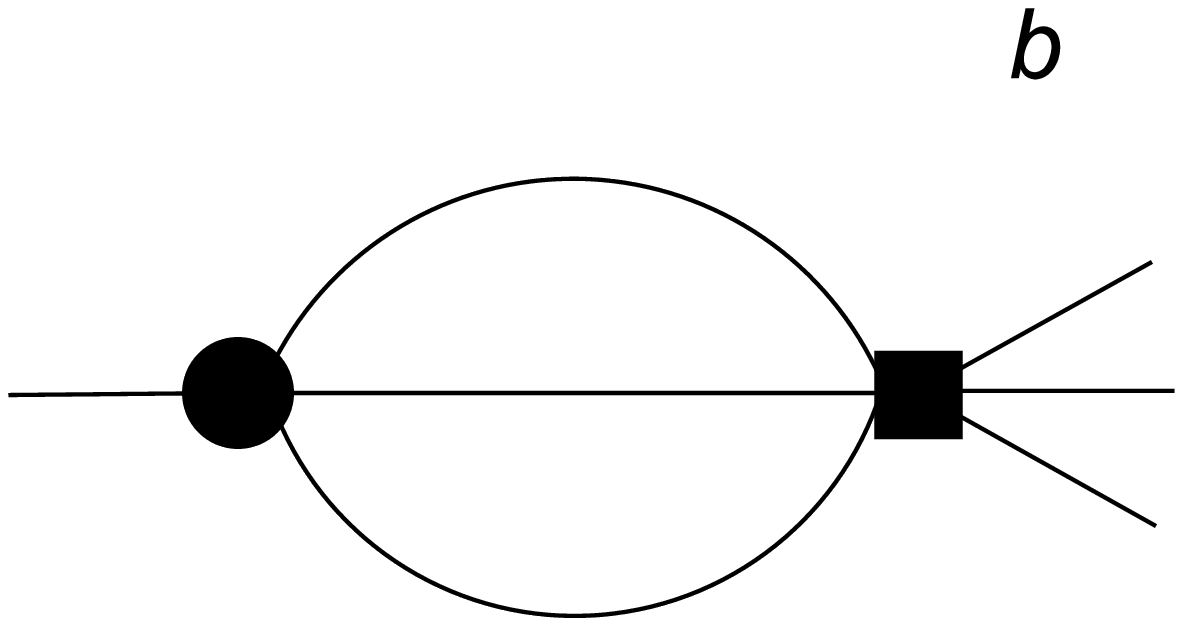}
    \hspace{5mm} 
    \epsfxsize=3.0cm
    \epsfbox{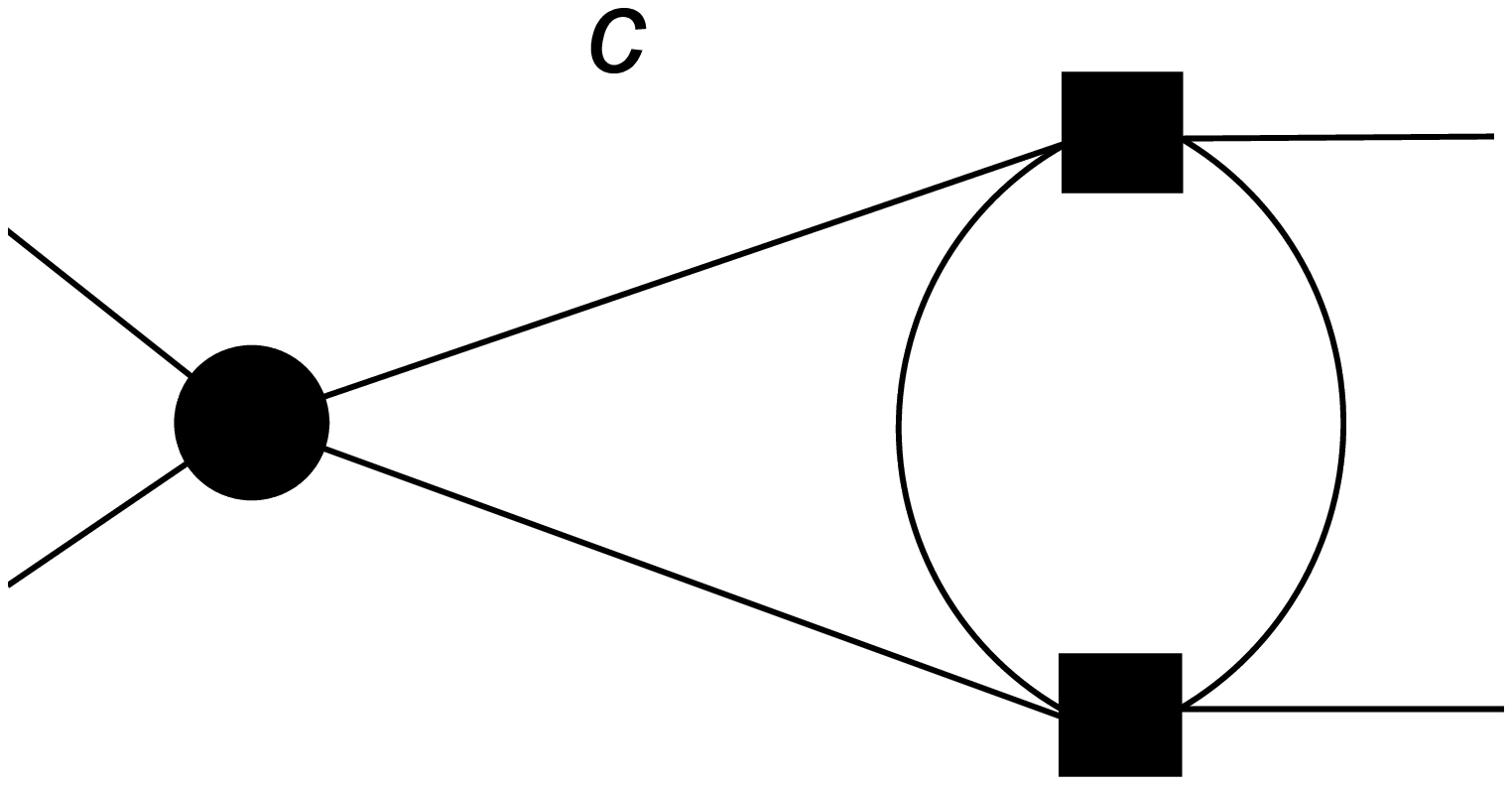}
    \vskip 1.0cm
    \epsfxsize=3.0cm
    \epsfbox{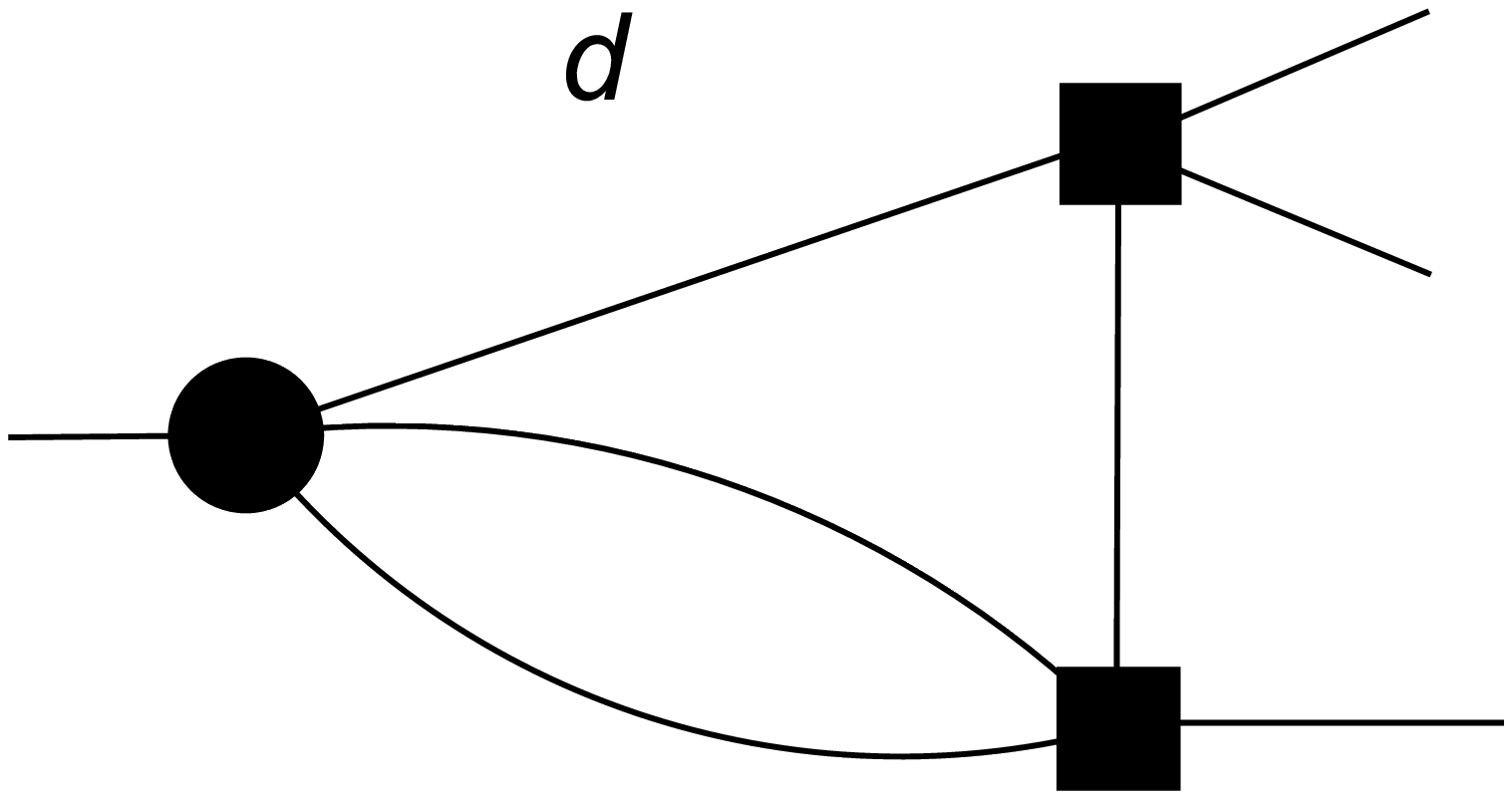}
    \hspace{5mm} 
    \epsfxsize=3.0cm
    \epsfbox{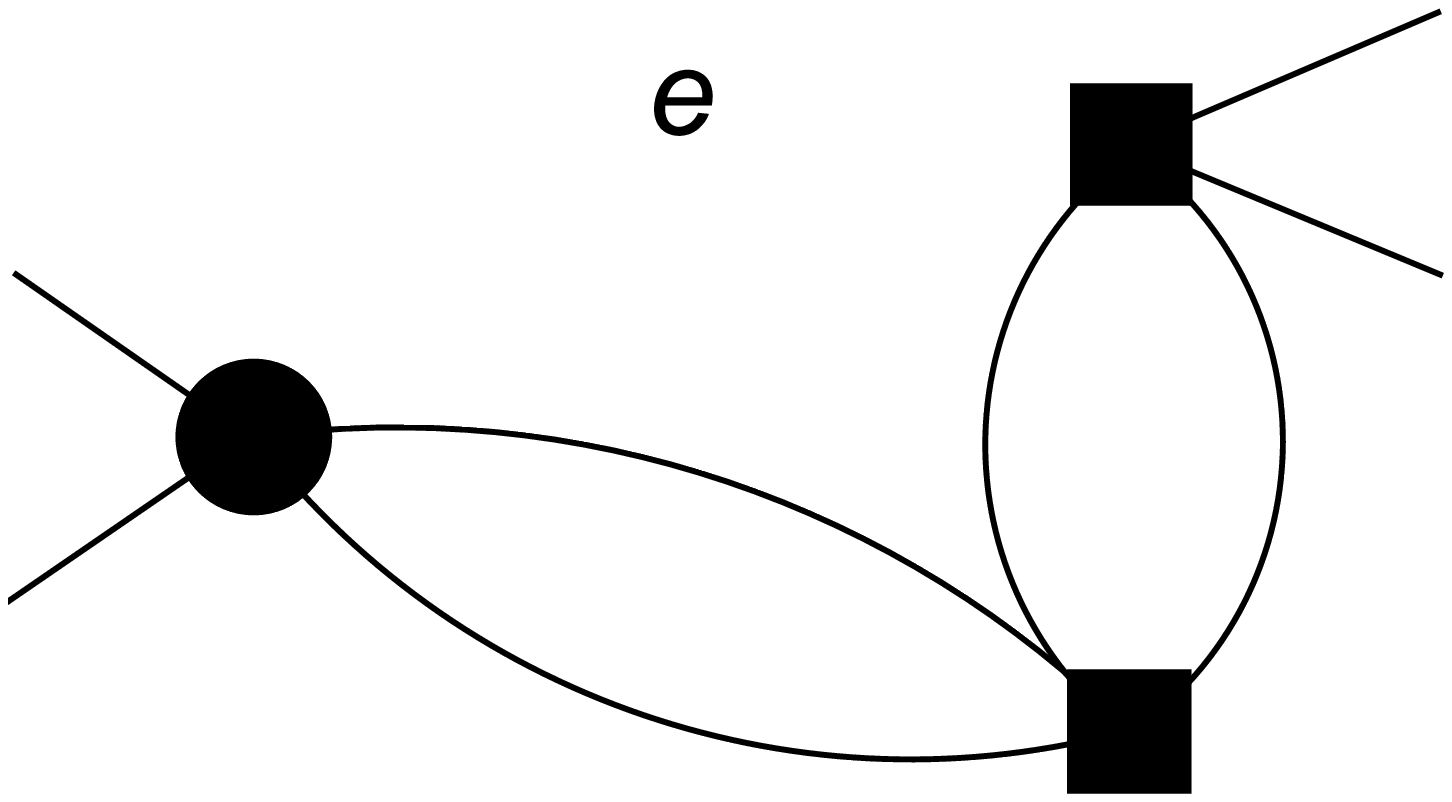}
    \hspace{5mm}
    \epsfxsize=2.3cm
    \epsfysize=2.1cm
    \epsfbox{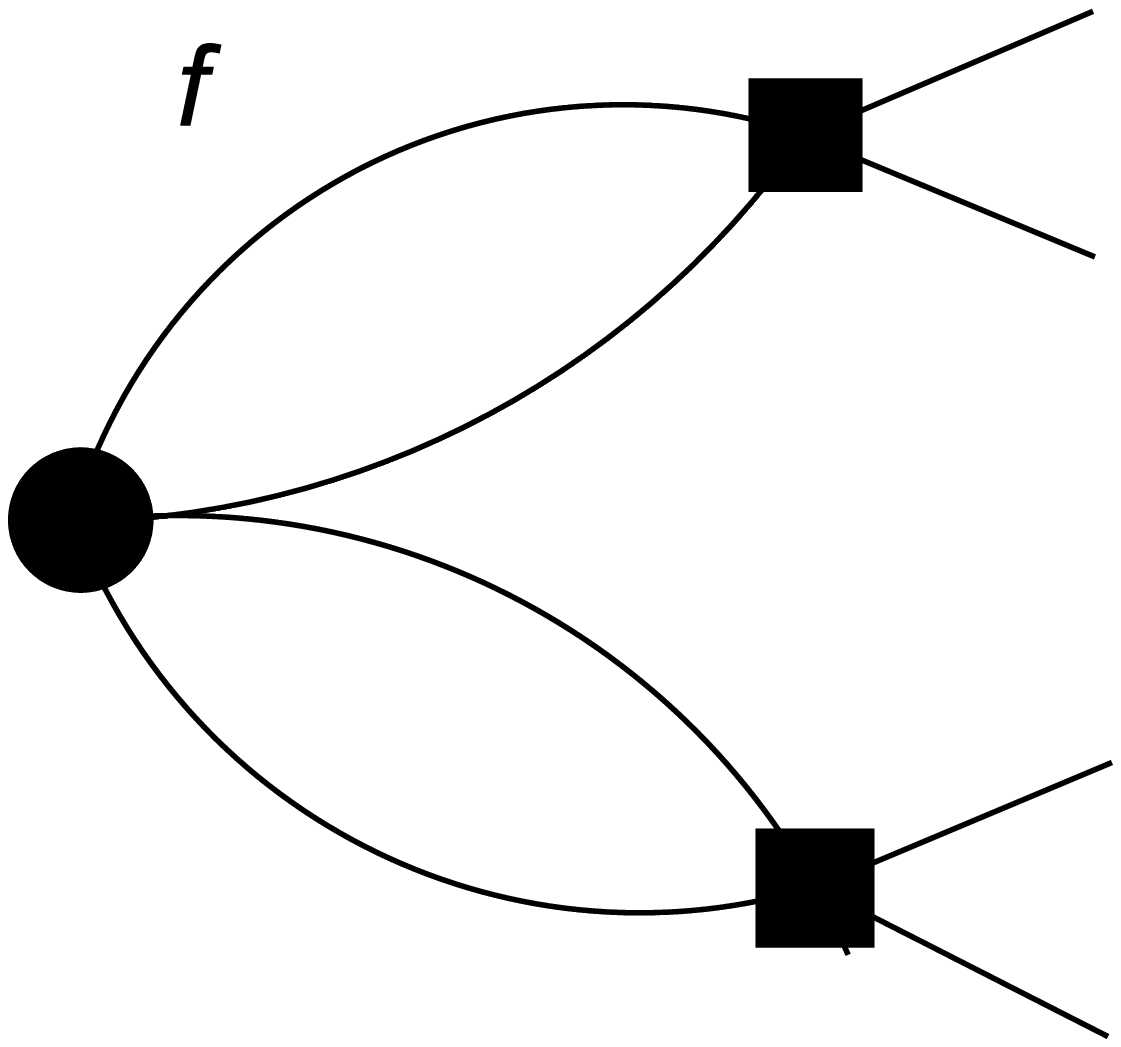}
    \vskip 1.0cm
    \epsfxsize=2.8cm
    \epsfbox{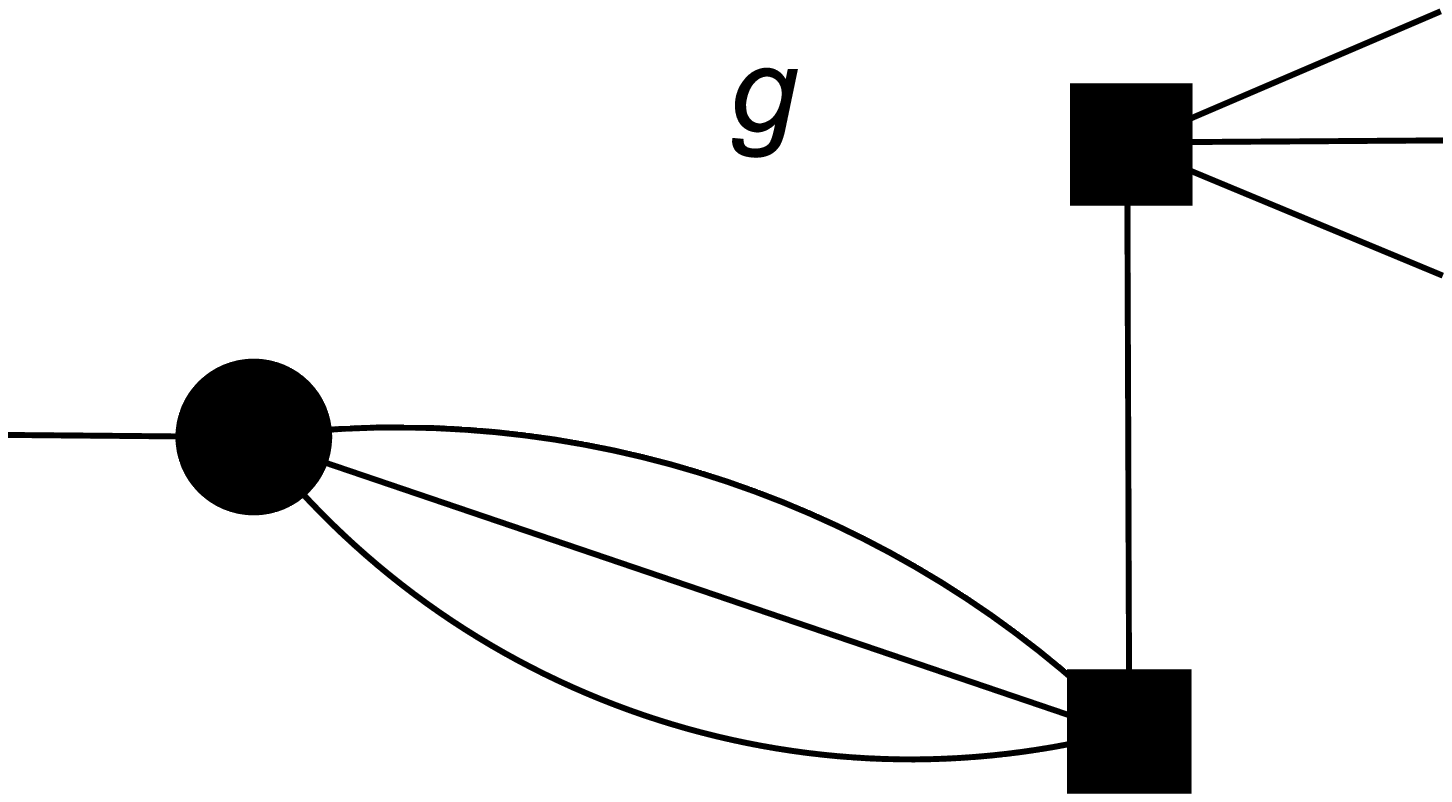}
    \hspace{7mm} 
    \epsfxsize=2.8cm
    \epsfbox{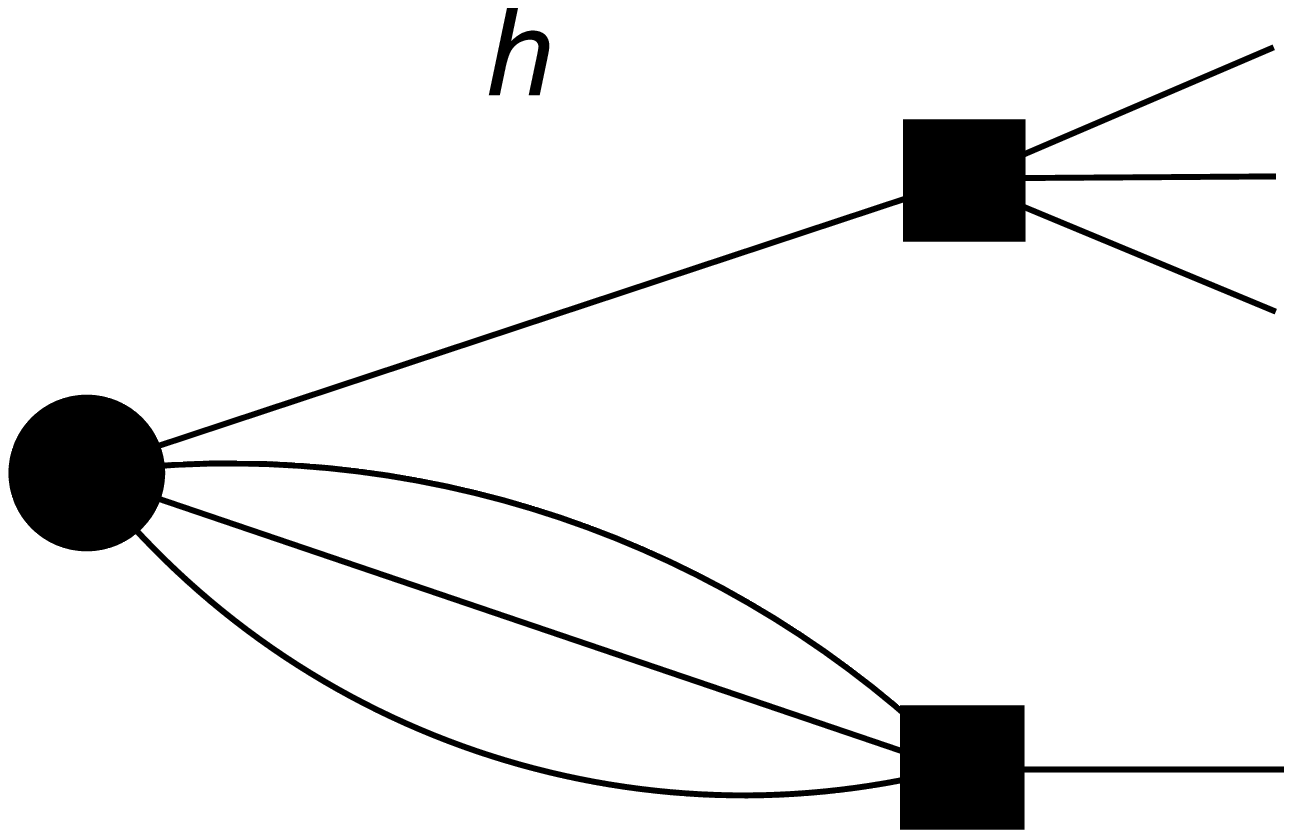}
    \hspace{7mm} 
    \epsfxsize=1.6cm
    \epsfbox{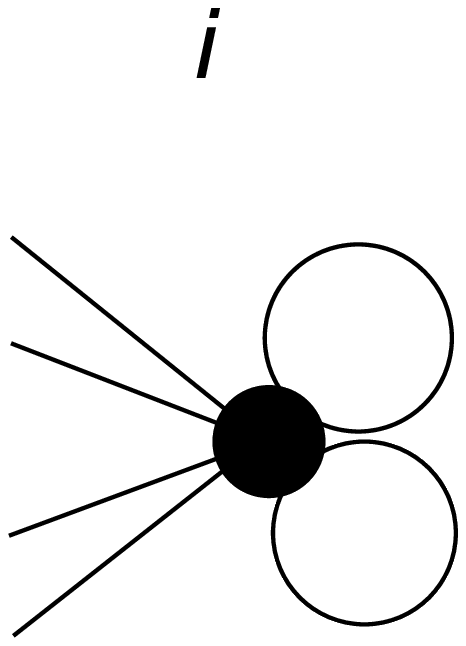}
    \vskip 1.0cm
    \epsfxsize=3.4cm
    \epsfbox{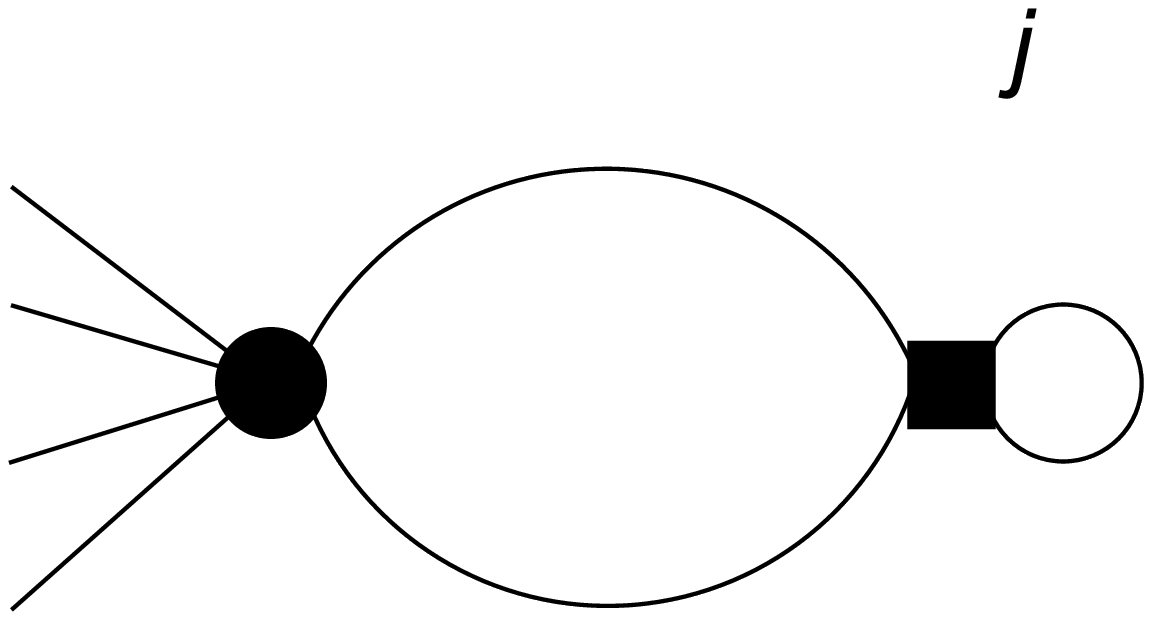}
    \hspace{3mm} 
    \epsfxsize=3.4cm
    \epsfbox{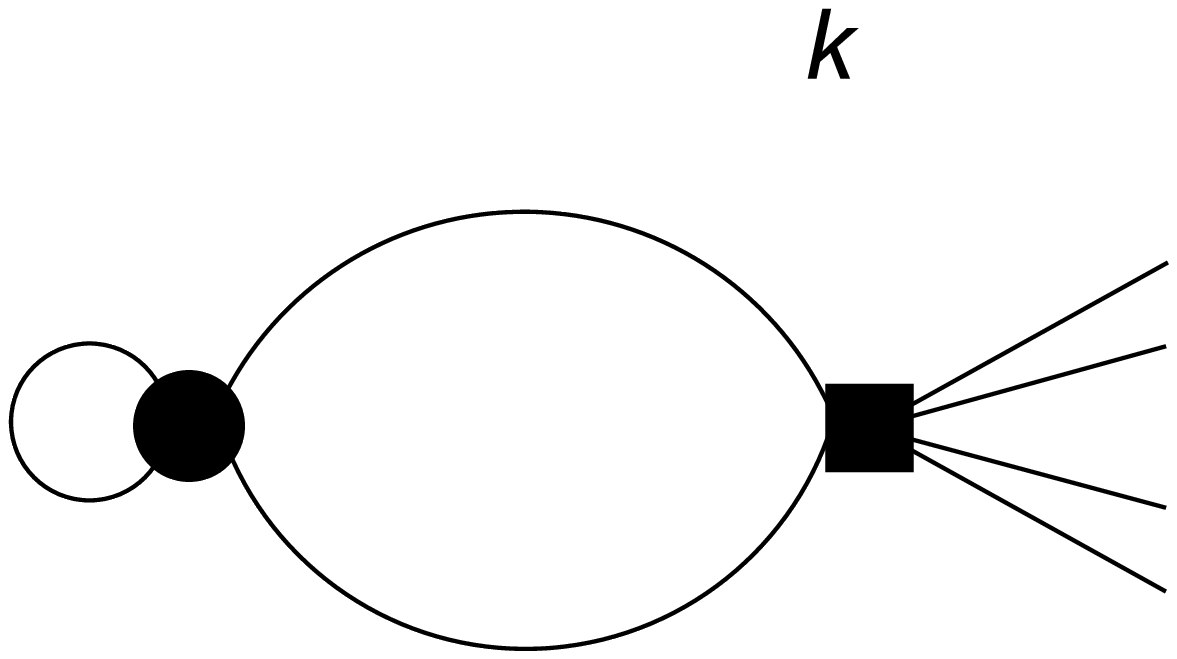}
    \hspace{3mm} 
    \epsfxsize=3.4cm
    \epsfbox{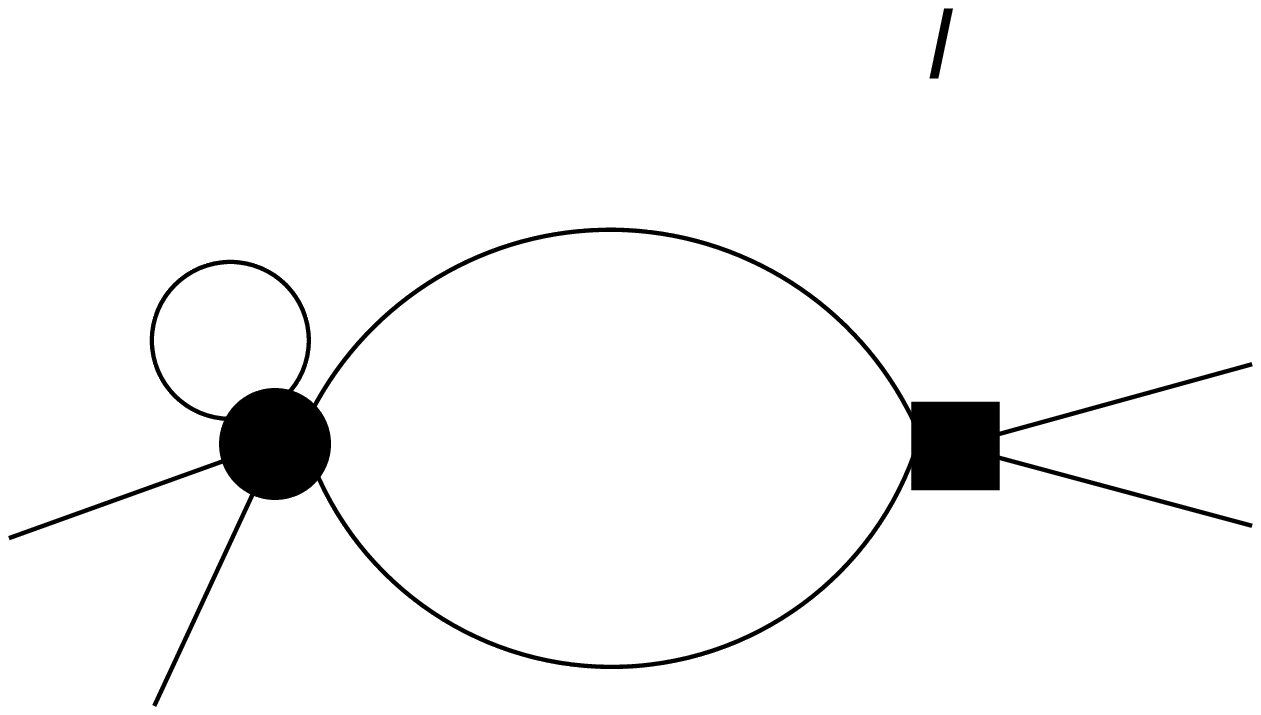}
    \vskip 0.8cm
    \hspace{5mm} 
    \epsfxsize=3.9cm
    \epsfbox{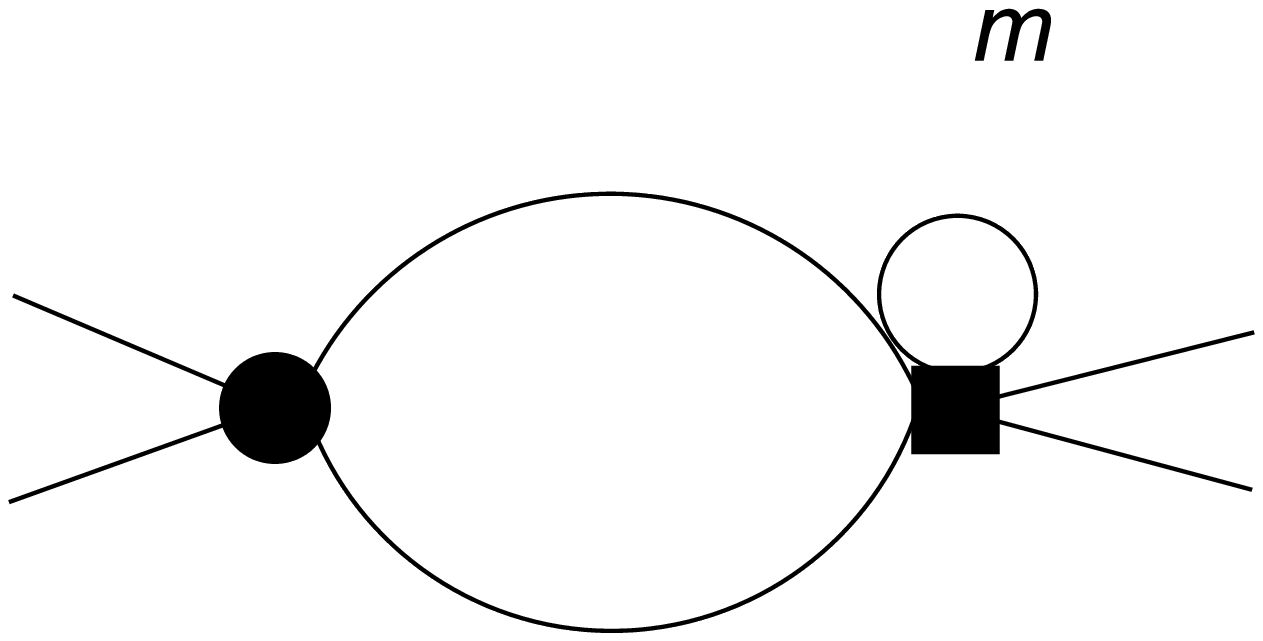}
    \hspace{15mm} 
    \epsfxsize=3.0cm
    \epsfbox{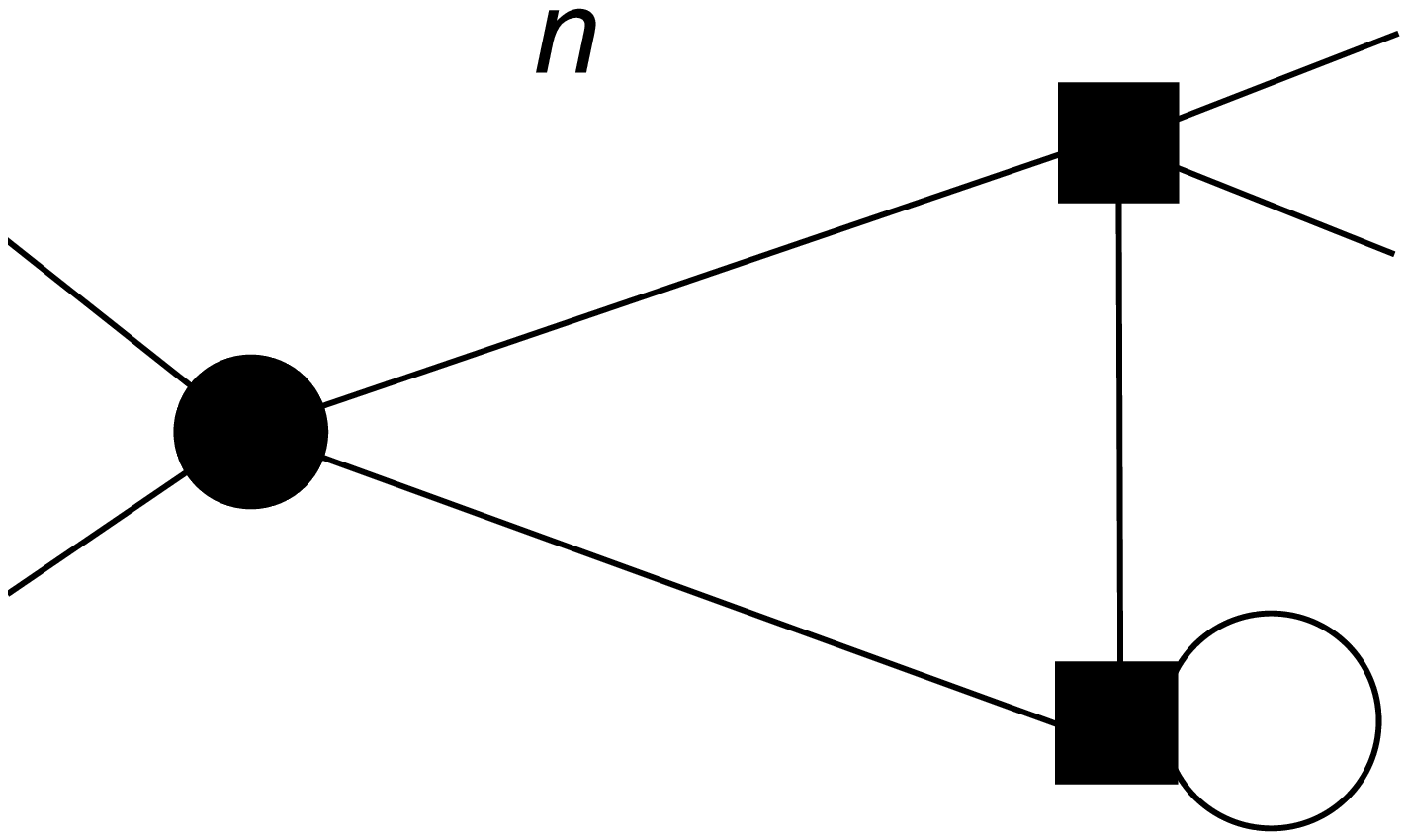}
  \end{center}
  \caption{
    2-loop contributions to the renormalization of operators ${\cal O}_1$
    and ${\cal O}_2$. Full circles represent the (4, 6 or 8-leg part) 
    of the operators, full squares stand for the (4-point or 6-point)
    interaction vertex of the model. 
  }
  \label{2LoopGraphs}
\end{figure}

The computation is again too lengthy 
to present the details here; we just list the 
relevant Feynman diagrams in Fig.~\ref{2LoopGraphs}.
Note that the 1-loop contribution corresponding to the diagram $b$ in 
Fig.~\ref{1LoopGraphs} is only a wave function renormalization.
Analogous contributions (with wave function renormalization on
one or two external legs) are not shown among the diagrams of
Fig.~\ref{2LoopGraphs}. Diagram $d$ in Fig.~\ref{1LoopGraphs}
contains an internal contraction line and thus vanishes in the
limit $m\to0$. Diagrams containing similar internal contractions
are not shown in Fig.~\ref{2LoopGraphs} either.

We checked that the $1/\varepsilon^2$
divergences were as expected from the RG considerations, 
and from the $1/\varepsilon$ divergences we finally obtained
\begin{equation}
  \begin{split}
    (2\pi)^2 p_{1j} & = \left\{ \frac{7n}{2}-11 \,, -9n+19 \,, \frac{n}{4} \,, 
      \frac{n-3}{2} \,, -n+1  \right\}\,,  \\
    (2\pi)^2 p_{2j} & = \left\{ -\frac{13n}{4}+7 \,, \frac{3n-13}{2} \,,
      \frac{n+2}{8} \,, -\frac{n-1}{2}\,, -\frac{n(n-1)}{2} \right\} \,.
  \end{split}
\end{equation}
This yields
\begin{equation}
  \begin{split}
    (2\pi)^2\nu_{11}^{(2)} & = -\frac{3(2n^2-5n+5)}{2(n-1)}\,,
    \\
    (2\pi)^2\nu_{21}^{(2)} &= \frac{n(n-4)}{(n-1)}\,.
  \end{split}
\end{equation}

\appendix
\renewcommand{\thesection}{Appendix~B: Lattice perturbation theory}
\section{}
\renewcommand{\thesection}{B}

\subsection{Perturbation theory with periodic bc} 

We consider a volume $V=N^2$ with periodic boundary conditions
in each direction
\begin{equation}
  S(x+N\hat{\mu})=S(x)\,,\,\,\mu=1,2\,.
\end{equation}
Let ${\cal O}$ be an O$(n)$ invariant observable. 
For  $\lambda_0\to0$ zero modes appear
due to the O$(n)$ invariance. To avoid this Hasenfratz 
\cite{PH} used the Fadeev-Popov trick:
\begin{equation}
  \langle{\cal O}\rangle=
  \frac{1}{Z}\int\rmd^nr\,
  \int [\rmd S]\rme^{-{\cal A}}\delta\left(r-\sum_x S(x)\right){\cal O}[S]\,.
\end{equation}
Because of the O$(n)$ invariance the inner integral is independent
of the direction of $r$. Take $r=|r|n_0\,,\,\,n_0=(0,0,\dots,0,1)$
and set
\begin{align} 
  S(x) & = \left(\lambda_0\pi(x),\sigma(x)\right)  \\
  \sigma(x) & = \sqrt{1-\lambda_0^2\pi(x)^2}
\end{align} 
where $\pi(x)$ has $n-1$ components $\pi^j(x)\,,j=1,\dots,n-1$\,.
Then 
\begin{equation} 
  \begin{split} 
    \langle{\cal O}\rangle & =
    \frac{1}{Z}\int\left[\prod_x\rmd^{n-1}\pi(x)\,\frac{1}{\sigma(x)}\right]
    \left(\frac{1}{V}\sum_x\sigma(x)\right)^{n-1}
    \delta\left(\sum_x\pi(x)\right)
    {\cal O}\,\rme^{-{\cal A}} \\
    &
    \\
    & = \frac{1}{Z_{\rm eff}}\int\left[\prod_x\rmd^{n-1}\pi(x)\right]
    \delta\left(\sum_x\pi(x)\right)
    {\cal O}\,\rme^{-{\cal A}_{\rm eff}}\,,
  \end{split} 
\end{equation} 
with
\begin{equation}
  {\cal A}_{\rm eff}={\cal A}
  +\sum_x\ln\sigma(x)-(n-1)\ln\left(\frac{1}{V}\sum_x\sigma(x)\right)\,.
\end{equation}
Then when one expands for $\lambda_0\to0$ no zero modes occur because of the 
delta function.

In the following we restrict results to the infinite volume limit.


\subsection{1-loop connected 4-point function}  

The leading orders of the connected 4-point function have been
given in subsect.~2.2.2. Here we give the result for $\widetilde{C}_2$
and consider the coefficients of powers of $(n-1)$ separately:
\begin{equation}
  \widetilde{C}_2=\sum_{r=0}^2(n-1)^r\widetilde{C}_2^{[r]}\,.
\end{equation}

First for $\widetilde{C}_2^{[2]}$ we obtain simply:
\begin{equation}
  \widetilde{C}_2^{[2]}(p_1,q_1;p_2,q_2)=
  \frac12\frac{1}{K_{p_1}K_{q_1}K_{p_2}K_{q_2}}K_{p_1+q_1}F(p_1+q_1)\,.
\end{equation}
Next for $\widetilde{C}_2^{[1]}$ we get:
\begin{equation} 
  \begin{split} 
    & \widetilde{C}_2^{[1]}(p_1,q_1;p_2,q_2)=
    \frac12\frac{1}{K_{p_1}K_{q_1}K_{p_2}K_{q_2}}
    \left[K_{p_1+p_2}F(p_1+p_2)+K_{p_1+q_2}F(p_1+q_2)\right]
    \\
    &-\frac12\left[F(p_1)+F(q_1)\right]\widetilde{C}_1^{(1)}(p_1,q_1;p_2,q_2)
    -\frac12\left[F(p_2)+F(q_2)\right]\widetilde{C}_1^{(2)}(p_1,q_1;p_2,q_2)
    \\
    &+\frac12\frac{1}{K_{p_2}K_{q_2}}\left[\frac{F(p_1)}{K_{p_1}}
      +\frac{F(q_1)}{K_{q_1}}\right]
    +\frac12\frac{1}{K_{p_1}K_{q_1}}\left[\frac{F(p_2)}{K_{p_2}}
      +\frac{F(q_2)}{K_{q_2}}\right]
    \\
    & +\frac{1}{K_{p_1}K_{q_1}K_{p_2}K_{q_2}}
    \left[k_{p_1}+k_{q_1}+k_{p_2}+k_{q_2}\right]
    \\
    & -\frac{1}{K_{p_1}K_{q_1}K_{p_2}K_{q_2}}K_{p_1+q_1}
    \left[\frac{k_{p_1}}{K_{p_1}}+\frac{k_{q_1}}{K_{q_1}}
      +\frac{k_{p_2}}{K_{p_2}}+\frac{k_{q_2}}{K_{q_2}}\right]
    \\
    & +\frac{K_{p_1+q_1}}{K_{p_1}K_{q_1}K_{p_2}K_{q_2}}
    \left[\overline{J}_1(p_1,q_1)
      +\overline{J}_1(p_2,q_2)\right]
    \\
    & +\frac{1}{K_{p_1}K_{q_1}K_{p_2}K_{q_2}}
    \left[\overline{Z}_4(p_1,p_2,q_1)
      +\overline{Z}_4(p_1,q_2,q_1)\right]
    \\
    & -K_{p_1+q_1}\left[
      \frac{\overline{T}_1(p_1,q_1)}{K_{p_2}K_{q_2}}
      +\frac{\overline{T}_1(p_2,q_2)}{K_{p_1}K_{q_1}}\right]\,,
  \end{split} 
\end{equation} 
where
\begin{align} 
  \overline{Z}_2(p,q) & =\int_k\left[\frac{K_{k-p-q}}{K_k K_{k-p}}
    -\frac{K_{p+q}}{K_k K_p}-\frac{K_q}{K_pK_{k-p}}\right]\,,
  \\
  \overline{J}_1(p,q) & =\overline{Z}_2(-p-q,p)=\overline{Z}_2(-p-q,q)
  \nonumber \\
  & =\int_k\left[\frac{K_k}{K_{k-p}K_{k+q}}
    -\frac{K_p}{K_{p+q}K_{k-p}}-\frac{K_q}{K_{p+q}K_{k+q}}\right]\,,
\end{align} 
\begin{align} 
  \overline{Z}_4(p,q,r) & =\int_k\left[\frac{K_{k-p}K_{k+r}}{K_k K_{k-p-q}}
    -1-\frac{K_p K_r}{K_k K_{p+q}}
    -\frac{K_{p+q+r}K_{q}}{K_{p+q}K_{k-p-q}}\right]\,,
  \\
  \overline{T}_1(p,q) & =\int_k\left[\frac{1}{K_k K_{k-p}K_{k+q}}
    -\frac{1}{K_k K_p K_q}-\frac{1}{K_pK_{k-p}K_{p+q}}
    -\frac{1}{K_q K_{p+q} K_{k+q}}\right]\,.
\end{align} 

Finally for $\widetilde{C}_2^{[0]}$ we get:
\begin{equation} 
  \begin{split} 
    & \widetilde{C}_2^{[0]}(p_1,q_1;p_2,q_2)=
    \frac{1}{K_{p_2}K_{q_2}}
    \left[\frac{F(p_1)}{K_{p_1}}+\frac{F(q_1)}{K_{q_1}}\right]
    +\frac{1}{K_{p_1}K_{q_1}}
    \left[\frac{F(p_2)}{K_{p_2}}+\frac{F(q_2)}{K_{q_2}}\right]
    \\
    & +\left[\frac{1}{K_{p_2}K_{q_2}}+\frac{1}{K_{p_1}K_{q_1}}\right]
    \left[\frac{k_{p_1+p_2}}{K_{p_1+p_2}^2}
      +\frac{k_{p_1+q_2}}{K_{p_1+q_2}^2}\right]
    \\
    & +\left[\frac{1}{K_{p_1+p_2}}+\frac{1}{K_{p_1+q_2}}\right]
    \left[\frac{1}{K_{p_2}K_{q_2}}
      \left\{\frac{k_{q_2}}{K_{q_2}}+\frac{k_{p_2}}{K_{p_2}}\right\}
      +\frac{1}{K_{p_1}K_{q_1}}
      \left\{\frac{k_{q_1}}{K_{q_1}}+\frac{k_{p_1}}{K_{p_1}}\right\}\right]
    \\
    & +\frac{2}{K_{p_1}K_{q_1}K_{p_2}K_{q_2}}\left[
      k_{p_1}+k_{q_1}+k_{p_2}+k_{q_2}\right]
    \\
    & -\frac{1}{K_{p_1}K_{q_1}K_{p_2}K_{q_2}}\left[K_{p_1+p_2}
      +K_{p_1+q_2}\right]
    \left[ \frac{k_{p_1}}{K_{p_1}}+\frac{k_{q_1}}{K_{q_1}}
      +\frac{k_{p_2}}{K_{p_2}}+\frac{k_{q_2}}{K_{q_2}}\right]
    \\
    & -\frac{1}{K_{p_2}K_{q_2}}\left[
      \frac{1}{K_{p_1+p_2}}\left\{\overline{Z}_2(p_1,p_2)
        +\overline{Z}_2(q_1,q_2)\right\}
      +\frac{1}{K_{p_1+q_2}}\left\{\overline{Z}_2(p_1,q_2)
        +\overline{Z}_2(q_1,p_2)\right\}\right]
    \\
    & -\frac{1}{K_{p_1}K_{q_1}}\left[
      \frac{1}{K_{p_1+p_2}}\left\{\overline{Z}_2(p_2,p_1)
        +\overline{Z}_2(q_2,q_1)\right\}
      +\frac{1}{K_{p_1+q_2}}\left\{\overline{Z}_2(p_2,q_1)
        +\overline{Z}_2(q_2,p_1)\right\}\right]
    \\
    & +\frac{1}{K_{p_1}K_{q_1}K_{p_2}K_{q_2}}\Bigl[
    K_{p_1+p_2}\left\{\overline{J}_1(p_1,p_2)
      +\overline{J}_1(q_1,q_2)\right\}
    \\
    & +K_{p_1+q_2}\left\{\overline{J}_1(p_1,q_2)
      +\overline{J}_1(q_1,p_2)\right\}\Bigr]
    \\
    & +\frac{1}{K_{p_1}K_{q_1}K_{p_2}K_{q_2}}\Bigl[
    \overline{Z}_4(p_1,q_1,p_2)+\overline{Z}_4(p_1,q_1,q_2)
    +\overline{Z}_4(p_1,p_2,q_2)+\overline{Z}_4(p_1,q_2,p_2)\Bigr]
    \\
    & -\frac{1}{K_{p_2}K_{q_2}}\left[\overline{T}_2(p_1,q_1,p_2)
      +\overline{T}_2(p_1,q_1,q_2)\right]
    -\frac{1}{K_{p_1}K_{q_1}}\left[\overline{T}_2(p_2,q_2,p_1)
      +\overline{T}_2(p_2,q_2,q_1)\right]
    \\
    & +\overline{B}(p_1,q_1,p_2)+\overline{B}(p_1,q_1,q_2)
    +\overline{B}(p_1,q_2,p_2)\,,
  \end{split} 
\end{equation} 
where 
\begin{multline} 
  \overline{T}_2(p,q,r)=\int_k\Bigl[
  \frac{K_{k-p-r}}{K_k K_{k-p}K_{k+q}}-\frac{K_{p+r}}{K_k K_pK_q}
  \\
  -\frac{K_r}{K_p K_{k-p}K_{p+q}}-\frac{K_{p+q+r}}{K_q K_{p+q}K_{k+q}}\Bigr]\,,
\end{multline} 

\begin{multline} 
  \overline{B}(p,q,r)=\int_k\Bigl[\frac{1}{K_kK_{k-p}K_{k+q}K_{k-p-r}}
  -\frac{1}{K_kK_pK_qK_{p+r}}
  \\
  -\frac{1}{K_pK_{k-p}K_{p+q}K_r}
  -\frac{1}{K_qK_{p+q}K_{k+q}K_{p+q+r}}
  -\frac{1}{K_{p+r}K_rK_{p+q+r}K_{k-p-r}}\Bigr].
\end{multline} 

\subsection{Lattice integrals} \label{LattIntegrals.tex}

\begin{table}[ht] 
  \centering 
  \begin{tabular}[t]{c|c|c|c|c} 
    \hline 
    const.$C$      & ST  & D(1/3) & D($-1/4$) & SYM \\ 
    \hline \hline 
    $A_0$          & $\frac14$              & $\phantom{-}0.37898718$  & $\phantom{-}0.17746732$ & $\phantom{-}0.17029804$ \\
    $A_1$          & $-\frac{1}{48}$        & $\phantom{-}0.03291132$  & $-0.05105528$ & $\phantom{-}0.01237582$ \\
    $A_2$          & $0$                    & $-0.06449359$  & $\phantom{-}0.03626634$ & $0$           \\
    $c_1$          & $-5\ln 2$              & $-4.56434819$  & $-3.06027079$ & $-2.87298102$ \\
    $c_2$          & $\frac{5}{16}\ln 2 + %
    \frac18$       & $\phantom{-}0.35045294$  & $\phantom{-}0.48105013$ & $\phantom{-}0.03014466$ \\
    $F_{10}$       & $-\frac{5}{4\pi}\ln 2$ & $-0.55231383$  & $-0.16155304$ & $-0.15765008$ \\ 
    $F_{15}^{(A)}$ & $\frac{1}{2\pi} %
    \left[-5\ln2+\pi-1\right]$  & $\phantom{-}0.09572176$  & $-0.39020365$ & $-0.29701963$ \\ 
    $F_{15}^{(B)}$ & $0$                    & $-0.07444697$  & $\phantom{-}0.03297538$ & $0$            \\ 
    $F_{17}^{(A)}$ & $\frac{1}{2\pi} %
    \left[\pi-2\right]$    & $-0.15764297$  & $\phantom{-}0.31124154$ & $\phantom{-}0.13997926$ \\ 
    $F_{17}^{(B)}$ & $\frac{1}{2\pi}$       & $\phantom{-}0.47746483$  & $-0.07957747$ & $\phantom{-}0.06287345$ \\ 
    $F_{17}^{(C)}$ & $0$                    & $\phantom{-}0.09153552$  & $\phantom{-}0.02817918$ & $0$           \\ 
    $F_{28}^{(A)}$ & $-0.07421002$          & $\phantom{-}0.02872250$  & $-0.18868881$ & $-0.07852019$ \\
    $F_{28}^{(B)}$ & $-0.05734358$          & $-0.10135050$  & $-0.00290322$ & $-0.03887267$ \\
    $F_{29}$       & $-0.07889581$          & $-0.11486836$  & $-0.05199331$ & $-0.06922970$ \\
    $F_{30}^{(A)}$ & $-0.05431565$          & $\phantom{-}0.00882813$  & $-0.13895289$ & $-0.07852019$ \\
    $F_{30}^{(B)}$ & $-0.07821415$          & $-0.07058178$  & $-0.13051244$ & $-0.04738831$ \\
    $F_{30}^{(C)}$ & $0$                    & $-0.01798155$  & $\phantom{-}0.05047165$ & $ 0$ \\
    \hline
  \end{tabular} 
  \caption{Constants for various actions.}
  \label{constants} 
\end{table}

The expansion in the cutoff of the functions $F(p),k_p$ and others 
appearing in $\widetilde{C}_2$ (which will be considered in the next 
subsection) involve various lattice integrals that we will first define here. 
They depend on the specific lattice action and hence on the associated 
function $K_p$ (\ref{Kp}). We first define
\begin{equation}  \label{defA0}
  A_0\delta_{\mu\nu}=
  -\frac12\int_s\frac{1}{K_s}\left[K_{\mu\nu}(s)-K_{\mu\nu}(0)\right]\,,
\end{equation} 
\begin{equation}    \label{defA12}
  A_1\delta_{\mu\nu\lambda\rho}+\frac13 A_2 s_{\mu\nu\lambda\rho}
  =-\frac{a^{-2}}{24}\int_s\frac{1}{K_s}
  \left[K_{\mu\nu\rho\lambda}(s)-K_{\mu\nu\rho\lambda}(0)\right]\,,
\end{equation} 
where $K_{\mu_1...\mu_n}(s)=\partial_{\mu_1}^s\dots\partial_{\mu_n}^s K_s$,
$\delta_{\mu\nu\lambda\rho}=\delta_{\mu\nu}\delta_{\nu\lambda}
\delta_{\lambda\rho}$ and
\begin{equation}
  s_{\mu\nu\lambda\rho}=\delta_{\mu\nu}\delta_{\lambda\rho}
  +\delta_{\mu\lambda}\delta_{\nu\rho}
  +\delta_{\mu\rho}\delta_{\lambda\nu}\,.
\end{equation}
Note
\begin{equation} 
  \begin{split} 
    K_{\mu\nu}(0) & = 2\delta_{\mu\nu}\,,
    \\
    K_{\mu\nu\rho\lambda}(0) & = 8a^2\left[3\kappa_1\delta_{\mu\nu\lambda\rho}
      +\kappa_2 s_{\mu\nu\lambda\rho}\right]\,.
  \end{split} 
\end{equation} 
The values of $A_i$ and other constants are given
in Table~\ref{constants} for various actions.

Introducing $\theta_B(k)$, which restricts the momenta to the Brillouin zone:
\begin{equation}
  \theta_B(k)=\prod_{\mu=1}^2\theta\left(\pi-a\vert k_\mu\vert\right)
\end{equation}
we define
\begin{align} 
  c_1 & = -4\pi\int_\infty\left\{
    \frac{\theta_B(k)}{K_k}-\frac{1}{k^2(a^2k^2+1)}\right\}\,,
  \label{defc1}
  \\
  F_6 & = a^{-2}\int_\infty\left\{
    \frac{\theta_B(k)}{K_k^2}-\frac{1}{(k^2)^2}
    +\frac{2a^2r(k)}{[k^2]^2(a^2k^2+1)}\right\}\,,
  \label{defF6}
\end{align} 
where we have introduced the shorthand
\begin{equation}
  \int_\infty=\int_{-\infty}^\infty\frac{\rmd^2k}{(2\pi)^2}\,.
\end{equation}
Also introduce
\begin{equation}
  c_2\equiv2\pi F_6 -\frac{15}{8}\kappa_1\,.
\end{equation}

Further constants appearing are
\begin{equation}
  F_{10}=a^{-2}\int_\infty
  \left[\frac{\theta_B(k)\left\{K_{\mu\mu}(k)-K_{\mu\mu}(0)\right\}}{K_k^2}
    -a^2\frac{(12\kappa_1+16\kappa_2)}{k^2(a^2k^2+1)}\right]\,,
\end{equation}
\begin{equation} 
  \begin{split} 
    F_{15\mu\nu\lambda\rho} & = a^{-2}
    \int_\infty\left[\frac{\theta_B(k)K_{\mu\nu\lambda}(k)K_\rho(k)}{K_k^2}
      -2K_{\mu\nu\lambda\tau}(0)\frac{k_\tau k_\rho}{[k^2]^2(a^2k^2+1)}\right]
    \\
    & = F_{15}^{(A)}\delta_{\mu\nu\lambda\rho}+F_{15}^{(B)}s_{\mu\nu\lambda\rho}\,,
  \end{split} 
\end{equation} 
and
\begin{equation} 
  \begin{split} 
    F_{17\mu\nu\lambda\rho} & = a^{-2}
    \int_k\frac{1}{K_k^2}\left[K_{\mu\nu}(k)-K_{\mu\nu}(0)\right]
    \left[K_{\lambda\rho}(k)-K_{\lambda\rho}(0)\right]
    \\
    & = F_{17}^{(A)}\delta_{\mu\nu\lambda\rho}
    +F_{17}^{(B)}\delta_{\mu\nu}\delta_{\lambda\rho}
    +F_{17}^{(C)}s_{\mu\nu\lambda\rho}\,.
  \end{split} 
\end{equation} 
\begin{equation} 
  \begin{split} 
    F_{28\mu\nu\lambda\rho} & = a^{-2}\int_\infty\Bigl[
    \frac{\theta_B(k)K_\mu(k)K_\nu(k)K_\lambda(k)K_\rho(k)}{K_k^4}
    -\frac{16k_\mu k_\nu k_\lambda k_\rho}{[k^2]^4}
    \\
    & -\frac43 K_{\alpha\beta\gamma\tau}(0)
    \frac{k_\alpha k_\beta k_\gamma}{[k^2]^5(a^2k^2+1)}\left\{
      k^2\left[ \delta_{\tau\mu} k_\nu k_\lambda k_\rho +3\,\,
        {\rm perms}\right]
      -2 k_\tau k_\mu k_\nu k_\lambda k_\rho\right\}\Bigr]
    \\
    & =  F_{28}^{(A)} \delta_{\mu\nu\lambda\rho} 
    + F_{28}^{(B)} s_{\mu\nu\lambda\rho}\,, 
  \end{split}
\end{equation}

\begin{equation} 
  F_{29} =  a^{-2}\int_\infty\Bigl[
  \frac{\theta_B(k)K_\mu(k)K_\mu(k)}{K_k^3}
  -\frac{4}{[k^2]^2}-\frac{4a^2r(k)}{[k^2]^2(a^2k^2+1)} \Bigr]\,,
\end{equation}

\begin{equation}
  \begin{split}
    F_{30\mu\nu\lambda\rho} & = a^{-2}\int_\infty\Bigl[
    \frac{\theta_B(k)K_\mu(k)K_\nu(k)}{K_k^3}
    \left\{K_{\lambda\rho}(k)-K_{\lambda\rho}(0)\right\} 
    \\
    & \qquad   -2K_{\alpha\beta\lambda\rho}(0)
    \frac{k_\alpha k_\beta k_\mu k_\nu}{[k^2]^3(a^2k^2+1)}\Bigr]  
    \\
    & = F_{30}^{(A)} \delta_{\mu\nu\lambda\rho} +
    F_{30}^{(B)} \delta_{\mu\nu} \delta_{\lambda\rho} + 
    F_{30}^{(C)} s_{\mu\nu\lambda\rho}\,.
  \end{split}
\end{equation} 

Note that there are various relations among the integrals defined above.
We have not attempted to determine all of them but note
the following identities
\begin{align} 
  & 3F_{28}^{(B)}-F_{29}-3F_{30}^{(C)}-F_{30}^{(B)}=
  \\
  & \qquad -8A_2-F_{15}^{(B)}+\frac{9}{16\pi}\kappa_1+
  \frac{1}{2\pi}\left(7-4c_1\right)\kappa_2\,,
  \nonumber \\
  & F_{30}^{(A)}-F_{28}^{(A)}=-\frac{3}{4\pi}\kappa_1\,,
  \\
  & 8A_1+\frac13 F_{15}^{(A)}
  +\frac{2}{\pi}\left(c_1-1\right)\kappa_1=0\,,
  \\
  & 8A_2+ F_{15}^{(B)}
  +\frac{2}{\pi}\left(c_1-1\right)\kappa_2=0\,.
\end{align} 
These appeared as consistency relations in the course of our
calculation and also serve as useful checks on the numerical evaluation,
and one can check are satisfied by the values given in Table~\ref{constants}.

\subsection{Integral expansions} \label{Lattintegral_expansions.tex} 

Here we give, without derivation, the expansion of the lattice 
integrals to $\rmO(a^2)$. 
\begin{multline}
  F(p) = \frac{1}{2\pi}\left[L(ap)+c_1\right] \\
  + a^2p^2\frac{1}{2\pi}
  \left[ \left(\frac34 \kappa_1+\kappa_2\right)L(ap)
    +\frac52\kappa_1\frac{p^4}{(p^2)^2}+c_2\right]+\rmO(a^4)\,,
\end{multline}
\begin{equation}
  k_p = A_0 p^2 +a^2\left[A_1 p^4 +A_2 (p^2)^2\right]+\rmO(a^4)\,.
\end{equation}
The constants in \eqref{C1pa} are thus given by
\begin{align}
  c_3 & = 4\pi A_1+\kappa_1\left(\frac52-c_1-8\pi A_0\right)\,,
  \\ 
  c_4 & = c_2+4\pi A_2 -\kappa_2(c_1+8\pi A_0)\,.
\end{align}

\begin{align}
  \overline{Z}_2(p,q) & \sim  
  q(p+q)\left[\frac{F(p)}{K_p}+\frac14 F_{10}\right]
  \nonumber\\
  & + \frac{a^2}{96\pi p^2}K_{\mu\nu\rho\lambda}(0)q_\mu\Bigl[
  -3(p+q)_\nu\left(2p_\rho p_\lambda
    +\delta_{\rho\lambda}p^2\left\{L(ap)-1\right\}\right)
  \nonumber\\
  & + 2\left\{L(ap)+c_1\right\}
  \left\{ (2p_\nu+3q_\nu)p_\rho p_\lambda+q_\nu q_\rho(2p_\lambda+q_\lambda) 
  \right\}\Bigr]\,.
\end{align}

To consider $\overline{Z}_4$ we first define the function $\overline{Z}_5$ 
through:               
\begin{equation}
  \overline{Z}_4(p,q,r)=\overline{Z}_5(p+q,-p,r)\,.
\end{equation}
Its expansion is given by
\begin{align}
  \overline{Z}_5(p,q,r) & \sim -A_0(p+q+r)^2
  +q_\mu r_\nu\left\{4F_{13\mu\nu}(p)
    -\delta_{\mu\nu}\frac{(c_1-1)}{2\pi}\right\}
  \nonumber\\
  & + \left[(r^2+pr)(q^2+pq)+(pr)(pq)\right]\frac{F(p)}{K_p}
  \nonumber\\
  & - a^2\left[A_1\left\{(p+q)^4+(p+r)^4-p^4\right\}
    +A_2\left\{[(p+q)^2]^2+[(p+r)^2]^2-[p^2]^2\right\}\right]
  \nonumber\\
  & + \frac{1}{24}K_{\mu\nu\lambda\rho}(0)
  \Big\{4\left[pr q_\mu + pq r_\mu\right]p_\nu p_\lambda p_\rho
  \nonumber\\
  & + \left[6\left(r_\mu r_\nu pq+q_\mu q_\nu pr\right)
    +2\left(r^2 q_\mu +q^2 r_\mu\right)p_\nu\right]p_\lambda p_\rho
  \nonumber\\
  & + 2\left[r_\mu r_\nu r_\lambda(q^2+2pq)
    +q_\mu q_\nu q_\lambda (r^2+2pr)\right]p_\rho
  \nonumber\\
  & + r_\mu r_\nu r_\lambda r_\rho(q^2+pq)
  +q_\mu q_\nu q_\lambda q_\rho(r^2+pr)\Bigr\}
  \frac{1}{2\pi p^2}\left[L(ap)+c_1\right]
  \nonumber\\
  & + q_\mu r_\nu\left\{\frac12 F_{19\mu\nu}(p)+F_{22\mu\nu}(p)\right\}
  +\frac14\left[r_\mu r_\nu q_\lambda+q_\mu q_\nu r_\lambda\right]
  F_{18\mu\nu\lambda}(p)
  \nonumber\\
  & + \frac{1}{6}
  \left[r_\mu r_\nu r_\lambda q_\rho+q_\mu q_\nu q_\lambda r_\rho \right]
  \left[2K_{\mu\nu\lambda\tau}(0)F_{13\tau\rho}(p)
    +a^2F_{15\mu\nu\lambda\rho}\right]\,,
  \nonumber\\
  & + \frac14 r_\mu r_\nu q_\lambda q_\rho F_{16\mu\nu\lambda\rho}(p)\,.
  \label{OZ5expansion}
\end{align}
The various expressions appearing here are given by
\begin{align}
  F_{13\rho\lambda}(p) & = \frac{1}{4\pi p^2}\left[-p_\rho p_\lambda
    +\frac12\delta_{\rho\lambda}p^2\left\{1-L(ap)\right\}\right]\,,
  \\
  F_{16\mu\nu\lambda\rho}(p) & = a^2\left[
    2\delta_{\mu\nu}\delta_{\lambda\rho}F_{10}+F_{17\mu\nu\lambda\rho}\right]
  \nonumber\\
  & + \frac{1}{4\pi}\left[\delta_{\mu\nu}K_{\lambda\rho\tau\sigma}(0)
    +\delta_{\lambda\rho}K_{\mu\nu\tau\sigma}(0)\right]
  \nonumber\\
  & \times\left\{\frac{p_\tau p_\sigma}{p^2}\left[L(ap)+c_1-1\right]
    -\frac12\delta_{\tau\sigma}\left[L(ap)-1\right]\right\}\,,
  \\
  F_{18\mu\nu\lambda}(p) & = 
  \delta_{\mu\nu}p_\rho K_{\lambda\rho\tau\sigma}(0)F_{13\tau\sigma}(p)
  \nonumber\\
  & + K_{\mu\nu\rho\tau}(0)\left[p_\lambda F_{13\rho\tau}(p)
    +p_\rho F_{13\lambda\tau}(p)+p_\tau F_{13\rho\lambda}(p)\right]
  \nonumber\\
  & + a^2p_\tau\left[2\delta_{\mu\nu}\delta_{\lambda\tau}F_{10}+
    F_{15\mu\nu\lambda\tau}+F_{17\mu\nu\lambda\tau}\right]\,,
  \\
  & 
  \nonumber\\
  F_{19\mu\nu}(p) & = \left[K_{\mu\rho\lambda\tau}(0)p_\nu 
    +K_{\nu\rho\lambda\tau}(0)p_\mu\right] p_\lambda F_{13\rho\tau}(p)
  \nonumber\\
  & + a^2p_\lambda p_\rho\left[ 
    2\delta_{\mu\lambda}\delta_{\nu\rho}F_{10}+F_{17\mu\lambda\nu\rho}\right]\,,
  \\
  & 
  \nonumber\\
  F_{22\mu\nu}(p) & = 
  \frac13K_{\sigma\lambda\rho\tau}(0)
  \left[\delta_{\nu\sigma}F_{24\mu\lambda\rho\tau}(p)
    +\delta_{\mu\sigma}F_{24\nu\lambda\rho\tau}(p)\right]
  \nonumber\\
  & + a^2\left[-4F_{25\mu\nu}(p)+\frac14 p_\lambda p_\rho 
    F_{26\mu\nu\lambda\rho}\right]\,,
\end{align}
with
\begin{align}
  F_{24\sigma\lambda\rho\tau}(p) & = \frac34\left[
    p_\rho p_\tau F_{13\sigma\lambda}(p)
    -p_\sigma p_\lambda F_{13\rho\tau}(p)\right]
  +F_{27\sigma\lambda\rho\tau}(p)\,,
  \\
  F_{25\mu\nu}(p) & = \frac{\kappa_1}{32\pi}\Bigl[\delta_{\mu\nu}\left\{
    p^2\left(L(ap)-\frac{59}{24}\right)
    -4p_\mu^2\left(L(ap)-\frac{5}{24}\right)+2\frac{p^4}{p^2}\right\}
  \nonumber\\
  & +p_\mu p_\nu\left\{2L(ap)-\frac{71}{12}
    +\frac{8}{3p^2}(p_\mu^2+p_\nu^2)-4\frac{p^4}{(p^2)^2}\right\}\Bigr]
  \nonumber\\
  & -\frac{\kappa_2}{32\pi}\left[p^2\delta_{\mu\nu}+6p_\mu p_\nu\right]\,,
  \\
  & 
  \nonumber\\
  F_{26\mu\nu\lambda\rho} & = 
  F_{28\mu\nu\lambda\rho}-\delta_{\lambda\rho}\delta_{\mu\nu}F_{29}
  -F_{30\mu\nu\lambda\rho}
  -\left[\delta_{\mu\lambda}\delta_{\nu\rho}
    +\delta_{\mu\rho}\delta_{\nu\lambda}\right]\left[2F_6+F_{10}\right]
  \nonumber\\
  & + F_{15\mu\nu\lambda\rho}
  -\frac12 F_{17\mu\lambda\nu\rho}-\frac12 F_{17\mu\rho\nu\lambda}\,,
  \\
  & 
  \nonumber\\
  F_{27\mu\nu\rho\lambda}(p) & = \frac{1}{96\pi}\Bigl[
  \left(L(ap)-\frac43\right)s_{\mu\nu\rho\lambda} p^2
  \nonumber\\
  & -\left(L(ap)-\frac56\right)\left\{p_\mu p_\nu\delta_{\rho\lambda}
    +5\,\,{\rm perms}\right\}
  -8\frac{p_\mu p_\nu p_\rho p_\lambda}{p^2}\Bigl]\,.
\end{align}

For the scaling part of $\overline{Z}_4$ this means ($p+q+r+s=0$)
\begin{align}
  & \overline{Z}_4(p,q,r)=
  -A_0(q+r)^2+\frac{1}{\pi(p+q)^2}\left[(pr)(qs)-(ps)(qr)\right]
  \nonumber\\
  & +\frac{1}{2\pi(p+q)^2}\left\{L(a(p+q))+c_1\right\}
  \left[(pq)(rs)-(pr)(qs)+(ps)(qr)\right]
  \nonumber\\
  & +\,\,\rmO(a^2)\,.
\end{align}

For the first triangle diagram integral we have:
\begin{equation}
  \overline{T}_1(p,q)\sim\frac{1}{2\pi p^2 q^2 (p+q)^2}
  \left[t_0(p,q)+a^2 t_1(p,q)+\rmO(a^4)\right]\,,
\end{equation}
with (here $r=-p-q$)
\begin{equation}
  t_0(p,q) = -qrL(ap)-prL(aq)-pq L(ar)+\frac12(p^2+q^2+r^2)c_1\,,
\end{equation}
\begin{equation}
  \begin{split}
    t_1(p,q) & = -\frac12 c_1\Bigl[
    \left\{\frac{p^4(q^2+r^2)}{p^2}+\frac{q^4(p^2+r^2)}{q^2}
      +\frac{r^4(p^2+q^2)}{r^2}\right\}\kappa_1
    \\
    & + 2(p^2q^2+q^2r^2+r^2p^2)\kappa_2\Bigr]
    \\
    & - \left(\frac34 \kappa_1+\kappa_2\right)
    \left[p^2 q^2 L(ar)+r^2 q^2 L(ap)+r^2 p^2 L(aq)\right]
    \\
    & + \frac{\kappa_1}{4}\Bigl[
    \left\{\frac{p^4}{(p^2)^2}-\frac34\right\}
    \left\{p^2 q^2 {\cal V}_5(p,q) + p^2 r^2 {\cal V}_5(p,r)\right\}
    \\
    & \phantom{\Bigl[}+\left\{\frac{q^4}{(q^2)^2}-\frac34\right\}
    \left\{q^2 p^2 {\cal V}_5(q,p) + q^2 r^2 {\cal V}_5(q,r)\right\}
    \\
    & \phantom{\Bigl[}+\left\{\frac{r^4}{(r^2)^2}-\frac34\right\}
    \left\{r^2 p^2 {\cal V}_5(r,p) + r^2 q^2 {\cal V}_5(r,q)\right\}\Bigr]\,.
  \end{split}
  \end{equation}
Here 
\begin{equation}
  \begin{split}
    {\cal V}_5(p,q) & = 
    \frac{1}{[p^2q^2-2(pq)^2]}\Bigl[\left\{(p^2)^2+(q^2)^2\right\}L(a(p+q))
    -(q^2)^2 L(aq)
    \\
    & - \left\{(p^2)^2+2p^2q^2-4(pq)^2\right\}L(ap)
    +4p^2q^2-2pq(p^2+q^2+4pq)\Bigr]\,.
  \end{split}
\end{equation}
Note for momenta such that $p^2q^2=2(pq)^2$ one has
$q^4/(q^2)^2-3/4=-p^4/(p^2)^2+3/4$ so that $t_1$ isn't singular in this 
case \footnote{indeed we have the identity
  $\frac{p^4}{(p^2)^2}+\frac{q^4}{(q^2)^2}-\frac32=
  \frac{\left[p^2q^2-2(pq)^2\right]}{(p^2q^2)^2}
  \times\Bigl[\frac12 p^2q^2+(pq)^2
  -\frac16\left\{(p+q)^4+(p-q)^4-2p^4-2q^4\right\}\Bigr]$}.

For the second triangle diagram integral we first define
\begin{equation}
  \overline{T}_2(p,q,r)=\overline{T}_6(p,q,-p-r)\,.
\end{equation}
Then
\begin{equation}
  \begin{split}
  \overline{T}_6(p,q,r) & \sim 
  \frac{F(p+q)}{K_{p+q}}+r_\mu \overline{T}_{6\mu}(p,q)
  +\frac12 r_\mu r_\nu \overline{T}_{6\mu\nu}(p,q)
  \\
  & + \frac16 r_\mu r_\nu K_{\mu\nu\lambda\rho}(0)
  r_\lambda\Bigl\{\frac{1}{4\pi(p+q)^2}\Bigl[
  \frac{p_\rho}{p^2}\left\{L(a(p+q))-L(aq)+L(ap)+c_1\right\}
  \\
  & - (p\leftrightarrow q)\Bigr]
  +\frac 14 r_\rho \overline{T}_1(p,q)\Bigr\}\,,
  \end{split}
\end{equation}
with
\begin{equation}
  \begin{split}
  \overline{T}_{6\mu\nu}(p,q) & \sim 2\delta_{\mu\nu}\overline{T}_1(p,q)
  \\
  & + \frac{1}{(p+q)^2}K_{\mu\nu\rho\lambda}(0)\Bigl[
  \frac{1}{8\pi}\Bigl\{\left(
    \frac{p_\rho p_\lambda}{p^2}-\frac12\delta_{\rho\lambda}\right)
  \left(L(ap)-1\right)+c_1\frac{p_\rho p_\lambda}{p^2}\Bigr\}
  \\
  & + \Bigl\{p\rightarrow q\Bigr\}-G_{4\rho\lambda}(p,q)\Bigr]\,,
  \end{split}
\end{equation}
where
\begin{equation}
  \begin{split}
    & G_{4\mu\nu}(p,q)=\frac{1}{8\pi(p+q)^2}\Bigl[
    \delta_{\mu\nu}\Bigl\{(p+q)^2-(p^2+q^2+4pq)L(a(p+q))
    \\
    & -\frac12{\cal V}_2(p,q)-\frac12{\cal V}_2(q,p)
    +\frac12(p^2-pq){\cal V}_3(p,q)+\frac12(q^2-pq){\cal V}_3(q,p)\Bigr\}
    \\
    & +\frac{p_\mu p_\nu}{p^2}\left\{-2p^2+2pq L(a(p+q))+
      {\cal V}_2(p,q)-(p^2-2pq){\cal V}_3(p,q)\right\}
    \\
    & +\frac{q_\mu q_\nu}{q^2}\left\{-2q^2+2pq L(a(p+q))+
      {\cal V}_2(q,p)-(q^2-2pq){\cal V}_3(q,p)\right\}
    \\
    & -(p_\mu q_\nu+q_\mu p_\nu)\left\{2+\frac12{\cal V}_3(p,q)
      +\frac12{\cal V}_3(q,p)\right\}\Bigr]\,,
  \end{split}
\end{equation}
where
\begin{align}
  {\cal V}_2(p,q) & = (p^2+q^2)L(a(p+q))-(p^2+2pq)L(ap)-q^2 L(aq)\,,
  \\
  {\cal V}_3(p,q) & = \frac{1}{\left[p^2q^2-4(pq)^2\right]}
  \Bigl[\{4(pq)^2+2p^2(pq)-p^2q^2-(q^2)^2\}L(a(p+q)) 
  \nonumber\\
  & +\{p^2q^2-2p^2(pq)-4(pq)^2\}L(ap)+(q^2)^2L(aq)
  \nonumber\\
  & +4(pq)^2+2q^2(pq)-2p^2q^2\Bigr]\,.
\end{align}
Finally
\begin{equation}
  \overline{T}_{6\mu}(p,q)=\frac{1}{K_{p+q}}\left[
    G_{6\mu}(p,q)-4G_{3\mu}(p,q)+
    \overline{Z}_{2\mu}(p)-\overline{Z}_{2\mu}(q)\right]\,,
\end{equation}
where
\begin{equation}
  \begin{split}
  \overline{Z}_{2\mu}(p) & = 
  \frac14 p_\mu F_{10}+\frac12\frac{K_\mu(p)}{K_p}F(p)
  \\
  & +\frac{p_\nu}{16\pi p^2}K_{\mu\nu\rho\lambda}(0)
  \left[-p_\rho p_\lambda
    +\frac12\delta_{\rho\lambda}p^2\left\{1-L(ap)\right\}\right]\,.
  \end{split}
\end{equation}
\begin{equation}
  G_{3\mu}(p,q) = -\frac{1}{8\pi}
  \left\{\frac{p_\mu}{p^2}\left[L(a(p+q))-L(aq)\right]
    -(p\leftrightarrow q)\right\}\,,
\end{equation}
\begin{equation}
  \begin{split}
    G_{6\mu}(p,q) & = G_{7a\mu}(p,q)+G_{7b\mu}(p,q)+G_{7c\mu}(p,q)
    \\
    & - \frac12(p-q)_\mu\left\{F_{10}+4F_6-F_{29}\right\}\,,
  \end{split}
\end{equation}
with ($t=p+q)$):
\begin{align}
  G_{7a\mu}(p,q) & = \frac{1}{24\pi}K_{\mu\nu\lambda\rho}(0)
  \Bigl[\frac{p_\nu p_\lambda p_\rho}{p^2}L(at)
  \nonumber\\
  & + \left\{\frac{p_\nu p_\lambda p_\rho}{p^2}
    -\frac14 s_{\nu\lambda\rho\sigma}p_\sigma\right\}
  \left\{{\cal V}_3(p,q)-L(ap)+\frac32\right\}
  \nonumber\\
  & - (p\leftrightarrow q)\Bigr]\,,
  \\
  & 
  \nonumber\\
  G_{7b\mu}(p,q) & = \frac{1}{12}K_{\nu\lambda\rho\tau}(0)
  \Bigl[-t^2M_{\mu\nu\lambda\rho\tau}(p,q)
  \nonumber\\
  & +\Bigl\{-\frac{1}{96\pi}\left(s_{\mu\nu\lambda\rho}p_\tau
    +4\,\,{\rm perms}\right)\left(L(ap)-\frac56\right)
  \nonumber\\
  & +\frac{1}{24\pi p^2}\left(\delta_{\mu\nu}p_\lambda p_\rho p_\tau
    +9\,\,{\rm perms}\right)
  +\frac{1}{4\pi (p^2)^2} p_\mu p_\nu p_\lambda p_\rho p_\tau
  \left(L(ap)-\frac{17}{6}\right)
  \nonumber\\
  & - (p\leftrightarrow q)\Bigr\}\Bigr]\,,
  \\
  & 
  \nonumber\\
  G_{7c\mu}(p,q) & = \frac{1}{12}K_{\nu\lambda\rho\tau}(0)\Bigl[
  t^2\left\{M_{\mu\nu\lambda\rho\tau}(p,-t)
    -p_\mu G_{5\nu\lambda\rho\tau}(p,-t)\right\}
  \nonumber\\ 
  & +\frac{t_\nu t_\lambda t_\rho t_\tau}{4\pi t^2}
  \frac{p_\mu}{p^2}\left\{L(at)+L(ap)-L(aq)\right\}
  \nonumber\\
  & +\frac{1}{96\pi}\left\{s_{\nu\lambda\rho\tau}p_\mu+4\,\,
    {\rm perms}\right\}
  \left\{L(ap)+\frac16\right\}
  \nonumber\\
  & -\frac{1}{24\pi p^2}\left\{\delta_{\mu\nu}p_\lambda p_\rho p_\tau
    +9\,\,{\rm perms}\right\}
  +\frac{p_\mu}{16\pi p^2}\left\{\delta_{\nu\lambda} p_\rho p_\tau
    +5\,\,{\rm perms}\right\}
  \nonumber\\
  & -\frac{1}{32\pi}s_{\nu\lambda\rho\tau}p_\mu
  \left\{L(ap)-\frac12\right\}
  +\frac{1}{12\pi}\frac{p_\mu p_\nu p_\lambda p_\rho p_\tau}{(p^2)^2}
  \nonumber\\
  & -(p\leftrightarrow q)\Bigr]\,,
\end{align}
where
\begin{align}
  M_{\mu\nu\rho\lambda\tau}(p,q) & = \frac{1}{4\pi(p+q)^2}\Bigl\{
  -\frac{1}{24}
  \left[s_{\mu\nu\rho\lambda}p_\tau+4\,\,{\rm perms}\right]{\cal V}_3(p,q)
  \nonumber\\
  & +I_{\mu\nu\rho\lambda\tau}(p){\cal V}_{10}(p,q)
  +\frac{p_\mu p_\nu p_\rho p_\lambda p_\tau}{(p^2)^2}
  \left\{{\cal V}_3(p,q)+L(a(p+q))\right\}
  \nonumber\\
  & -(p\leftrightarrow q)\Bigr\}\,,
\end{align}
where $I_{\mu\nu\rho\lambda\tau}(p)$ is the totally symmetric traceless
tensor:
\begin{multline}
  I_{\mu\nu\rho\lambda\tau}(p)=
  \frac{p_\mu p_\nu p_\rho p_\lambda p_\tau}{(p^2)^2}
  -\frac{1}{8p^2}\left[\delta_{\mu\nu}p_\rho p_\lambda p_\tau
    +9\,\,{\rm perms}\right] \\
  +\frac{1}{48}\left[s_{\mu\nu\rho\lambda}p_\tau+4\,\,{\rm perms}\right]\,.
\end{multline}
and
\begin{equation}
  \begin{split}
    & {\cal V}_{10}(p,q)=\frac{-1}{\left[p^2q^2-4(pq)^2\right]
      \left[(p^2q^2)^2-12p^2q^2(pq)^2+16(pq)^4\right]}
    \Bigl\{  \\
    & (q^2)^2\left[(p^2q^2)^2-2p^2(q^2)^2pq
      -12p^2q^2(pq)^2+8q^2(pq)^3+16(pq)^4\right]L(aq) \\
    & -p^2\Bigl[(p^2)^3(q^2)^2+2(p^2q^2)^2pq
    -8(p^2)^2q^2(pq)^2-24p^2q^2(pq)^3 \\
    & +16p^2(pq)^4+32(pq)^5\Bigr]L(ap) \\
    & +(p+q)^2\Bigl[(p^2q^2)^2\{p^2-q^2\}+2p^2(q^2)^3pq
    +8p^2q^2(pq)^2\{q^2-p^2\} \\
    & -8q^2(pq)^3(p^2+q^2)+16p^2(pq)^4\Bigr]L(a(p+q)) \\
    & -2pq\Bigl[(p^2)^3(q^2)^2+p^2(q^2)^2pq\{2q^2+5p^2\}
    +p^2q^2(pq)^2\{6q^2-8p^2\} \\
    & -8q^2(pq)^3\{q^2+5p^2\}-8(pq)^4\{q^2-2p^2\}+48(pq)^5\Bigr]\Bigr\}\,,
  \end{split}
\end{equation}
and
\begin{multline}
  G_{5\mu\nu\rho\lambda}(p,q)
  = \frac{1}{8\pi pq(p+q)^2}\Bigl[ \frac12 s_{\mu\nu\rho\lambda}pq L(a(p+q))
  \\
  -\left(\frac{p_\mu p_\nu p_\rho p_\lambda}{(p^2)^2}
    -\frac18 s_{\mu\nu\rho\lambda}\right){\cal V}_2(p,q)
  -\left(\frac{q_\mu q_\nu q_\rho q_\lambda}{(q^2)^2}
    -\frac18 s_{\mu\nu\rho\lambda}\right){\cal V}_2(q,p)
  \\
  +\left\{H_{\mu\nu\rho\lambda}(p)
    +H_{\mu\nu\rho\lambda}(q)\right\}{\cal V}_4(p,q)\Bigr]\,,
\end{multline}
where $H_{\mu\nu\rho\lambda}(p)$ is the totally symmetric traceless tensor:
\begin{equation}
  H_{\mu\nu\rho\lambda}(p)=\frac{p_\mu p_\nu p_\rho p_\lambda}{(p^2)^2}
  -\frac{1}{6p^2}\left\{p_\mu p_\nu \delta_{\rho\lambda}+5\,\,
    {\rm perms}\right\}
  +\frac{1}{24}s_{\mu\nu\rho\lambda}\,,
\end{equation}
and
\begin{multline}
  {\cal V}_4(p,q)  = -\frac{1}{[p^2q^2-2(pq)^2]}
  \Bigl[(p+q)^2\left\{pq(p^2+q^2)-p^2q^2\right\}L(a(p+q))
  \\
  + p^2\left\{p^2q^2-pq(p^2+2pq)\right\}L(ap)
  +q^2\left\{p^2q^2-pq(q^2+2pq)\right\}L(aq)
  \\
  + 2pq\left\{2p^2q^2-pq(p^2+q^2+4pq)\right\}\Bigr]\,.
\end{multline}

Writing
\begin{equation}
  \overline{T}_2=\overline{T}_2^{(a^0)}+a^2 \overline{T}_2^{(a^2)}+...
\end{equation}
we have for the scaling part
\begin{equation}
  \overline{T}_2^{(a^0)}(p,q,r)=T_2^\prime(p,q,r)
  +\frac{c_1}{4\pi}\Bigl\{\frac{(p+r)^2}{p^2q^2} 
  +\frac{r^2}{p^2(p+q)^2}+\frac{(p+q+r)^2}{q^2(p+q)^2}\Bigr\}\,,
\end{equation}
where $T_2^\prime$ is the corresponding continuum triangle 
integral defined by
\begin{multline}
  T_2^\prime(p,q,r)=\int_\infty\Bigl[\frac{(k-p-r)^2}{k^2(k-p)^2(k+q)^2}
  \\
  -\frac{(p+r)^2}{k^2p^2q^2}-\frac{r^2}{p^2(k-p)^2(p+q)^2}
  -\frac{(p+q+r)^2}{q^2(p+q)^2(k+q)^2}
  \\
  +\left\{\frac{(p+r)^2}{p^2q^2}+\frac{r^2}{p^2(p+q)^2}
    +\frac{(p+q+r)^2}{q^2(p+q)^2}\right\}\frac{1}{(k^2+a^{-2})}\Bigr]\,.
\end{multline} 

For the box diagram we first define
\begin{equation}
  \overline{B}(p,q,r)=\overline{B}_1(p,-q,p+r)\,.
\end{equation}
Then writing
\begin{equation}
  \overline{B}_1=\overline{B}_1^{(a^0)}+a^2 \overline{B}_1^{(a^2)}+...
\end{equation}
we have for the scaling part
\begin{multline}
  \overline{B}_1^{(a^0)}(p,q,r)=b'(p,q,r)+
  \frac{c_1}{4\pi}\Bigl\{\frac{1}{p^2q^2r^2}
  \\
  +\frac{1}{r^2(r-p)^2(q-r)^2}+\frac{1}{q^2(p-q)^2(q-r)^2}+
  \frac{1}{p^2(r-p)^2(p-q)^2}\Bigr\}\,,
\end{multline}
where $b'$ is the continuum integral 
\begin{multline}
  b'(p,q,r)=\int_\infty\Bigl\{\frac{1}{k^2(k-p)^2(k-q)^2(k-r)^2}
  -\frac{1}{p^2q^2r^2}\left[\frac{1}{k^2}-\frac{1}{k^2+a^{-2}}\right]
  \\
  -\left(\frac{1}{p^2(p-q)^2(p-r)^2}\left[\frac{1}{(k-p)^2}
      -\frac{1}{k^2+a^{-2}}\right]+2\,\,{\rm perms}\right)\Bigr\}\,.
\end{multline}
For the leading scale-breaking piece one gets
\begin{equation}
  \overline{B}_1^{(a^2)}(p,q,r)=
  \frac{-1}{2\pi}\left\{\kappa_1 \overline{B}_1^{(3)}(p,q,r)
    +\left(\kappa_2+\frac34\kappa_1\right)
    \overline{B}_1^{(6)}(p,q,r)\right\}\,,
\end{equation}
with
\begin{multline}
  \overline{B}_1^{(i)}(p,q,r)  = 
  {\cal B}^{(i)}(p,q,r)+{\cal B}^{(i)}(p,p-q,p-r)
  \\
  + {\cal B}^{(i)}(q,q-r,q-p)+{\cal B}^{(i)}(r,r-p,r-q)\,,
\end{multline}
where
\begin{multline}
  {\cal B}^{(3)}(p,q,r)=\frac{1}{(p-q)^2(q-r)^2(r-p)^2}
  \\
  \left\{(p-q)(p-r)\left[L(a(q-r))+c_1\right]+2\,\,{\rm perms}\right\}\,,
\end{multline}
and
\begin{multline}
  {\cal B}^{(6)}(p,q,r)
  =\frac{c_1}{2}\left\{\frac{\left[R(p)-\frac34\right]}{(p-q)^2(p-r)^2}
    +2\,\,{\rm perms}\right\}\
  \\
  +2\pi\int_\infty\Bigl\{\frac{R(k)-\frac34}{(k-p)^2(k-q)^2(k-r)^2}
  \\
  -\left(\frac{R(p)-\frac34}{(p-q)^2(p-r)^2}\left[\frac{1}{(k-p)^2}
      -\frac{1}{k^2+a^{-2}}\right]
    +2\,\,{\rm perms}\right)\Bigr\}\,.
\end{multline}

\appendix
\renewcommand{\thesection}{Appendix~C: 1-loop 4-point functions with   
  ${\cal O}_3$ or ${\cal O}_6$ insertions.}
\section{}
\renewcommand{\thesection}{C}

\begin{figure}[htb]
  \begin{center}
    \vskip 5mm
    \leavevmode
    \epsfxsize=3.0cm
    \epsfbox{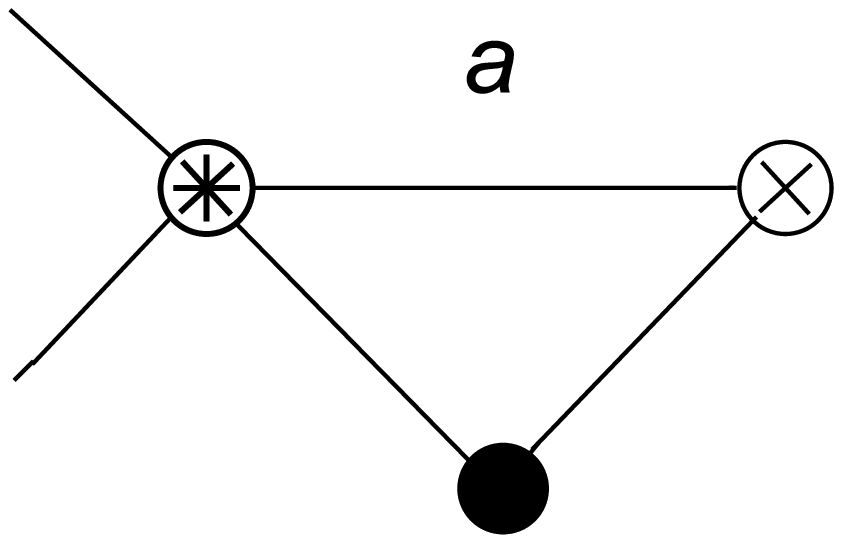}
    \hspace{5mm} 
    \epsfxsize=3.0cm
    \epsfbox{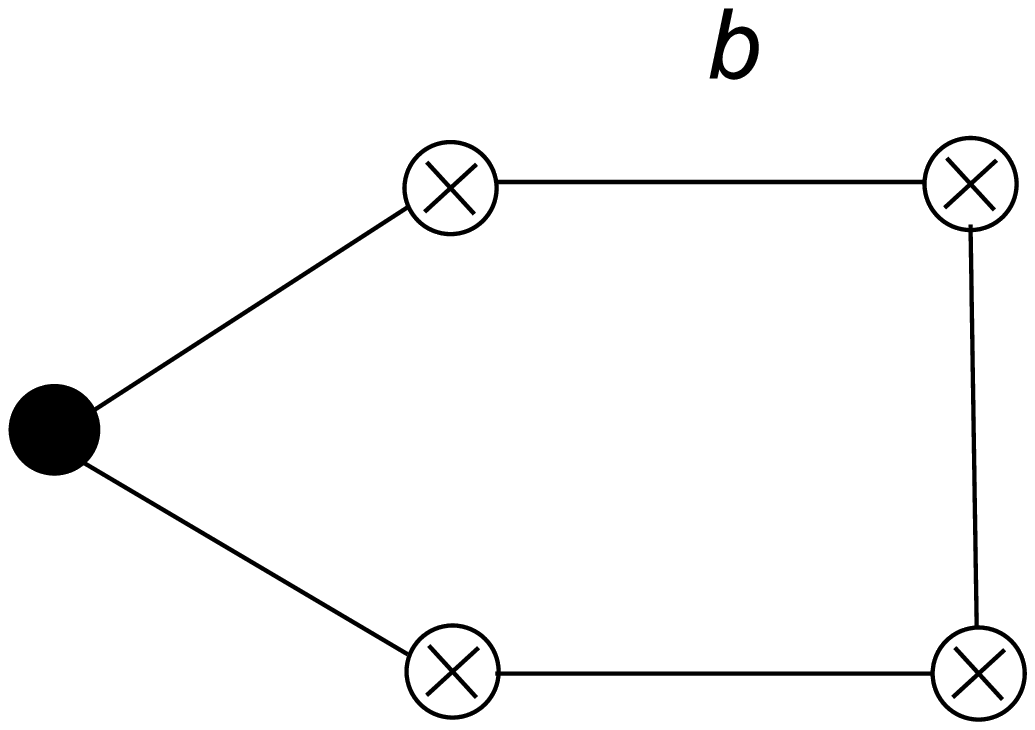}
    \hspace{5mm} 
    \epsfxsize=3.0cm
    \epsfbox{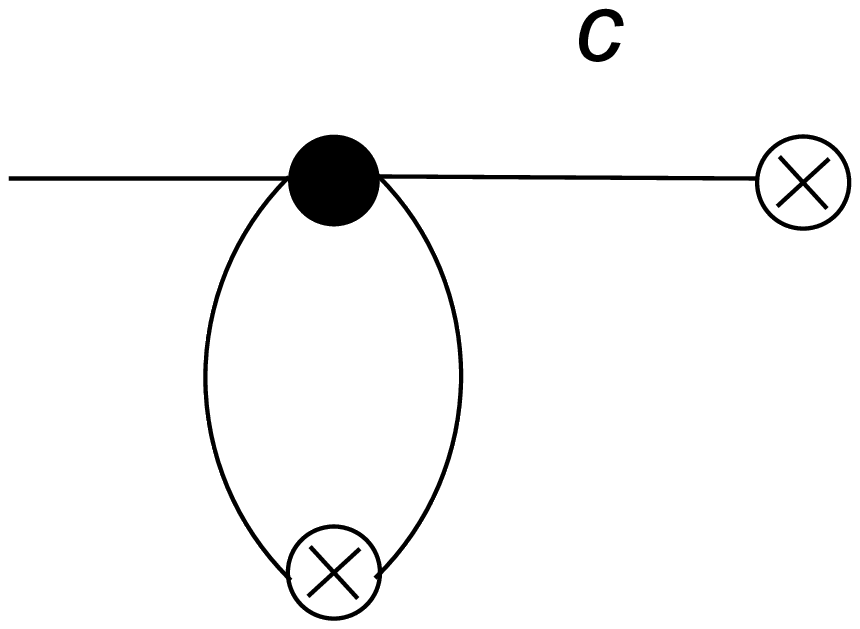}
    \vskip 2.0cm
    \epsfxsize=3.0cm
    \epsfbox{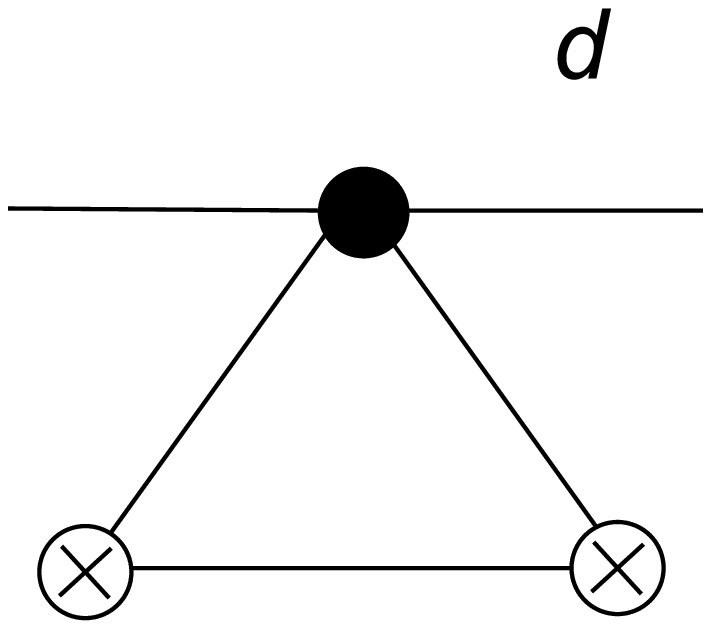}
    \hspace{5mm} 
    \epsfxsize=2.6cm
    \epsfbox{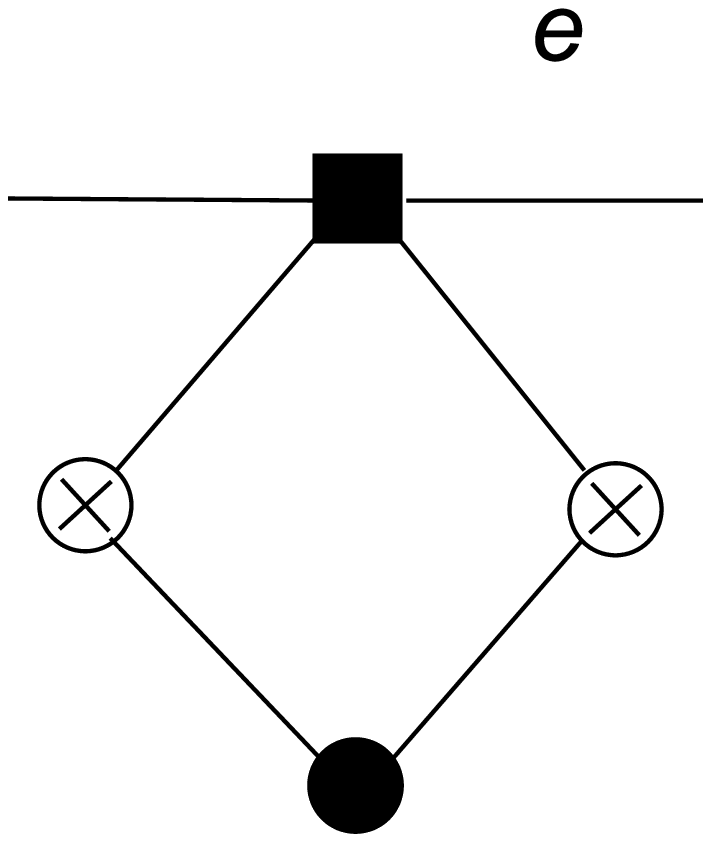}
    \hspace{5mm} 
    \epsfxsize=2.6cm
    \epsfbox{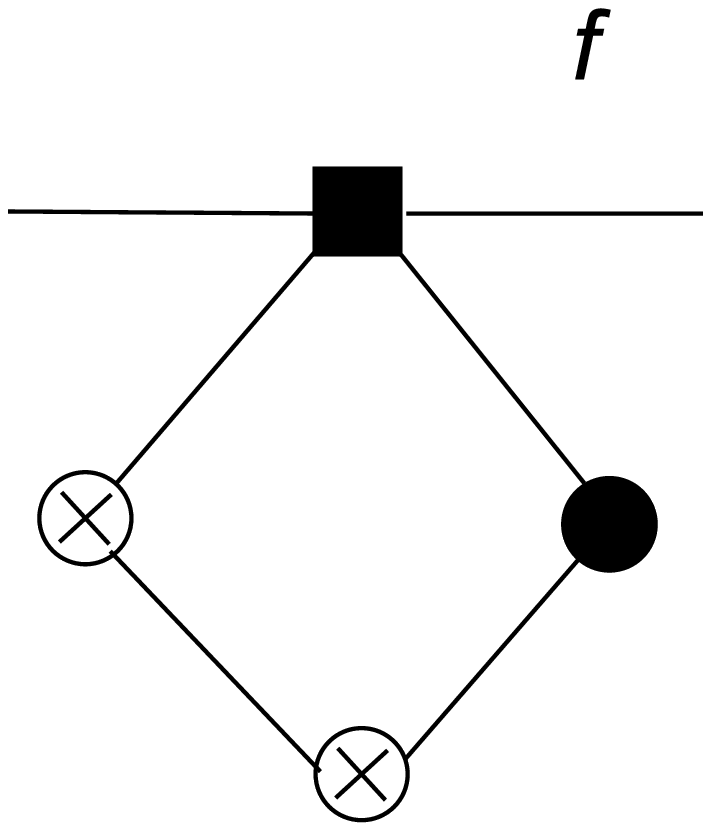}
    \vskip 2.0cm
    \epsfxsize=3.6cm
    \epsfbox{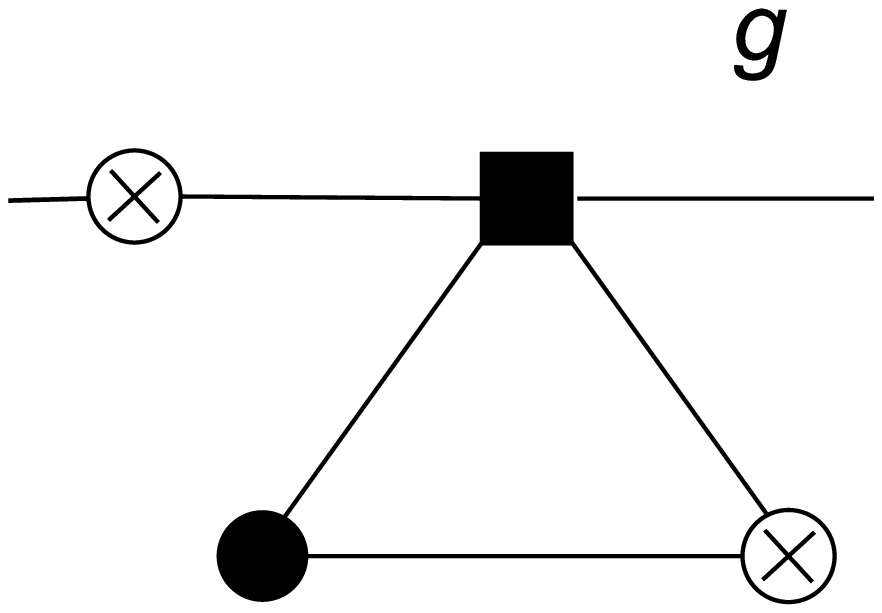}
    \hspace{5mm} 
    \epsfxsize=3.9cm
    \epsfbox{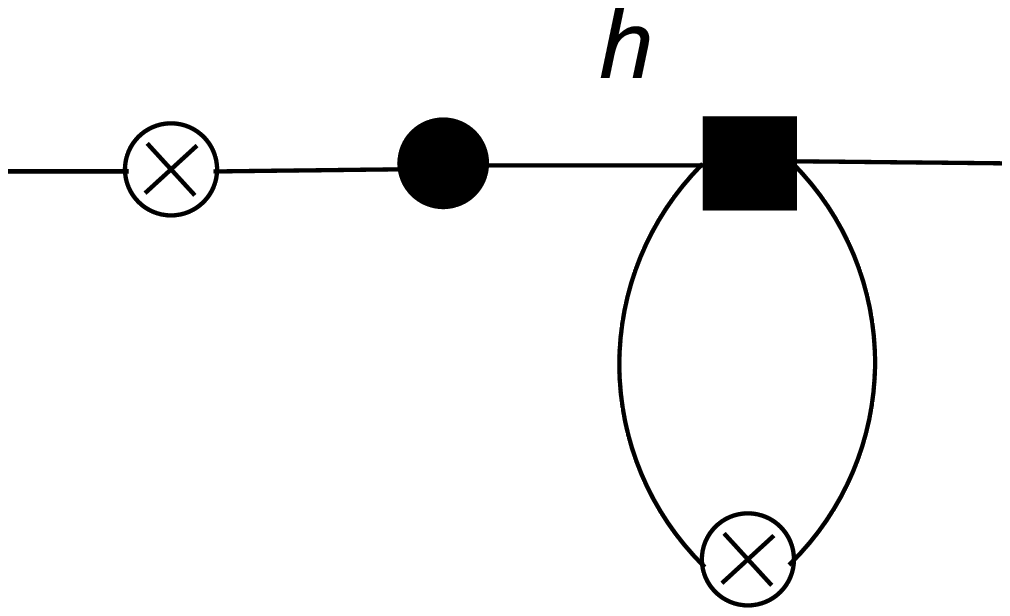}
    \hspace{5mm} 
    \epsfxsize=2.1cm
    \epsfbox{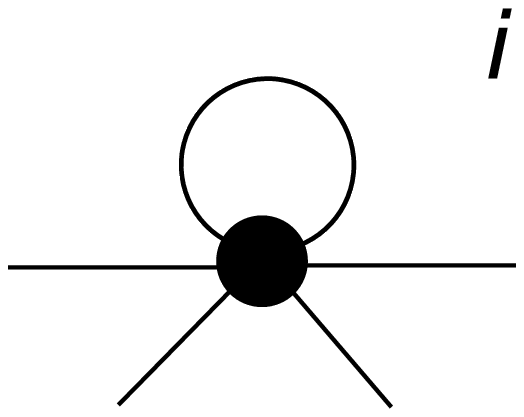}
    \vskip 1.6cm
    \hspace{5mm} 
    \epsfxsize=4.8cm
    \epsfbox{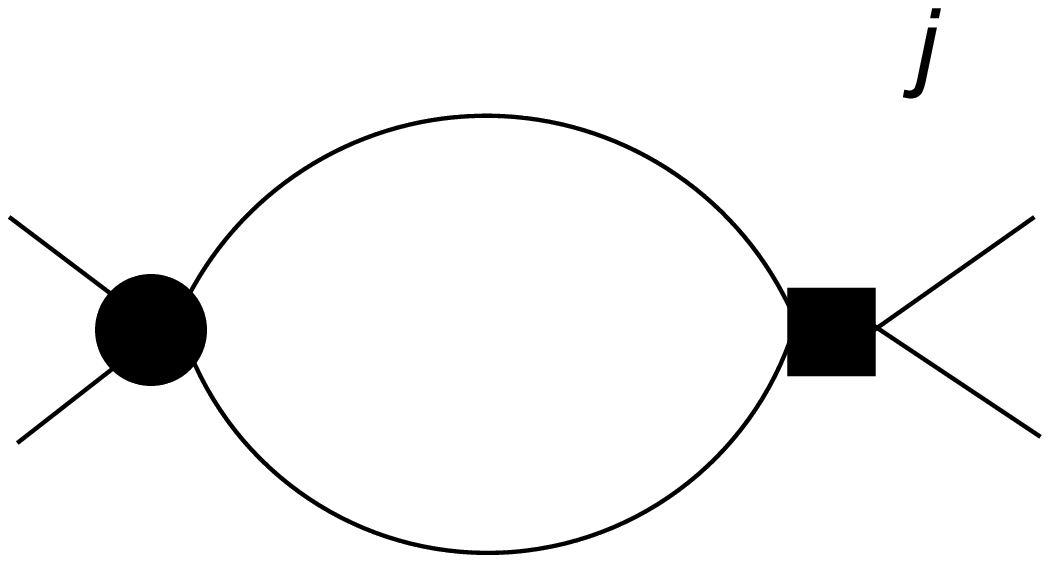}
    \hspace{5mm} 
    \epsfxsize=4.8cm
    \epsfbox{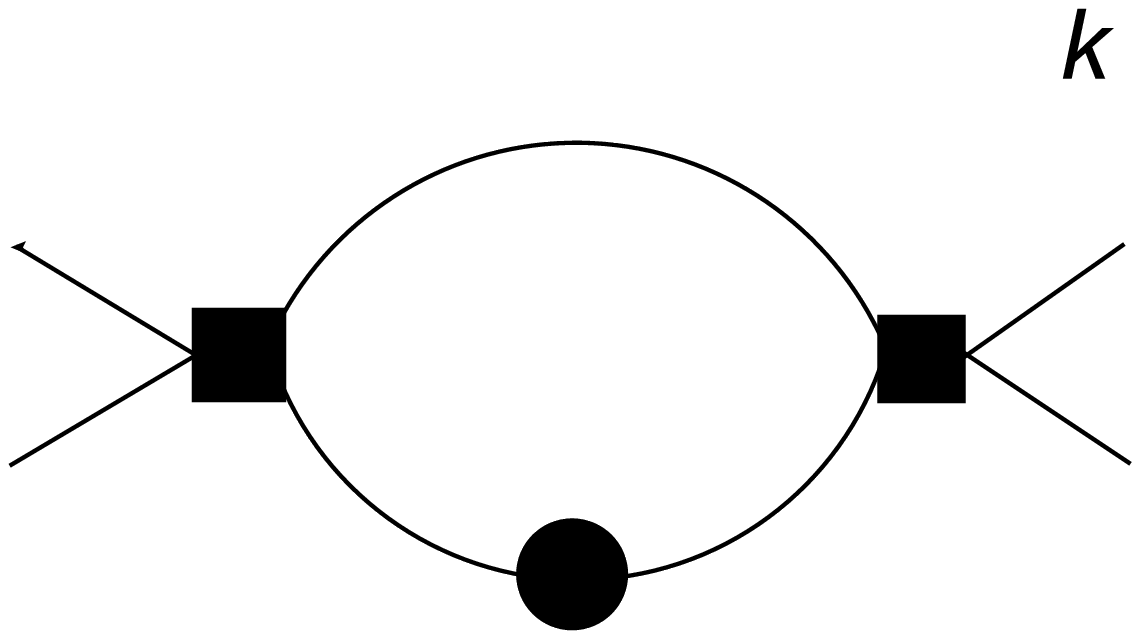}
  \end{center}
  \caption{{}
    The 11 irreducible contributions to the 1-loop 4-point function with
    insertion of the operator ${\cal O}_3$ or ${\cal O}_6$. Full circles
    represent the (2, 4 or 6-leg part) of the operators, full squares stand
    for the 4-point interaction vertex of the model. The crossed vertices
    represent the insertion of $\pi^2$, and the starred vertex of diagram
    $a$ represents the insertion of $(\pi^2)^2$.
  }
  \label{IrrGraphs}
\end{figure}

There are many contributions to the 1-loop 4-point functions with
insertion of the operators ${\cal O}_3$ or ${\cal O}_6$. Some
simple contributions are of the form of (vertex or propagator)
correction to the corresponding tree level 4-point functions with operator
insertion. An other category of simple contributions is where the operators 
are inserted on external lines of the ordinary 1-loop 4-point function 
of the model. (Some of the contributions belong to both of the above type.)
There remain 11 \lq\lq irreducible" contributions, which do not belong to
either of the above sets of simple graphs. They are shown in 
Fig.~\ref{IrrGraphs}. Each of these contributions is a complicated
function of the four momenta and since they are topologically distinct,
it is difficult to automate the calculation.

We note that it is sufficient to compute the 4-point functions with 
insertions for special momenta $p_1+q_1=0=p_2+q_2$. 
Since some diagrams are singular for this configuration
one computes with $m\ne0$ and then takes the limit $m\to0$ 
(which is non-singular) in the final result. 
We checked that the limits $m\to0$ and the limit to the special 
momenta configurations commute, a fact that was (and still is) 
to us not obviously the case. 

\clearpage

\appendix
\renewcommand{\thesection}{Appendix~D: solution of the RG equations 
  (\ref{Uij})}
\section{}
\renewcommand{\thesection}{D}

In this Appendix we will analyse the solution of the RG equations (\ref{Uij})
in detail. It is well-known that if there are also integer numbers in the 
set of differences\break
$\{\Delta_i-\Delta_j\}$ ($i\not=j$) then in general terms 
proportional to powers of the logarithm of the renormalized coupling may 
occur in the solution. We will show that nevertheless all coefficients 
we need in our analysis here are actually free of these terms.

Our starting point is the expansion (\ref{Qij}), where the coefficients
$k^{(\ell)}_{ij}$ may logarithmically depend on the coupling through
\begin{equation}
  \omega=\ln\left(2\beta_0g^2\right)\,.
\end{equation}
In fact, the coefficients are finite polynomials in $\omega$:
\begin{equation}
  k^{(\ell)}_{ij}(\omega)=\sum_{\sigma=0}^\kappa\,k^{(\ell,\sigma)}_{ij}
  \omega^\sigma,
\end{equation}
where the order $\kappa$ is not larger than the largest integer in the
set $\left\{\Delta_i-\Delta_j\right\}_{i,j=1}^7$\footnote{Note that
  this maximal power in $\omega$ first occurs at higher loop orders. 
  The actual power of $\omega$ is always smaller than $\ell$, the loop order.}.
Substituting the ansatz (\ref{Qij}) into (\ref{Uij}) gives the following set
of algebraic equations for the numerical coefficients:
\begin{equation}
  \begin{split}
    \left(\Delta_j-\Delta_i-\ell+1\right)&k^{(\ell,\sigma)}_{ij}-(\sigma+1)
    k^{(\ell,\sigma+1)}_{ij}\\
    &=\frac{1}{2}\delta_{\sigma,0}\,\rho^{(\ell)}_{ij}+
    \frac{1}{2}\sum_{m=2}^{\ell-1}\,\rho^{(\ell+1-m)}_{is}\,
    k^{(m,\sigma)}_{sj}
  \end{split}
\end{equation}
for $\ell=2,3,\dots$ and $\sigma=0,1,\dots,\kappa$. 
($k^{(\ell,\kappa+1)}_{ij}=0$ by convention.)

These equations can be solved order by order in the loop expansion. 
More precisely, if we already found the solution up to $\ell-1$ loop order
then in the case of $\Delta_j-\Delta_i\not=\ell-1$ we can find a unique
solution for all the $\ell$-loop coefficients $k^{(\ell,\sigma)}_{ij}$.
If $\Delta_j-\Delta_i=\ell-1$ then we can still solve for all 
$k^{(\ell,\sigma)}_{ij}$ but $k^{(\ell,0)}_{ij}$, which is not occurring
in the equations in this case. It remains arbitrary: we will make
the solution unique by putting it equal to zero in this case.

We get for the two-loop coefficients
\begin{equation}
  \Delta_j-\Delta_i\not=1\,:
  \hspace{-2.4cm}
  \begin{split}
    k^{(2,1)}_{ij}&=0\,,\\
    k^{(2,0)}_{ij}&=\frac{1}{2(\Delta_j-\Delta_i-1)}\rho^{(2)}_{ij}
  \end{split}
\end{equation}
and
\begin{equation}
  \Delta_j-\Delta_i=1\,:
  \hspace{-2.4cm}
  \begin{split}
    k^{(2,0)}_{ij}&=0\,,\\
    k^{(2,1)}_{ij}&=-\frac{1}{2}\rho^{(2)}_{ij}\,.
  \end{split}
\end{equation}
The three-loop coefficients are
\begin{equation}
  \Delta_j-\Delta_i\not=2\,:
  \hspace{-2.4cm}
  \begin{split}
    k^{(3,2)}_{ij}&=0\,,\\
    k^{(3,1)}_{ij}&=\frac{1}{2(\Delta_j-\Delta_i-2)}\rho^{(2)}_{is}
    k^{(2,1)}_{sj}\,,
  \end{split}
\end{equation}
\begin{equation}
  k^{(3,0)}_{ij}=\frac{1}{(\Delta_j-\Delta_i-2)}\left\{
    k^{(3,1)}_{ij}+\frac{1}{2}\rho^{(3)}_{ij}+\frac{1}{2}
    \rho^{(2)}_{is}k^{(2,0)}_{sj}\right\}
\end{equation}
and
\begin{equation}
  \Delta_j-\Delta_i=2\,:
  \hspace{-2.4cm}
  \begin{split}
    k^{(3,0)}_{ij}&=0\,,\\
    k^{(3,1)}_{ij}&=-\frac{1}{2}\rho^{(3)}_{ij}-\frac{1}{2}
    \rho^{(2)}_{is}k^{(2,0)}_{sj}\,,\\
    k^{(3,2)}_{ij}&=-\frac{1}{4}
    \rho^{(2)}_{is}k^{(2,1)}_{sj}\,.
  \end{split}
\end{equation}

Now the general coefficients $v^{(\ell)}_i$ in (\ref{vig}) depend on $\omega$.
This dependence is inherited from the $\omega$-dependence of the
coefficients $k^{(\ell)}_{s1}$ in (\ref{expcoffs}). However, the coefficients
we actually need (the 2-loop coefficients for general $n$ and also the
3-loop coefficients for $n=3$) are actually free of this dependence.
We can see this by inspecting the above formulae and remembering the 
triangular structure of the anomalous dimension matrix (and its expansion 
coefficients). The needed terms are
\begin{equation}
  k^{(2)}_{s1}(\omega)=k^{(2,0)}_{s1}=\frac{1}{2(\Delta_1-1-\Delta_s)}
  \rho^{(2)}_{s1}\,,
\end{equation}
which is different from zero for $s=1,2$ only and for the
case of $n=3$ (where we need it)
\begin{equation}
  k^{(3)}_{s1}(\omega)=k^{(3,0)}_{s1}=\frac{1}{1-\Delta_s}
  \left\{\frac{1}{2}\rho^{(3)}_{s1}+\frac{1}{2}\sum_{r=1}^2
    \rho^{(2)}_{sr}k^{(2,0)}_{r1}\right\}\,,
\end{equation}
different from zero only for $s=1,2$.

\appendix
\renewcommand{\thesection}{Appendix~E:
  The 1d improved estimator} 
\section{}
\renewcommand{\thesection}{E}

It is well known that improved estimators can decrease the statistical
error significantly. 
Hasenbusch \cite{Hasenbusch} proposed an improved estimator for the strip 
geometry with free boundary conditions in the time direction, which reduces
the statistical error considerably 
for the case when $\xi(L) \gtrsim L$, i.e. $u=m(L)L \lesssim 1$.
It uses the fact that the situation is nearly one-dimensional,
i.e. the spins within a time-slice are strongly correlated.

We give here a short overview of the method and 
add a useful modification used in these simulations.

Consider a new configuration $S'$ where the old spins $S(x,t')$ 
for $t' > t$ are rotated globally, by the same O($n$) matrix $X(t)$, 
while the spins on the time slices $t' \le t$ are unchanged. 
Performing such rotations for all time slices, $1 \le t \le N_t-1$, 
the resulting spin configuration becomes
\begin{equation}
  S'(x,t) = X(1) X(2) \cdots X(t-1) S(x,t) \,.
\end{equation}
The new spin configuration $S'$ can be described 
(equivalently but redundantly)
by the original spins $S$ and the rotation matrices $X$.

Consider the MC updates of only the $X$ variables, with fixed $S$ variables.
For the standard O($n$) spin action the action governing the dynamics 
of $X(t)$ degrees of freedom is
\begin{equation} \label{TrXQ}
  {\cal A}_X(X | S) = 
  - \sum_{t=1}^{N_t-1} \mathrm{tr}\left( X(t) Q^T(t) \right) \,,
\end{equation}
where the $n\times n$ matrix $Q(t)$ is given by the spins of the corresponding
two neighboring time slices,
\begin{equation} \label{QQij}
  Q_{ij}(t) = \beta \sum_x S_i(x,t) S_j(x,t+1) \,.
\end{equation}
(The interaction terms within a given time slice are O($n$) invariant,
hence do not depend on $X$ and are not written out in \eqref{TrXQ}.)
The improved estimator introduced by Hasenbusch
relies on the fact that the variables $X(t)$ are independent.
Obviously, 
\begin{multline} \label{impest}
  \langle S'(x,t_0) S'(y,t_1)\rangle_X 
  = \langle S(x,t_0) X(t_0)\cdots X(t_1-1) S(y,t_1)\rangle_X  \\
  = S(x,t_0)  \langle X(t_0)\rangle_X \cdots \langle X(t_1-1)\rangle_X  
  S(y,t_1) \,,
\end{multline}
where $\langle \ldots \rangle_X$ denotes the average over the $X(t)$
variables, each with its own Boltzmann factor 
\begin{equation} \label{weight}
  P(X(t)) \propto \exp( \mathrm{tr}(X(t)Q^T(t))) \,.
\end{equation}

Note that updating only the $X$ variables is not 
ergodic\footnote{with the exception of the 1d case}
-- the relative orientations of the spins within a given time slice
remain unchanged -- hence the procedure has to be supplemented by
other updates, e.g. the cluster algorithm. 
One does not even need to actually update the spin configuration
in this way -- the main advantage of the procedure is to yield
a very good improved estimator:
for $Lm(L) \lesssim 1$ it suppresses the statistical errors
considerably better than the cluster improved estimator \cite{Hasenbusch}.

The improved estimator \eqref{impest} boils down 
to calculate (or measure) the expectation values $\langle X(t) \rangle$
with the weight \eqref{weight},
\begin{equation} \label{Xav}
  \langle X \rangle = 
  \frac{1}{Z} \int \rmd X {\rm e}^{\mathrm{tr}(X Q^T )} X \,,
\end{equation}
where $\rmd X$ is the Haar measure over SO($n$).
It is useful to make a singular value decomposition
\begin{equation} \label{svd}
  Q = U q V\,,
\end{equation}
where $U,V \in \mathrm{SO}(n)$ and $q$ is a diagonal 
matrix\footnote{{}Standard SVD routines return a {\em non-negative}
  diagonal matrix and two O($n$) matrices. One has to make sure
  that in this decomposition $\det U = \det V = 1$ by changing properly
  the signs in $q$ if needed.}.
Introducing the new integration variable $Y$ by  $X=U Y V$
one gets
\begin{equation} \label{inty}
  \langle X \rangle = U \langle Y \rangle V \,,
\end{equation}
where
\begin{equation} \label{Yav}
  \langle Y \rangle = \int \rmd Y {\rm e}^{\mathrm{tr}( Y q )} Y \,,
\end{equation}
For $n \ge 3$ this integral is not known analytically,
and Hasenbusch \cite{Hasenbusch}
proposed to measure it stochastically, by an unbiased estimator.

At this point we slightly modify the procedure proposed in 
\cite{Hasenbusch}.
By making a MC update on $X$, one can start with
the value $X(t)=1$, since this corresponds to the original
spin configuration  $S'=S$ which was assumed to be an equilibrium
configuration. 
Making any number of updates (not necessarily many) starting from $X(t)=1$
one gets a new configuration $S'$ with the proper weight. 
The starting value $X_0=1$ corresponds to
\begin{equation} \label{Y0}
  Y_0 \equiv U^T V^T \,.
\end{equation}
On the other hand, $\mathrm{tr}(Yq)$ depends only on the diagonal elements 
of $Y$, hence the off diagonal elements in $\langle Y \rangle$
can be replaced by zero keeping the diagonal elements 
unchanged.\footnote{{}by averaging over properly
  chosen matrices Y obtained by diagonal SO($n$) transformations 
  of the form $[-1,-1,1,\ldots,1]$}
Therefore one can make the replacement
\begin{equation} \label{Xav1}
  \langle X \rangle \to U \left( \overline{Y} \right)_\mathrm{diag} V \,,
\end{equation}
where $\overline{Y}$ is an average over the given number of 
actually performed updates and 
$\left( ... \right)_\mathrm{diag}$ denotes the procedure 
of replacing the off-diagonal elements by zero.
In particular, it is possible to avoid any actual updates, just
use the estimator
\begin{equation} \label{Xav2}
  \langle X \rangle \to U \left( U^T V^T\right)_\mathrm{diag} V \,.
\end{equation}
It turns out that this trick alone reduces considerably 
the largest eigenvalue of $\langle X \rangle $ and 
thus the overall stochastic error.

The 1d improved estimator works better as $m(L)L$ decreases,
i.e. as the system becomes effectively more 1-dimensional.
We have compared the errors for different estimators in 
some of our runs at $u_0 \approx 1$:

\noindent\quad 1: 1d improved estimator, exact integration,

\noindent\quad 2: 1d improved estimator using eq.~\eqref{Xav2} with no updates,

\noindent\quad 3: 1d improved estimator with 100 updates for 
$\langle X \rangle$,

\noindent\quad 4: cluster improved estimator,

\noindent\quad 5: standard estimator.

\noindent The ratios of these errors were: 1 : 1.1 : 2.7 : 4 : 11.
This shows that the simple procedure described above is quite efficient.


\eject

\end{document}